\documentclass[reprint,
showpacs,  
aps,pra,
floatfix,superscriptaddress]{revtex4-1}
\usepackage{amsmath,amsfonts,amsthm}
\usepackage{dsfont}
\usepackage{braket}
\usepackage{graphicx}
\usepackage{epstopdf}
\usepackage{dcolumn}
\usepackage{bm}
\usepackage{float}
\usepackage[caption=false]{subfig}
\usepackage{color}
\usepackage{hyperref}

\begin{document}

\preprint{APS/123-QED}

\title{Topologically induced prescrambling and dynamical detection of topological phase transitions at infinite temperature}

\author{Ceren B. Da\u{g}}
\email{cbdag@umich.edu}
\affiliation{Department of Physics, University of Michigan, Ann Arbor, Michigan 48109, USA}
\author{L.-M. Duan}
\affiliation{Center for Quantum Information, IIIS, Tsinghua University, Beijing 100084, PR China}
\author{Kai Sun}
\affiliation{Department of Physics, University of Michigan, Ann Arbor, Michigan 48109, USA}
\date{\today}

\begin{abstract}
We report a numerical observation where the infinite-temperature out-of-time-order correlators (OTOCs) directly probe quantum phase transitions at zero temperature, 
in contrast to common intuition where low energy quantum effects are washed away by strong thermal fluctuations at high temperature. By comparing numerical simulations with exact analytic results, we determine that this phenomenon has a topological origin and is highly generic, 
as long as the underlying system can be mapped to a 1D Majorana chain. Using the Majorana basis, we show that the infinite-temperature OTOCs probe zero-temperature quantum phases via detecting
the presence of Majorana zero modes at the ends of the chain that is associated with 1D $Z_2$ topological order. Hence, we show that \emph{strong zero modes} also affect OTOCs and scrambling dynamics. Our results demonstrate an intriguing interplay between information scrambling and topological order, which leads to a new phenomenon in the scrambling of generic nonintegrable models: topological order induced prescrambling, paralleling the notion of prethermalization of two-time correlators, that defines a time-scale for the restricted scrambling of topologically-protected quantum information.

\end{abstract}

\pacs{}
\maketitle

%\tableofcontents
%%%%%%%%%%%%%%%%%%%%%%%%%%%%%%%%%%%%%%%%%%%%%%%%%%%%%%%%%%%%%%%%%%%%%%%%%%%
\section{Introduction \label{Intro}}

Out-of-time-order correlators (OTOCs) have become a widely-appreciated tool to measure the correlation build-up in space and time, and hence quantitatively characterize information scrambling in interacting many-body systems ~\cite{2008JHEP...10..065S,2013JHEP...04..022L,2017PhRvB..95f0201S,2017JHEP...10..138H,articleRey}. Started off as a theoretical tool to understand quantum information in a black hole \cite{2008JHEP...10..065S,2014JHEP...03..067S} its impact quickly expanded to a wide variety of subjects including but not limited to: quantum chaos \cite{Maldacena2016,PhysRevX.7.031011,doi:10.1002/andp.201600332,2017NJPh...19f3001B,2018arXiv180200801X},  many-body localization~\cite{2017PhRvB..95f0201S,doi:10.1002/andp.201600332,doi:10.1002/andp.201600318,2017PhRvB..95e4201H,PhysRevA.99.052322}, quantum integrability \cite{PhysRevX.7.031011,PhysRevB.97.144304,PhysRevA.97.042330,2018arXiv180200801X}, quantum criticality \cite{PhysRevB.96.054503} and recently symmetry-breaking quantum phase transitions \cite{PhysRevLett.121.016801,PhysRevLett.123.140602}. 

At temperature $T=1/\beta$, an OTOC is defined as,
\begin{eqnarray}
F(t) &=& \text{Tr}\left(e^{-\beta H} W^{\dag}(t) V^{\dag} W(t) V \right),
\label{OTOCEq}
\end{eqnarray}
where $W$ and $V$ are local quantum operators and $H$ is the Hamiltonian. At infinite temperature ($T=\infty$ and $\beta=0$),  the Boltzmann weight $e^{-\beta H}$ becomes an identity operator and thus the OTOC reads
\begin{eqnarray}
F(t)&=& \frac{1}{M}\sum_{n=1}^M\left\langle \psi_{n}| W^{\dag}(t) V^{\dag} W(t) V |\psi_{n}\right\rangle, \notag \\
&\approx &  \left\langle \psi_{h}| W^{\dag}(t) V^{\dag} W(t) V |\psi_{h}\right\rangle.
\label{OTOCEq2}
\end{eqnarray}
Here we sum over a complete basis of the Hilbert space of dimension $M$, while in the second line, we use a random state $\Ket{\psi_{h}} $ drawn from the Haar measure \cite{PhysRevB.96.020406,PhysRevA.99.052322} to approximate an infinite-temperature state in a correlation function, e.g. Eq.~\eqref{OTOCEq} \cite{PhysRevLett.96.050403,popescu2006entanglement,PhysRevLett.99.160404,PhysRevLett.108.240401,2017AnP...52900350L}. 
\begin{figure}
\centerline{\includegraphics[width=0.35\textwidth]{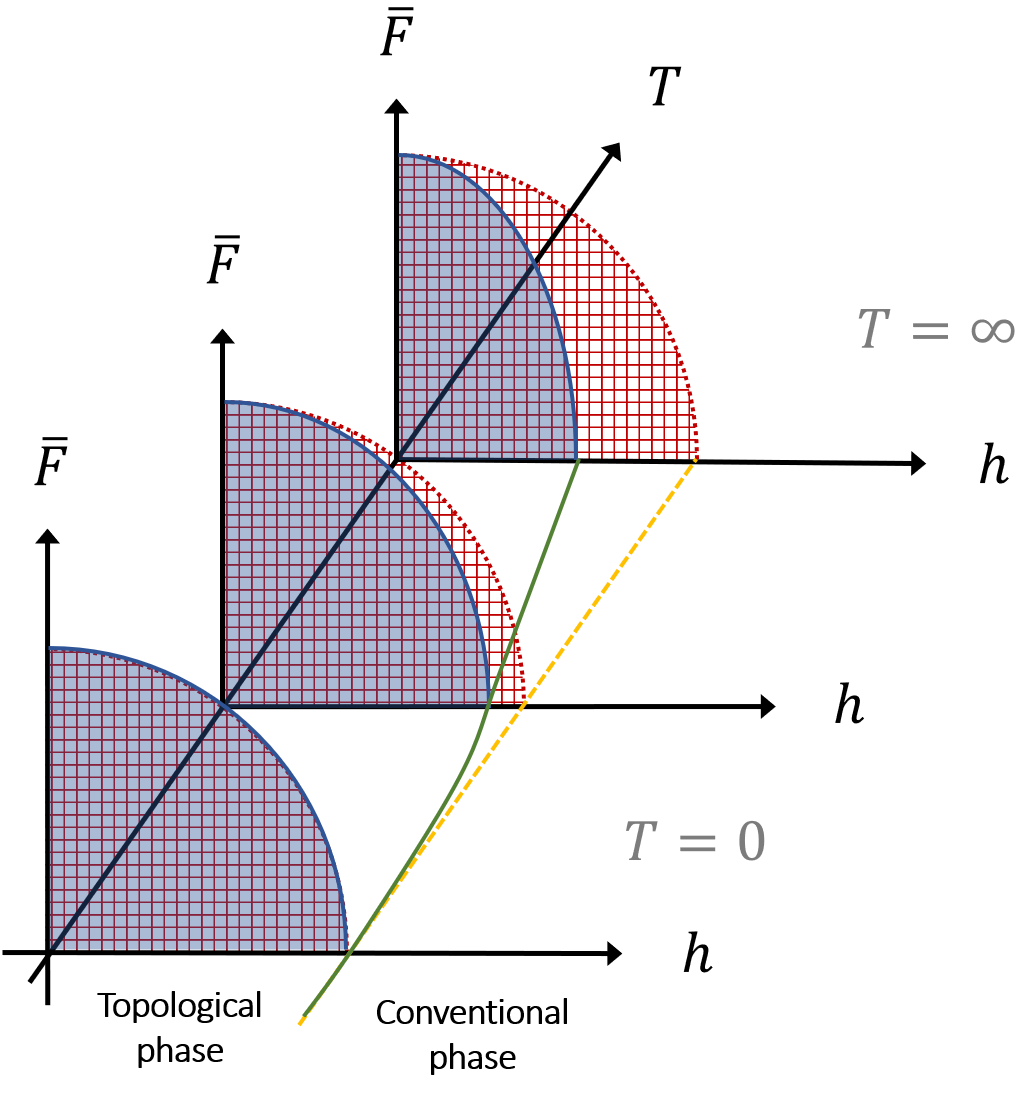}}
\caption{The schematic of dynamic phase boundaries determined by OTOC time-average $\bar{F}$ with respect to control parameter $h$ and temperature $T$. The system experiences a topological phase transition (TPT) defined at $T=0$ temperature from $Z_2$ topologically ordered phase to a trivial phase. The graphics with red-grids and solid-blue show how the topological phase survives in dynamics and at higher temperatures for integrable and generic nonintegrable models, respectively. While integrable models recover zero-temperature phase boundary at infinite temperature, nonintegrable models experience a shift that tends to destroy order quicker than at low temperature.}
\label{Fig0a}
\end{figure}

The OTOC of a generic system is expected to decay to zero fast where the rate of decay carries information on the chaotic properties of the system; and saturate at zero in long time dynamics. Saturation at zero indicates that the system scrambles information completely, whereas a finite saturation value points to a restricted scrambling \cite{Swingle:2018ekw}. In this manuscript, we focus on the regime starting shortly after the (initial) decay of OTOC and lasts for a time interval of $\mathcal{T}$. It has been recently found that the OTOC saturation value at zero-temperature exhibits order parameter-like behavior, and thus can directly probe the long-range quantum order 
and quantum phase transitions \cite{PhysRevLett.123.140602}. In contrast to the naive intuition, where thermal fluctuations wash away low energy quantum effects at high temperature, in this work we observe an emergent relation between \emph{infinite-temperature information scrambling} and \emph{zero-temperature $Z_2$ topological order} in the bulk in multiple model systems, e.g. non-interacting, interacting and/or nonintegrable. The effect is robust where the qualitative features remain invariant regardless of microscopic details, e.g. integrability and symmetries. In particular, by setting $W$ and $V$ as local degrees of freedom localized near the edge of the system, we find that the time-average of OTOC $\bar{F}=1/\mathcal{T} \int dt F(t)$ (or equivalently the saturation value, if the OTOC saturates) behaves like an order parameter (Fig.~\ref{Fig0a}). It is worthwhile to emphasize that the infinite temperature OTOCs are effective tools for detecting chaos that is based on the entire energy spectrum \cite{Maldacena2016,PhysRevX.7.031011,doi:10.1002/andp.201600332,2017NJPh...19f3001B,2018arXiv180200801X,PhysRevA.99.052322}. Hence it is surprising and highly not obvious that this correlator can also directly probe zero temperature physics of the ground state, such as quantum phase transitions. \emph{Then what is the underlying physics that allows the infinite temperature out-of-time-order correlator at the edge to accurately sense the bulk ground state physics and capture the bulk phase transition? Is this a generic feature?} 

\begin{figure}
\centerline{\includegraphics[width=0.4\textwidth]{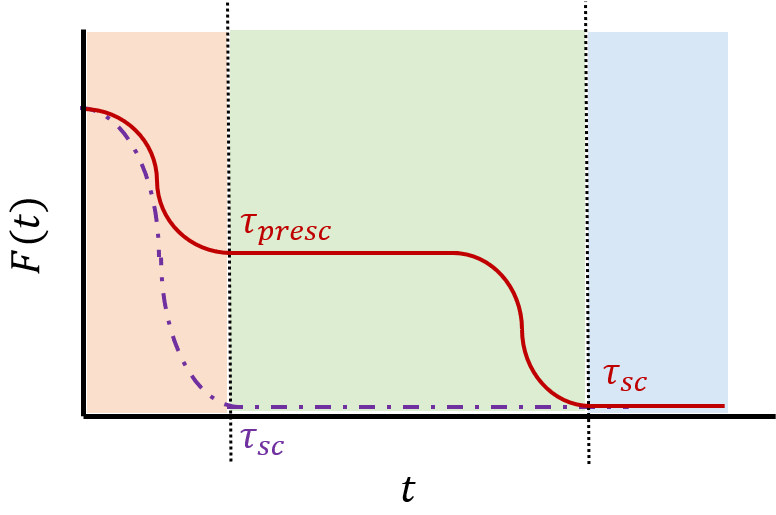}}
\caption{The schematic of infinite-temperature OTOC evolving in time $t$ for a quantum system with (solid-red line) and without (dotted-purple line) $Z_2$ topological order. A generic system with $Z_2$ topological order would exhibit topologically induced prescrambling $\bar{F} \neq 0$ before fully scrambles at scrambling time $\tau_{sc}$. We coin $\tau_{presc}$ for the prescrambling time-scale. Our study focuses on this prescrambling plateau (green panel), where the OTOC time-average exhibits order parameter like behavior (Fig.~\ref{Fig0a}).}
\label{Fig0}
\end{figure}

Through a careful analysis, we find that this connection arises universally as long as the quantum system can be mapped to a Majorana chain (1D superconductor) \cite{2001PhyU...44..131K}, and $\bar{F}$ value of edge operators serves as the $Z_2$ topological order parameter. It is known that $Z_2$ topological order results in a two-fold degeneracy for all energy eigenstates of the entire spectrum; and recently it is pointed out that this degeneracy structure of $Z_2$ topological order has a highly nontrivial impact on dynamics at any temperature, e.g. long coherence times for edge spins in Ref.~\cite{2017JSMTE..06.3105K} while the zero modes surviving in the dynamics is dubbed as \emph{strong zero modes}, and pre-thermalization effect in Ref.~\cite{PhysRevX.7.041062}. Our results extend this impact of $Z_2$ topological order to information scrambling and OTOCs, opens up new avenues to dynamically detect and study topological order through utilizing information scrambling as an order parameter. Paralleling the well-known prethermalization effect appearing in simpler correlators \cite{PhysRevLett.93.142002,2018JPhB...51k2001M,PhysRevX.7.041062}, we find that a new time-scale appears in information scrambling when $Z_2$ topological order \cite{RevModPhys.89.041004} exists. We name this phenomenon topologically induced prescrambling and hence define the time-scale as prescrambling time. Fig.~\ref{Fig0} shows a cartoon picture of prescrambling for a generic (nonintegrable) model with solid-red line where the system experiences restricted scrambling, $\bar{F} \neq 0$, forming a plateau at $\tau_{presc}$ for a period of time $\mathcal{T}$ after the first OTOC decay and preceding the full scrambling at $\tau_{sc}$ in a topological phase. On the other hand, the purple-dotted line in Fig.~\ref{Fig0} shows the expected rapid OTOC decay until scrambling time $\tau_{sc}$ for a generic system with no topological order. Prescrambling (green panel) plateau in Fig.~\ref{Fig0} survives at infinite-time in thermodynamic limit for systems with extensive number of symmetries, e.g. non-interacting and/or integrable limits, with no full scrambling occurring. Such systems might demonstrate $\bar{F} \neq 0$ in their trivial phases \cite{doi:10.1002/andp.201600332,FAN2017707,PhysRevA.99.052322}, nevertheless it is still possible to mark down the topological phase transition due to sharp transition signatures. We compare the infinite-temperature dynamic phase boundary with zero-temperature quantum phase boundary where topological order starts to develop in Fig.~\ref{Fig0a} and observe that they perfectly coincide with each other in integrable systems. Away from the integrability, the dynamical phase boundary significantly shifts away from the zero-temperature phase boundary, although the qualitative trend of $\bar{F}$ survives. 

%Previously, the OTOC of the integrable Ising model \cite{PhysRevB.97.144304} and XX-chain \cite{2019arXiv190109327B} are analytically studied, which required periodic boundary conditions \cite{sachdev2001quantum} and hence double-OTOCs \cite{PhysRevB.97.144304,2019arXiv190109327B}. Our study indicates that OTOCs of edge operators contain new information, which is not accessible by the bulk operators. 

The dynamical detection of topological order has been under intensive investigation \cite{PhysRevB.95.041105,2017JSMTE..06.3105K,PhysRevX.7.041062,PhysRevB.97.235134,PhysRevE.98.042128}. Furthermore, the topological insulators and superconductors have been studied \cite{PhysRevLett.115.236403,PhysRevA.96.023601,PhysRevLett.117.126803,2018NatPh..14..265F,2015NatCo...6E8336D} and classified \cite{PhysRevB.99.075148} according to their non-equilibrium dynamics rather in an analogy to the classification tables for topological states of matter \cite{doi:10.1063/1.3149495} superposed with the notion of dynamical quantum phase transitions \cite{2013PhRvL.110m5704H,PhysRevB.93.085416,PhysRevB.92.075114}. Thus, understanding if the information scrambling has fundamental restrictions when topological order exists is a puzzle left at the intersection of many sub-fields. 

In Sec.~II, we are going to detail our numerical observation around its corresponding Majorana chain and discuss about the connection between infinite temperature scrambling and $T=0$ topological order with quantitative arguments. Later in Sec.~III, we are going to show how the topological order is encoded in the saturation regime of OTOCs based on the analytical calculations in the non-interacting regime. In Sec.~IV, we extend the discussion to interacting and/or nonintegrable models and demonstrate topologically induced prescrambling. Later we show how topological order persists in two separate contributions to the coherence times of the prescrambling plateaus. This will help us to explore if and how strong zero modes affect the scrambling dynamics of OTOC different than the dynamics of two-time correlators. Finally we discuss the effect of prescrambling on dynamic phase diagrams. We conclude in Sec.~V and elaborate on possible questions to answer in the future.

\section{Demonstration of Topological Origin \label{sec2}}

\begin{figure}
\centerline{\includegraphics[width=0.45\textwidth]{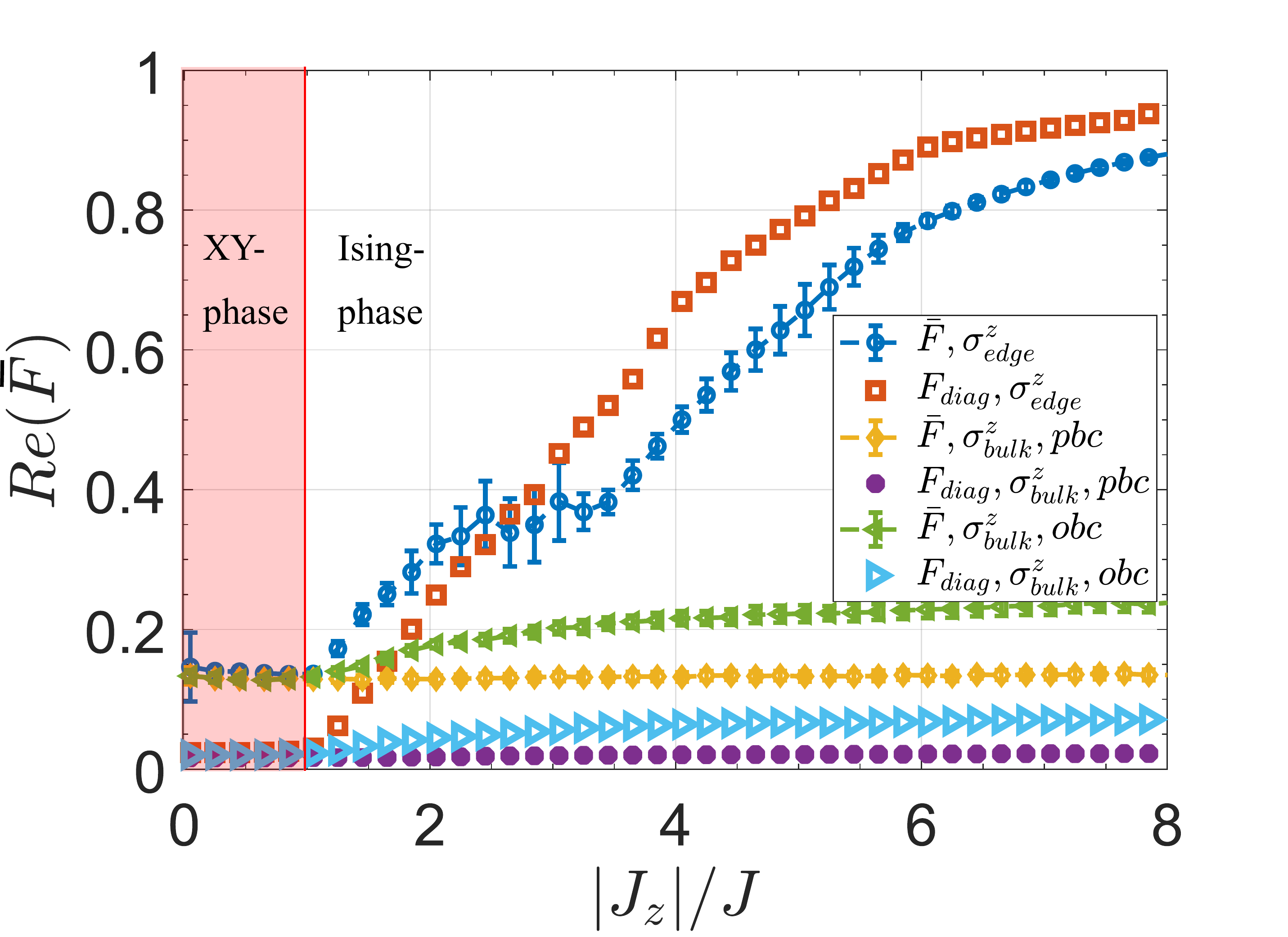}}
\caption{Long-time average of OTOC for XXZ model for edge-spin operators $W=V=\sigma^z_{\text{edge}}$ in blue circles and its (later explained) diagonal contribution in orange squares; for bulk-spin operators $\sigma^z_{\text{bulk}}$ with periodic boundary chain (pbc) in yellow diamonds and its diagonal contribution in purple dots; with open boundary chain (obc) in green left-pointing arrows and the diagonal contribution in light-blue right-pointing arrows. System size is $N=14$ and the time of averaging is $tJ=800$.}
\label{Fig1}
\end{figure}

It turns out that the connection between infinite-temperature information scrambling and quantum phases at zero temperature has a robust topological origin. Let us demonstrate how the topological origin reveals itself in the dynamics of OTOCs with an example on 1D XXZ chain,
\begin{eqnarray}
H&=&J \sum_i \left( \sigma_i^x \sigma_{i+1}^x + \sigma_i^y \sigma_{i+1}^y + \frac{J_z}{J} \sigma_i^z \sigma_{i+1}^z \right).\label{XXZHamiltonian}
\end{eqnarray}
At $T=0$, the model exhibits quantum phase transitions between a gapped Ising phase $|J_z|>1$ and a critical XY-phase $|J_z|<1$ where the spectrum is gapless \cite{2017LNP...940.....F}. We employ Haar-distributed random states $\Ket{\psi_{h}} $ and compute $\bar{F}$ shown in Fig.~\ref{Fig1}. 

If spin operators at the edge of the chain $W=V=\sigma^z_{\text{edge}}$ are utilized (blue-circles),  the infinite-temperature OTOC saturation value behaves like
an order parameter of the zero-temperature quantum phase transition, i.e., $\bar{F}\sim 0$ in the XY phase ($|J_z/J|<1$) and increases monotonically as we enter the 
Ising phases ($|J_z/J|>1$). In contrast, under periodic boundary conditions (yellow diamonds line) and for a bulk spin $W=V=\sigma^z_{\text{bulk}}$ (green left-pointing triangles), the OTOC no longer differentiates the two phases, and the transition point is smoothed out consistent with predictions from Ref.~\cite{PhysRevLett.123.140602}.

To demonstrate the role of topological order, we rewrite the Hamiltonian of the XXZ model in the Majorana basis. First, via the Jordan-Wigner (JW) transformation~\cite{sachdev2001quantum}
\begin{eqnarray}
\sigma^z_i &=& - \prod_{j<i} \left(1-2c_j^{\dagger}c_j \right)\left(c_i+c_i^{\dagger}\right),\label{mapping} \\
\sigma^x_i &=& 1-2c_i^{\dagger}c_i,\notag \\
\sigma^y_i &=& -i \prod_{j<i} \left(1-2c_j^{\dagger}c_j \right)\left(c_i-c_i^{\dagger}\right).\notag
\end{eqnarray}
the spin Hamiltonian is mapped to
\begin{eqnarray}
H &=& J \sum_i \bigg [ \left(1-2c_i^{\dagger}c_i \right) \left(1-2c_{i+1}^{\dagger}c_{i+1} \right) -  \left(c_i+c_i^{\dagger}\right) \notag \\
&\times & \left(c_{i+1}-c_{i+1}^{\dagger}\right) + \frac{J_z}{J} \left(c_i-c_i^{\dagger}\right)\left(c_{i+1}+c_{i+1}^{\dagger}\right) \bigg ],
\label{XXZFermions}
\end{eqnarray}
which can be written in terms of the Majorana fermions $a_{2j-1}=c_j+c_j^{\dagger}$ and $a_{2j}=-i\left(c_j-c_j^{\dagger}\right)$ \cite{2001PhyU...44..131K}:
\begin{eqnarray}
H &=& -J\sum_i \left(a_{2i-1}a_{2i}a_{2i+1}a_{2i+2}+i a_{2i-1}a_{2i+2}\right) \notag\\
&+&iJ_z\sum_i  a_{2i}a_{2i+1}.
\label{XXZMajorana}
\end{eqnarray}
In the Majorana basis, the spin system is mapped to an interacting Majorana chain. The XY (Ising) phase is mapped to a gapless (topological) phase, and the quantum phase transition becomes a topological transition. Same as the Kitaev chain, the topological phase in Eq.~\eqref{XXZMajorana} develops $Z_2$ topological order and is characterized by two Majorana zero-modes 
localized at the two ends of the chain~\cite{2001PhyU...44..131K}. 

The physics can be understood by considering the $J_z \gg J$ limit, where Eq.~\eqref{XXZMajorana} converges to the Kitaev model~\cite{2001PhyU...44..131K} with two zero-energy Majorana modes $\gamma_1=a_1$ and $\gamma_2=a_{2N}$ fully decoupled from the rest of the chain. Away from the $J_z \gg J$ limit, quartic terms in the Hamiltonian introduce interactions, but the zero-energy Majorana modes at the two ends of the chain remain topologically protected for the entire topological (Ising) phase. The existence of two Majorana modes at the two ends of the chain ($\gamma_1$ and $\gamma_2$) indicates that a zero-energy non-local fermion $d= \frac{\gamma_1+i\gamma_2}{\sqrt{2}}$ can be defined. Because of its zero-energy nature, for an eigenstate of the Hamiltonian $\Ket{\psi_0}$, another degenerate state $\Ket{\psi_1} = d \Ket{\psi_0}$ must exist with an opposite fermion parity. Therefore, in the topological phase, the edge modes are responsible of the degenerate subspaces forming not only in the ground state, but throughout the entire spectrum \cite{2001PhyU...44..131K,2017LNP...940.....F}. In other words, in contrast to a conventional (Landau-type) quantum phase transition, where across the phase boundary the ground state changes from non-degenerate (the disordered phase) to degenerate (the ordered phase), $Z_2$ topological order has a direct impact for the degeneracy of all eigenstates in the entire energy spectrum, i.e. two-fold degeneracy for the entire spectrum. The effect has a direct impact on measurements and dynamical quantities at any temperature \cite{2017JSMTE..06.3105K,PhysRevX.7.041062} and it is in sharp contrast to a conventional phase transition that can only be detected by zooming to the ground state at low-temperature. This is the key reason why the infinite-temperature OTOC is capable of detecting a zero-temperature topological order, but not a regular Landau-type quantum order (unless it can be mapped into a topological order).

\section{Topological Edge Physics Encoded in the Out-of-time-order correlators \label{sec3}} 

In this section, we study the non-interacting limit to provide analytical arguments in the demonstration of how infinite-temperature information scrambling of edge spins encodes the existence or absence of Majorana zero modes. Later we will mark the topological phase transition point via $\bar{F}$ in this non-interacting limit.

\subsection{Transverse-field Ising Model} 

\begin{figure*}
\centering
\subfloat[]{\label{Fig2a}\includegraphics[width=0.33\textwidth]{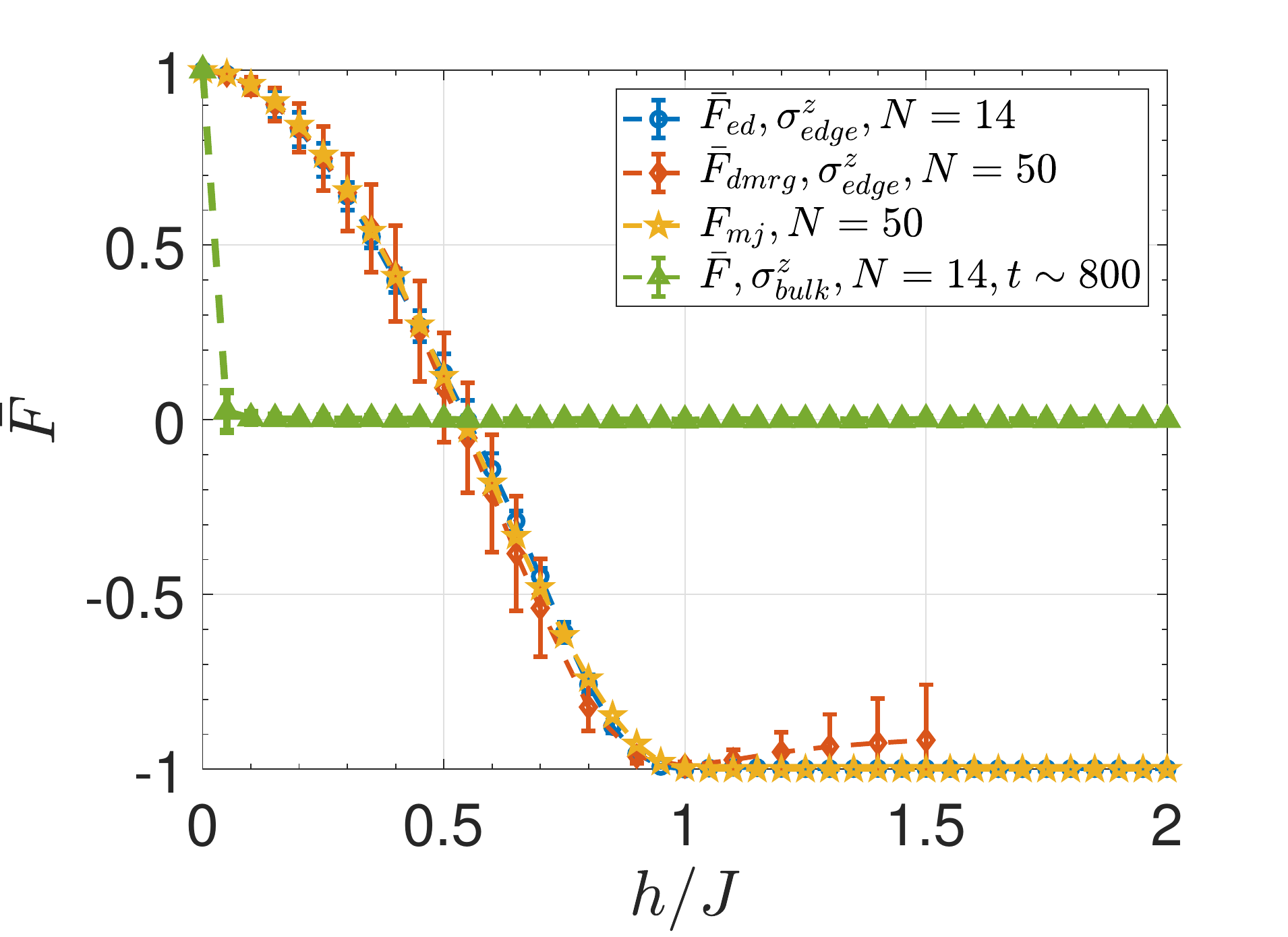}}\hfill 
\subfloat[]{\label{Fig2b}\includegraphics[width=0.33\textwidth]{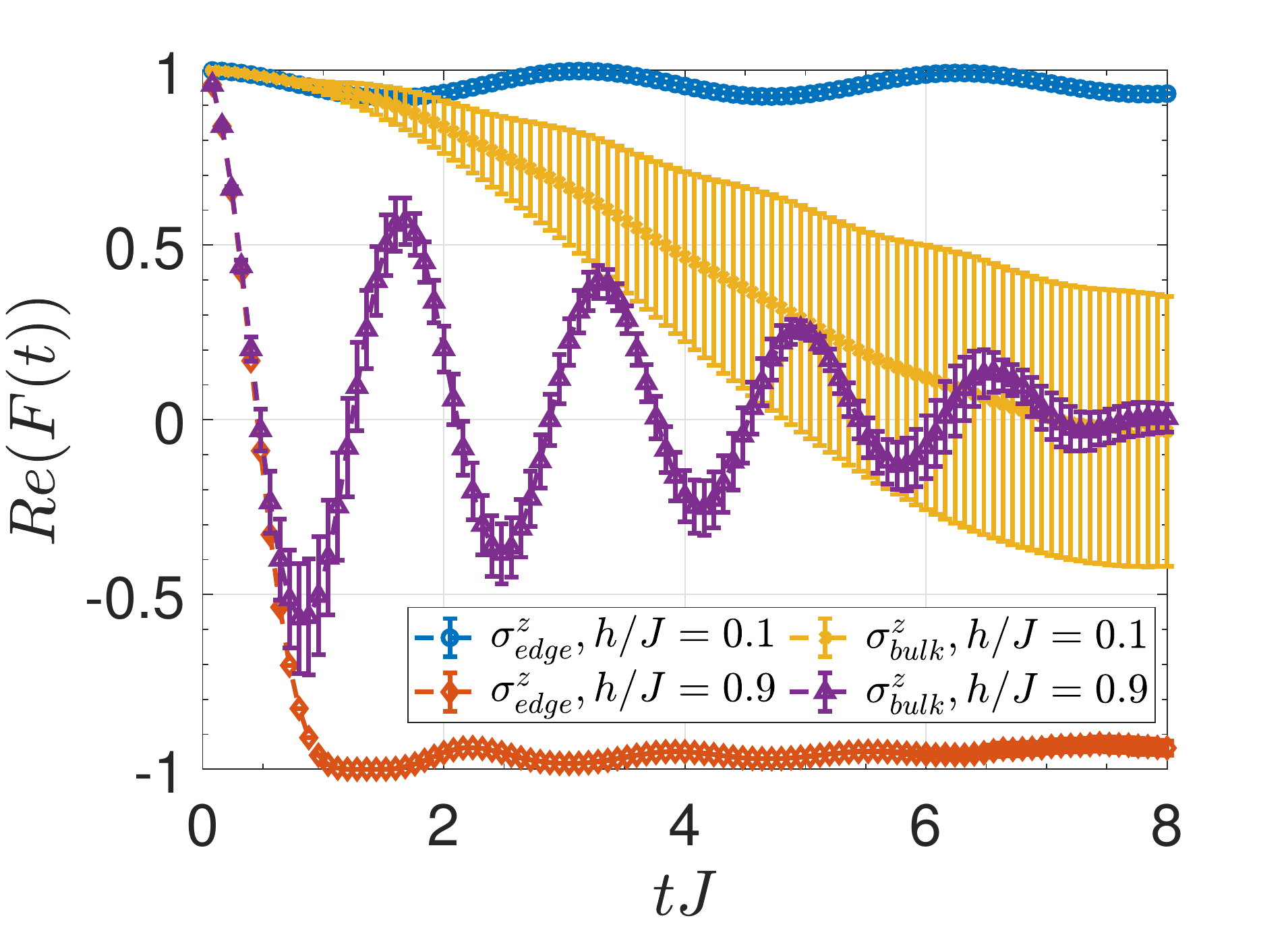}}\hfill 
\subfloat[]{\label{Fig2c}\includegraphics[width=0.33\textwidth]{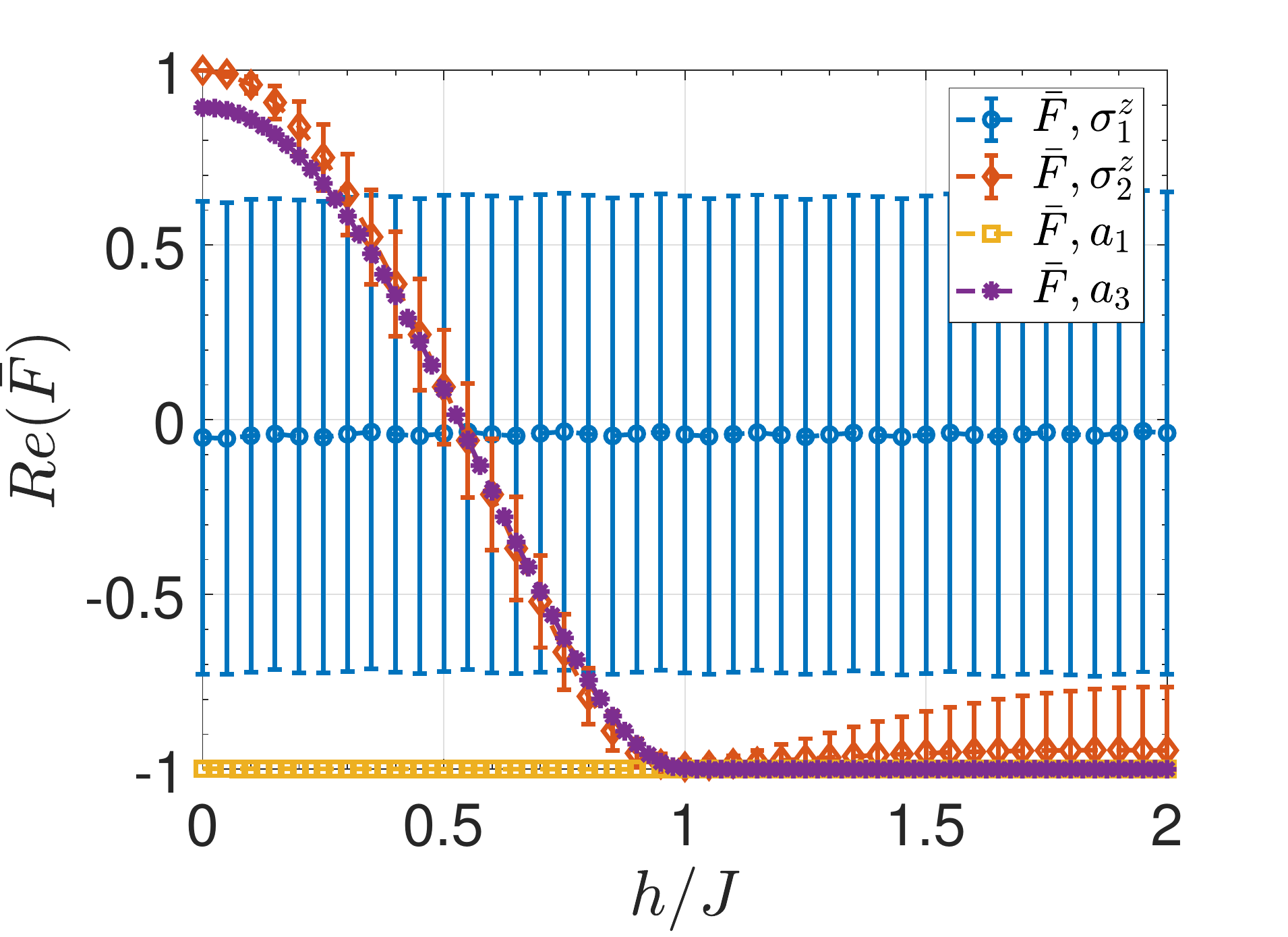}}\hfill 
\caption{Transverse-field Ising model at infinite-temperature. (a) The OTOC time-average of the edge spin operators $\sigma^z_1$ via real-time OTOC dynamics (blue circles) at $N=14$ and (orange diamonds) at $N=50$ where we used MPS (see Appendix A) for a time interval $tJ=\frac{\pi}{4}10\sim 7.85$. The yellow-pentagrams show $F_{11}$ based on Eq.~\eqref{OTOCEq3M} where the Majorana edge states are extracted from $H_{\text{BdG}}$ matrix at $N=50$ at infinite time limit for a comparison with other data. The green-triangles show the OTOC time-average of the bulk spin operator $\sigma^z_7$ at $N=14$ for a time interval $tJ=\frac{\pi}{4}10^3\sim 800$. (b) The OTOC dynamics $F(t)$ with respect to $tJ$. Blue-circle and orange-diamond lines are the OTOC of edge $\sigma^z_1$ operator for $h=0.1$ and $h=0.9$, respectively. Yellow-cross and purple-triangle lines are the OTOC of bulk $\sigma^z_{25}$ operator for $h=0.1$ and $h=0.9$, respectively. All curves are computed in t-DMRG for a system size of $N=50$, averaged over $10$ random product states to generate $\beta=0$ results. The error bars stand for $1\sigma$ variation of OTOC in this set of random states. (c) Robustness of order against changing the boundary conditions: a strong field is applied to the first spin only for $N=13$ and $tJ\sim 8$ (blue circles); and to the edge fermions in the non-interacting fermion chain for $N=50$ and $tJ \rightarrow \infty$ (yellow squares). The edge modes shifted to the nearest site that is free of pinning field, $\bar{F}$ of $\sigma^z_2$ spin (red-diamonds) and $\bar{F}_{33}$ of $a_3$ Majorana fermion (purple asterisks), respectively.}
\label{Fig2}
\end{figure*}

We consider a non-interacting, hence analytically solvable model and directly compute the contributions of Majorana zero-modes in the infinite-temperature OTOCs with edge operators. The Hamiltonian for the transverse-field Ising model with open boundary conditions is,
\begin{eqnarray}
H &=& -J \sum_{j=1}^{N-1} \sigma^z_{j}\sigma^z_{j+1}+h\sum_{j=1}^N\sigma^x_j. 
\label{HIsing}
\end{eqnarray}
Eq. \ref{HIsing} has a critical point at $h=1$ that separates a ferromagnetic ordered phase from a disordered phase. The time-average of OTOC $\bar{F}$ with $\sigma^z_1$ at $\beta=0$ is shown with the lines with blue-circles and orange-diamonds for $N=14$ and $N=50$, respectively in Fig. \ref{Fig2a}. The simulation with $N=50$ spins is performed with matrix product states (MPS) in a t-DMRG (time-dependent density matrix renormalization group) method, (see Appendix A for details). Here the error bars stand for the extend of oscillations in time, as we time-average the real part of the OTOC signal in a time interval of $tJ = \frac{\pi}{4}10 \sim 7.85$. For an edge spin operator $\sigma^z_1$, $\bar{F}$ behaves like an order parameter, which is $\bar{F}\sim -1$ in the disordered phase ($h>J$) and increase monotonically in the ordered phase ($h<J$). On the contrary, for a bulk spin operator, $\sigma^z_7$, this feature disappears (green-triangles in Fig. \ref{Fig2a}). This observation reflects that the physics captured by edge- and bulk-spin operators are different; a similar observation to what we presented for the XXZ model earlier. To further show how the real-time OTOC dynamics look like, we contrast time-evolving OTOC $F(t)$ of edge and bulk operators in Fig. \ref{Fig2b}. The OTOCs of the edge spin converge to different values at large times, depending on the value of $h/J$, while the OTOCs of bulk spins always converge to $0$ at large $t$, as long as $h \ne 0$. The $h=0$ limit is trivial for information scrambling, because the spin chain turns into the classical Ising model without quantum fluctuations or non-trivial dynamics, and thus information cannot scramble, $F(t)=1$.

\begin{figure}
\centering
\subfloat[]{\label{Fig3a}\includegraphics[width=0.24\textwidth]{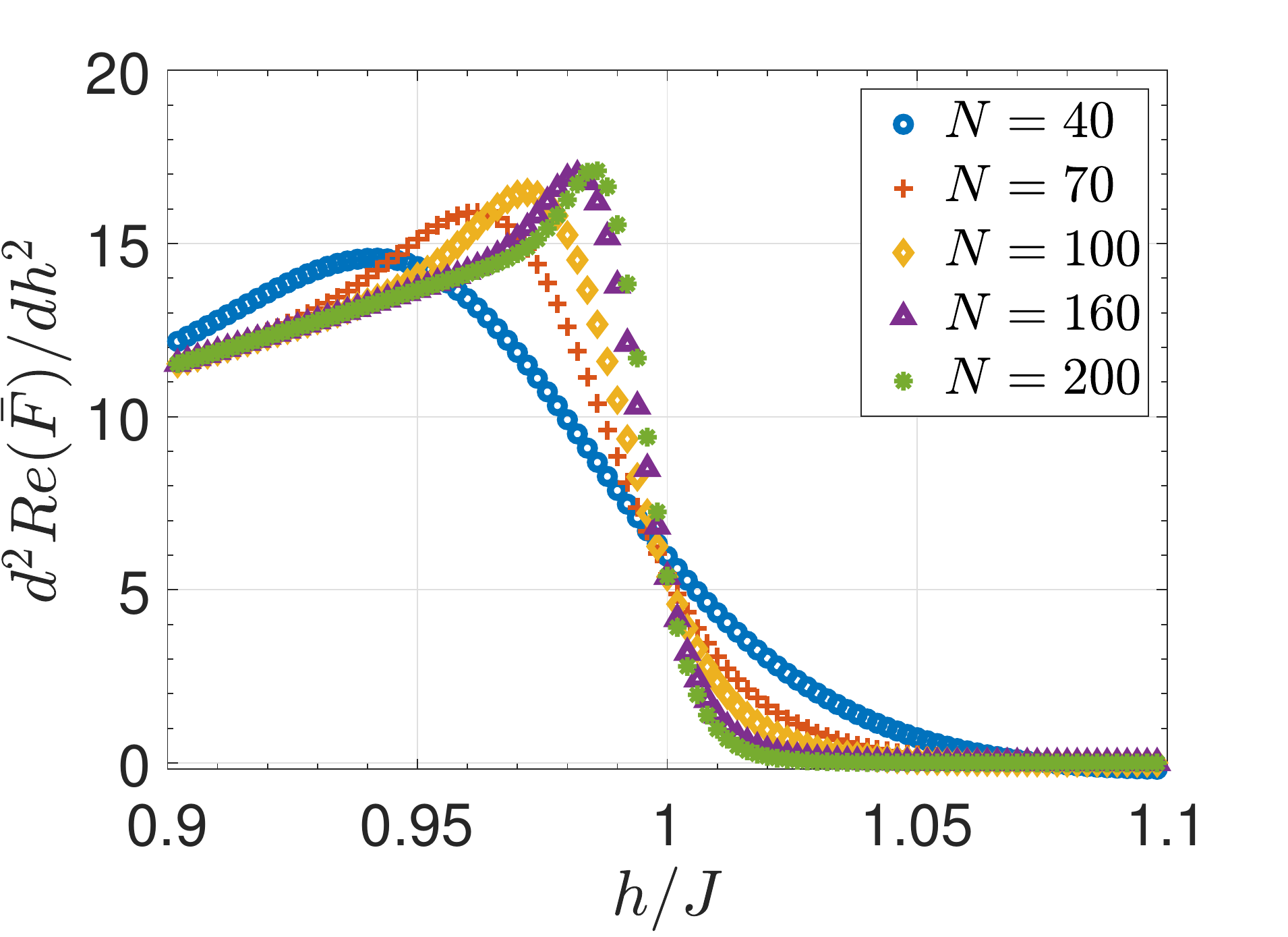}}\hfill 
\subfloat[]{\label{Fig3b}\includegraphics[width=0.24\textwidth]{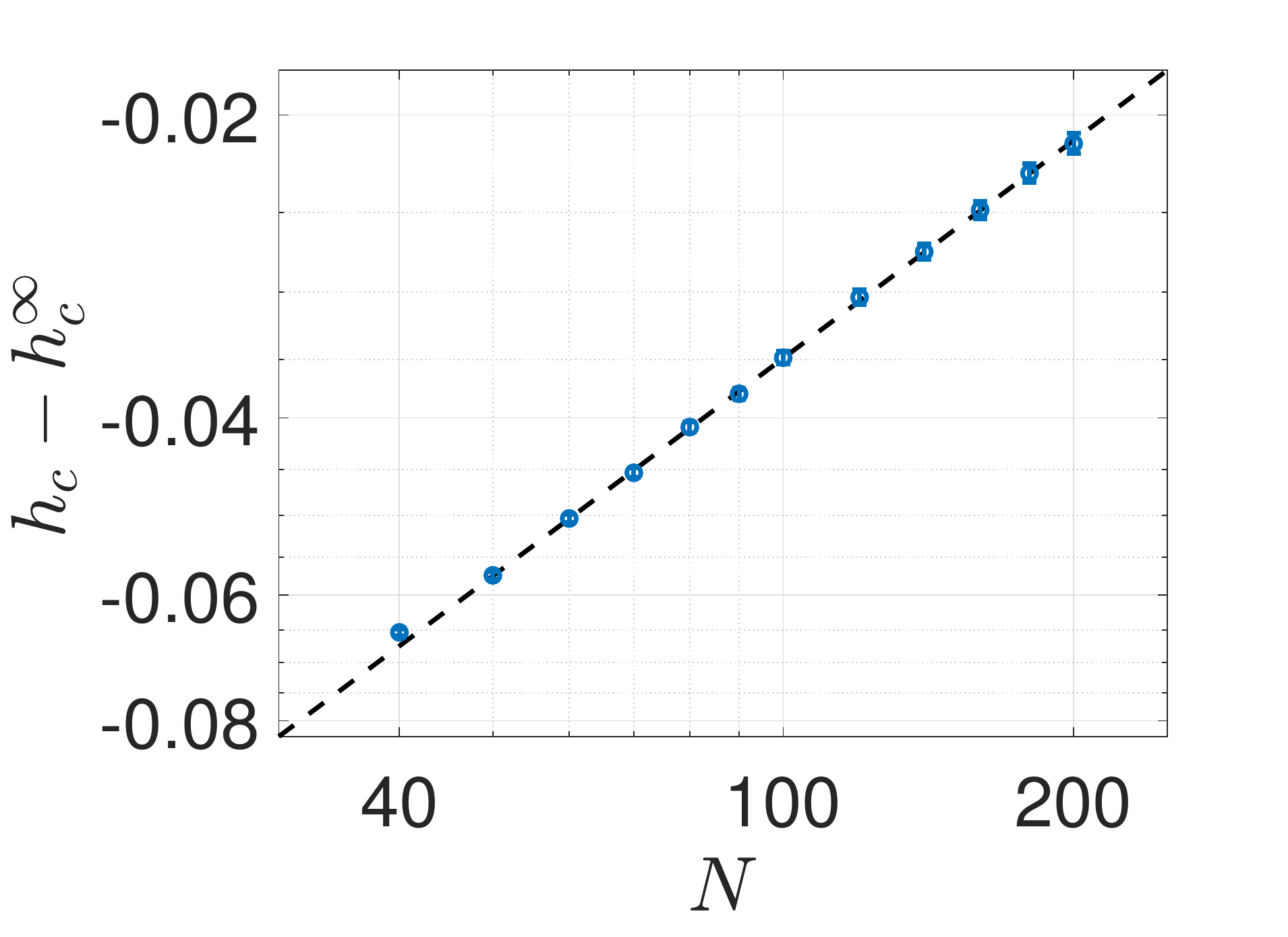}}\hfill 
\caption{(a) The second derivative of the OTOC time-average $d^2\bar{F}_{11}(t\rightarrow\infty)/dh^2$ pinpoints the phase transition point via its maximum. (b) The system-size scaling of the phase transition point gives $h_{dc} \sim N^{-0.7189} + 1.0069$ with $R^2=0.9996$, meaning in the thermodynamic limit the OTOC pinpoints the phase transition point as $h_{dc}^{\infty}=1.0069$. }
\label{Fig3}
\end{figure}
The results above can be easily understood by using the Majorana basis, which transforms the spin Hamiltonian into a non-interacting Majorana chain 
\begin{eqnarray}
H &=& -iJ\sum_{j=1}^{N-1}  a_{2j}a_{2j+1}-ih\sum_{j=1}^N  a_{2j-1}a_{2j},
\label{HMajorana}
\end{eqnarray} 
where we used Eqs.~\eqref{mapping}. In contrast to the XXZ model discussed above, Eq.~\eqref{HMajorana} only contains quadratic terms, hence non-interacting, and thus can be easily diagonalized, which enables us to compute infinite-temperature OTOC saturation values $\bar{F}$ exactly. This exact solution agrees perfectly with numerical simulations in Fig.~\ref{Fig2a}. More interestingly, as will be shown below, the analytical result exhibits that $F_{\infty}$ is solely contributed by Majorana zero modes, while the contributions from all other finite energy excitations fade away at large $t$.

\subsection{Exact solution \label{ssec:exact}}

We compute the OTOC of an edge spin using the Majorana basis in this section. In the Majorana basis, the OTOC of Majorana fermions can be defined as $F_{2i-1,2i-1}(t)=\text{Tr}\left( a_{2i-1}(t) a_{2i-1} a_{2i-1}(t) a_{2i-1} \right)/2^N$, where we set $W=V=a_{2i-1}=c_i+c_i^{\dagger}$. Since it can be easily showed that the OTOC of edge Majorana fermions must be identical to the OTOC of edge spins, $\sigma^z_1=\left(c_1+c_1^{\dag}\right)=\gamma_1$ and $\sigma^z_N=\mathbb{P}\left(c_N-c_N^{\dag}\right) = i\mathbb{P} \gamma_2$, where $\mathbb{P}=\prod_{j}^N\left(1-2c_j^{\dag}c_j\right)$ is the parity operator, here we focus on $F_{11}$ with $W=V=a_1$.

The Majorana-fermion OTOC $F_{2i-1,2i-1}(t)$ can be conveniently computed by utilizing the Bogoliubov-de Gennes (BdG) basis, as detailed in Appendix B. With fermion operators defined for a space of double spectrum, we write the BdG Hamiltonian and calculate $F_{2i-1,2i-1}(t)$ at site $i$,
\begin{eqnarray}
&F&_{2i-1,2i-1}(t) = \notag\\
&\hspace{0.1mm}& \left[\sum_{\alpha}^{2N} \left(|\psi_{\alpha,i}|^2+ \psi_{\alpha,i} \psi_{\alpha,i+N}^* \right) \cos \left(E_{\alpha}t\right)\right]^2 \label{OTOCEq3M}\\
&+& \left[\sum_{\alpha}^{2N} \left(|\psi_{\alpha,i+N}|^2+ \psi_{\alpha,i+N} \psi_{\alpha,i}^* \right) \cos \left(E_{\alpha}t\right)\right]^2 -1. \notag
\end{eqnarray}
where $E_{\alpha}$ and $\psi_{\alpha}$ are eigenenergy and eigenstate of the BdG Hamiltonian, while the sum goes over all energy eigenstates $\alpha=1$, $\ldots$ $2N$. In the long-time limit, only the non-oscillating terms (i.e., $E_{\alpha}=0$) contribute to the saturation value of $F_{2i-1,2i-1}(t)$, i.e., only zero modes need to be considered for $t \rightarrow \infty$. For $h<J$ in the Ising ordered phase, the BdG Hamiltonian describes a topological superconductor with Majorana zero modes at the two ends, and hence we only sum over the two Majorana zero modes, e.g. $\alpha=mj$. In the disordered phase ($h>J$), the BdG Hamiltonian describes a topologically-trivial superconductor without any zero modes. Thus in the absence of zero modes, $E_{\alpha}=0$, $F_{2i-1,2i-1}(t) \rightarrow -1$, explaining $\bar{F}$ approaching to $-1$ in the Ising model results (Figs.~\ref{Fig2}). By calculating Eq.~\eqref{OTOCEq3M} as $t\rightarrow \infty$, we plot $F_{11}=F_{mj}$ in Fig.~\ref{Fig2a} with orange-pentagrams, which matches well with the Ising model results. To conclude, the derived relation, e.g. Eq.~\eqref{OTOCEq3M} rigorously proves that the saturation value of an OTOC with Majorana fermions ($W=V=a_{2i-1}$) is contributed only by Majorana zero modes ($E_\alpha=0$), while the contributions from any excited states ($E_\alpha\ne 0$) vanish at long times. Since the Ising model can be exactly mapped to a 1D Majorana chain, the infinite-temperature OTOC of the edge spins directly probes the presence or absence of the Majorana zero modes. This is one of the key conclusions in our manuscript.

Motivated by this observation, we pinpoint the phase boundary of the topological phase transition in the following. Since the OTOC $F_{11}(t\rightarrow \infty)$ has a continuous transition from topologically non-trivial to trivial phase, we focus on its second derivative $d^2\bar{F}_{11}(t\rightarrow\infty)/dh^2$ with respect to external field $h$. The maximum of the second derivative pinpoints the transition point, Fig.~\ref{Fig3a}. Then the system-size scaling provides the transition point in the thermodynamic limit as $h_{dc}^{\infty}=1.0069$ in a power-law scaling $h_{dc} \sim N^{-0.7189} + 1.0069$ (Fig.~\ref{Fig3b}). For further details, see Appendix D. We note that the results obtained in the non-interacting limit (Ising model) are valid at the infinite time in the thermodynamic limit since topologically induced prescrambling plateau persists indefinitely (Appendix D). 

\subsection{Robustness against varying the boundary conditions} 

Although the phenomenon discussed above relies on utilizing edge degrees of freedom, all the key conclusions are robust against any local perturbations and independent of boundary conditions. Because, the physics is based on topological edge modes. To demonstrate this robustness, we vary the boundary condition of the transverse-field Ising chain by introducing a constant magnetic field (along the x direction) for the edge spin only, i.e. $h_1/J=h/J+6$ where $h_1$ is the strength of the transverse field for the first site, while the rest of the spins have the same transverse field $h$. This strong field at the edge site introduces a strong pinning to the first spin and hence $\bar{F}$ oscillates significantly, being featureless across the phase boundary (blue-circles in Fig.~\ref{Fig2c}). However, if we choose the spin operator at the second site instead, the physics discussed above is recovered as shown in Fig.~\ref{Fig2c} with orange-diamonds. This is because such a local field cannot destroy the Majorana zero mode, which is topologically protected by the nontrivial bulk. Instead, it can only move the location of the zero modes, and thus, utilizing the second site, the conclusion remains the same. We additionally show the results for non-interacting fermion chain with an additive field affecting only the fermion at the edge. Yellow-squares in Fig.~\ref{Fig2c} show $\bar{F}_{mj}$ (Eq.~\eqref{OTOCEq3M}), the OTOC of edge Majorana mode $\gamma_1$ at the infinite-time limit, hence demonstrating no transition point. Purple-asterisks, on the other hand, show $\bar{F}_{33}$, the OTOC of Majorana mode $a_3$ at site $i=2$ at the infinite-time limit, which is observed to match with $\bar{F}$ of the Ising model, implying an agreement between numerics and analytics.

\section{The Interplay between Topological Order and Scrambling \label{sec4}} 

The default expectation for generic systems in 1D is scrambling over a time interval where the OTOC decays fast or slow but saturates to a residue close to zero, both depending on the set of symmetries existing in the system and the size of the Hilbert space \cite{doi:10.1002/andp.201600332, doi:10.1002/andp.201600318,FAN2017707,PhysRevLett.123.010601,PhysRevA.99.052322}. An exception to this observation is the models that possess a symmetry-breaking phase transition with a long-range ordered phase at zero temperature regardless of the interactions \cite{PhysRevLett.123.140602} or the non-integrability \cite{PhysRevLett.121.016801}. However, could order in such generic systems be captured at higher temperatures, preferably at infinite temperature? Now we systematically study the detection of topological order in generic systems at infinite temperature, and show that the machinery for the detection of the topological order with simpler correlators can also be used for OTOCs. In fact, this encourages us to devise a method to show if and how the dynamical imprint of topological order on information scrambling could differ from the one on thermalization dynamics.

\subsection{Coherence times of prescrambling plateaus \label{ssec:preS}}

$Z_2$ topological degeneracy does not only slow down the scrambling process, but also temporarily freezes the dynamics for generic nonintegrable models, causing \emph{topologically induced prescrambling}. Hence we observe that the topological order has a profound effect on the dynamics of systems \cite{2017JSMTE..06.3105K,PhysRevX.7.041062}, suggesting a new time-scale for information scrambling in our case. In this section, we explore the coherence times of the prescrambling plateaus to understand the associated timescales in the thermodynamic limit.

\begin{figure}
\centering
\subfloat[]{\label{Fig5a}\includegraphics[width=0.24\textwidth]{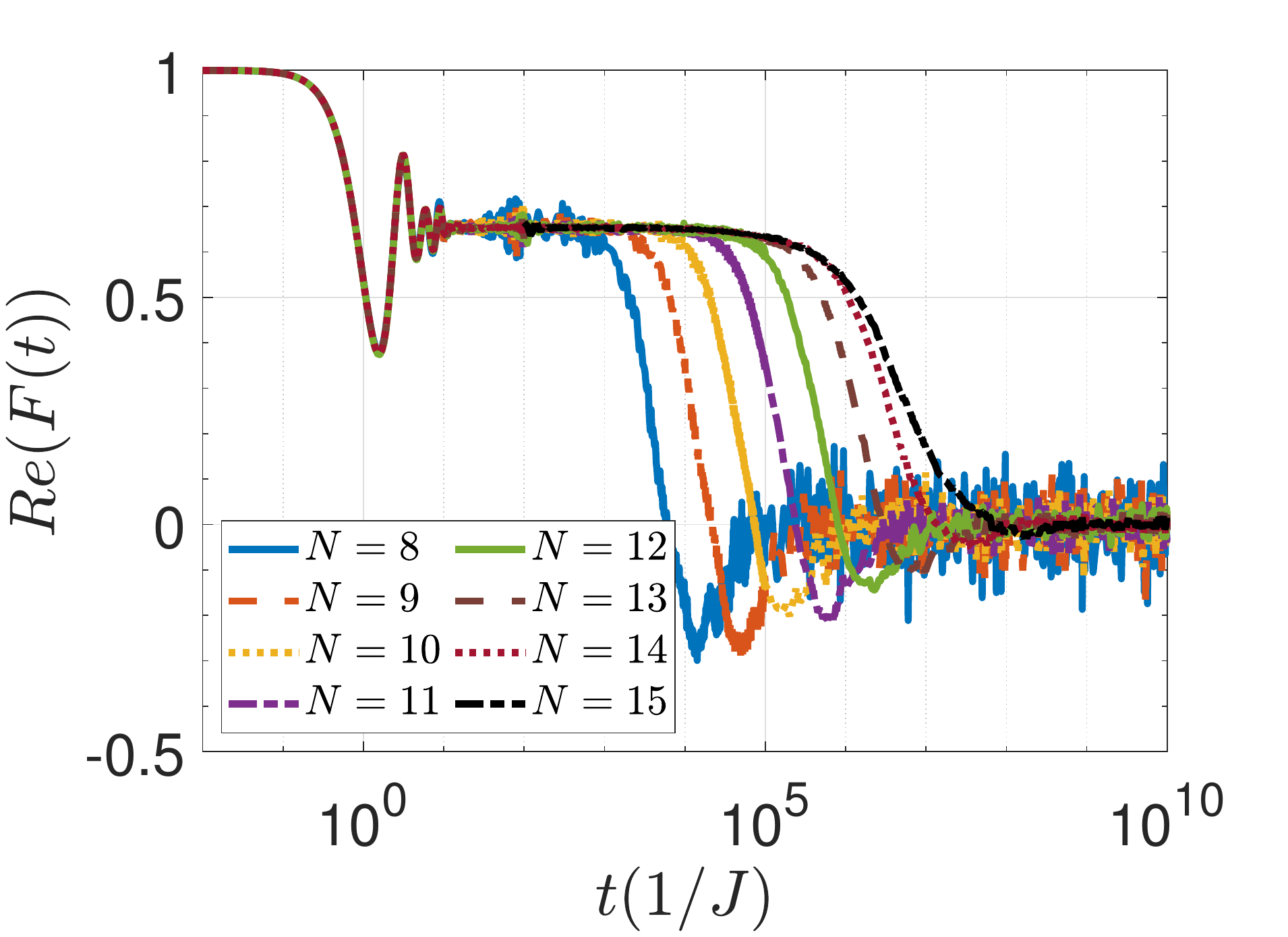}}\hfill  
\subfloat[]{\label{Fig5b}\includegraphics[width=0.24\textwidth]{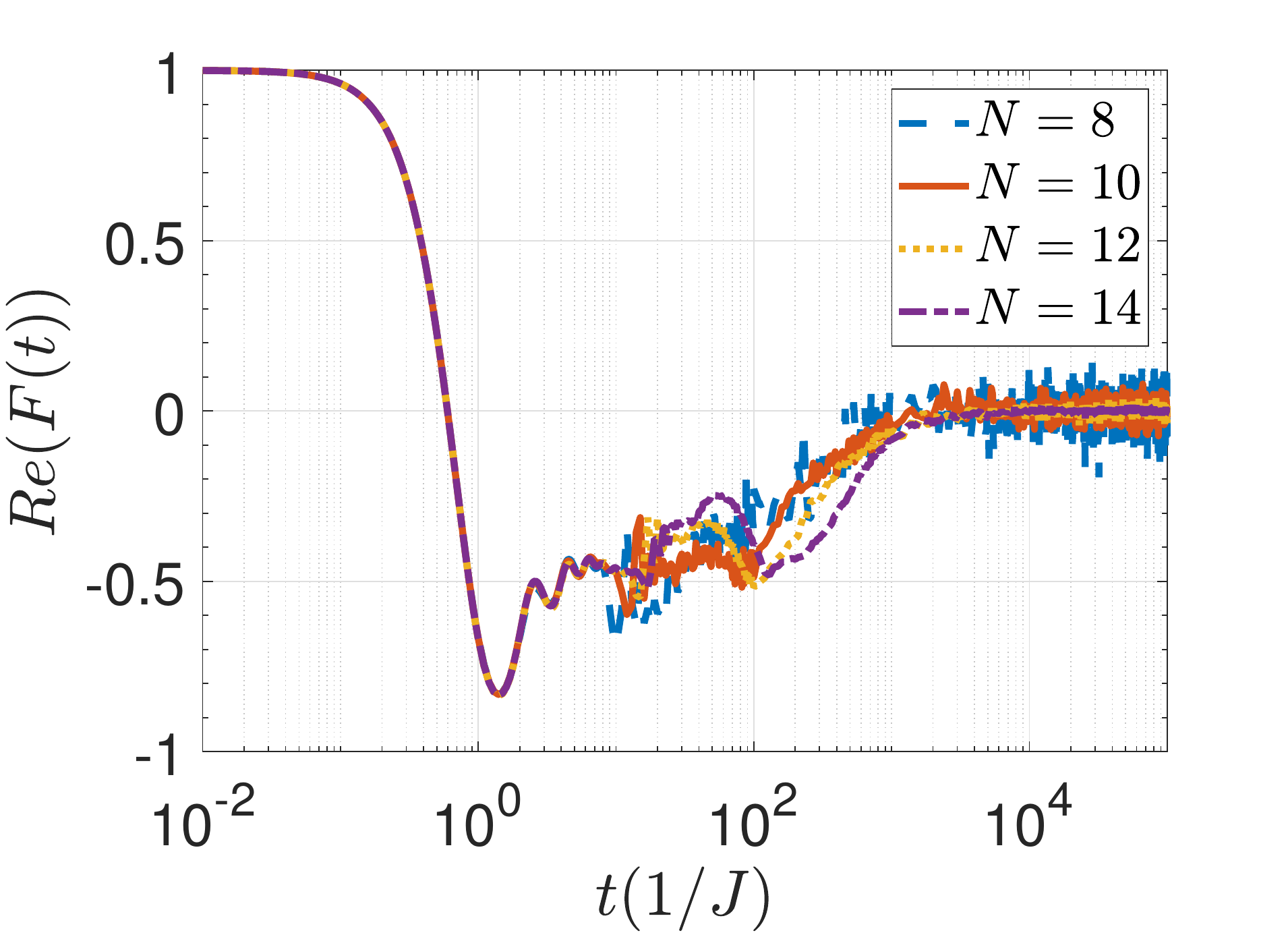}}\hfill  
\subfloat[]{\label{Fig5c}\includegraphics[width=0.24\textwidth]{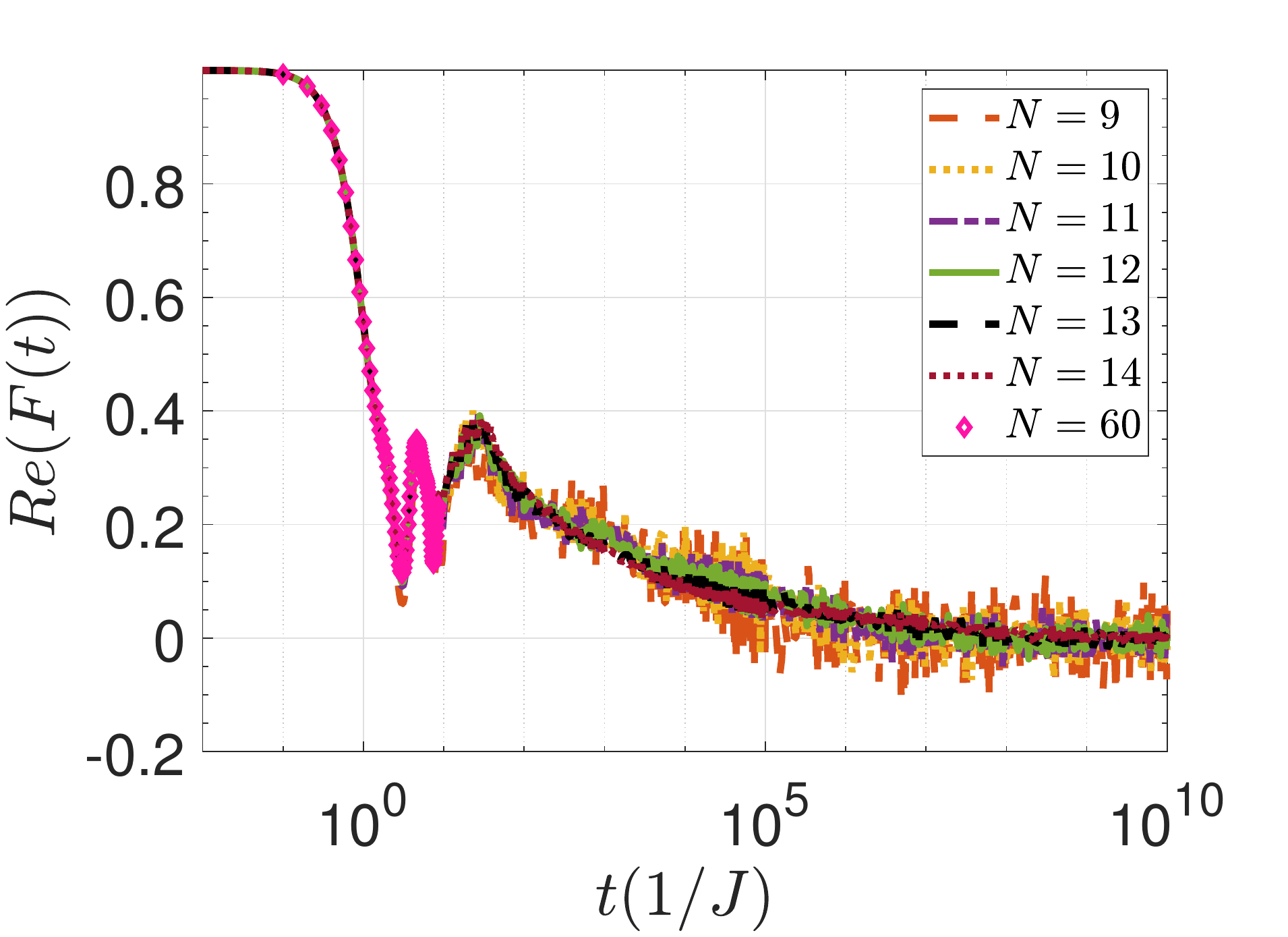}}\hfill 
\subfloat[]{\label{Fig5d}\includegraphics[width=0.24\textwidth]{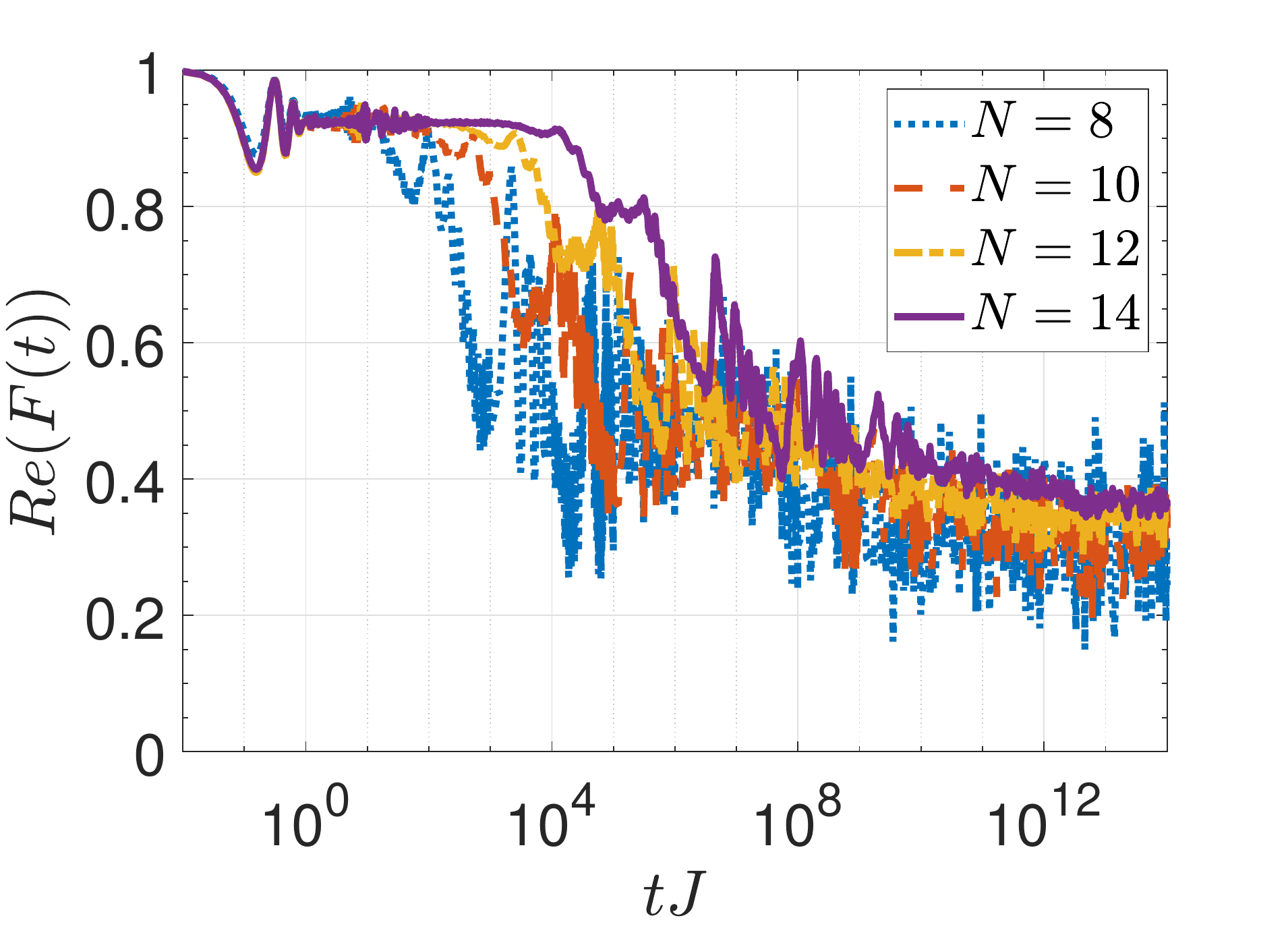}}\hfill 
\caption{Coherence times of prescrambling plateau at (a)-(b) $\Delta=-0.1$, (a) deep in the topologically non-trivial phase $h/J=0.3$ and (b) at $h/J=0.7$ showing negative prescrambling plateau values; (c) $\Delta=-0.5$ at $h/J=0.3$. $N=60$ is computed via t-DMRG with $25$ random initial states to have the infinite-temperature OTOC. (d) Prescrambling plateau deep in the topologically non-trivial phase of the XXZ model with $J_z/J=10$ persists indefinitely.}
\label{Fig5}
\end{figure}

Fig.~\ref{Fig5a} shows how the coherence times of the prescrambling plateau in a near-integrability model, see Eq.~\eqref{HNonIIsing}, ($\Delta/J=-0.1$) exponentially increase until around $N=15$ where the increase halts, suggesting that the curves of the systems with larger sizes possibly collapse on each other. Better examples can be seen in Figs.~\ref{Fig5b}-\ref{Fig5c} of $h/J=0.7$ of near-integrability model and deep in the non-trivial phase of the model with stronger interactions $\Delta/J=-0.5$, respectively. Therefore, prescrambling plateau has a finite lifetime in generic systems, including the vicinity of non-interacting limit. When the model becomes integrable, prescrambling plateau persists indefinitely, meaning that a system in thermodynamic limit never scrambles. Fig.~\ref{Fig5d} shows the exponential increase of full scrambling decay times in the XXZ model, thus implying that the observed scrambling is a finite-size effect. Similar behavior can be found for different $J_z/J$ parameter (Appendix E), as well as the non-interacting limit (Appendix C).

A natural question is how a generic system could host a prescrambling plateau for mostly long but finite amount of time. Finite coherence times of edge-spin two-time correlators in generic systems have been recently based on spectrum characteristics \cite{2017JSMTE..06.3105K}. Hence these findings should be applicable to information scrambling. The notion of easy spin flips are introduced by Ref.~\cite{2017JSMTE..06.3105K} to demonstrate that these spin flip processes destroy the perfect pairing of energy eigenstates that are caused by zero modes. Such perfect pairing, meaning exponentially close eigenstates, happen in the integrable case and is dubbed as \emph{strong zero modes}. When integrability breaking interactions are introduced, due to the poles appearing in the perturbation theory, also called resonances, degeneracies are no longer exponentially close, but polynomially in system size. Hence there is not perfect pairing anymore, and strong zero modes turn into \emph{almost-strong zero modes} as called by Ref.~\cite{2017JSMTE..06.3105K}. The processes of easy spin flips are the reason behind such a change in the degeneracy structure. Due to the poles in the perturbation theory, certain basis states with spin flips are equally energetically favorable with the Kramer partner. When the external transverse field is on, these states mix and one ends up with eigenstates that are comprised of not only a state and its Kramer partner as expected in a doubly-degenerate spectrum, but a state, its ‘easy spin partners’ and the Kramer partners of all. These now polynomially close eigenstates, depending on the external field strength as well as where the poles are, could cause bigger regions of degeneracy compared to double degeneracy. However we stress on the fact that these degeneracies are, so to speak, weaker than the degeneracies when there are no integrability breaking interactions, hence they indeed deserve the name \emph{almost-strong zero modes}. Again we emphasize that these eigenstates are still Kramer partners of each other, as would be expected from a system that obeys $Z_2$ symmetry. Hence the $Z_2$ topological imprint is not lost, but instead reduced to a signature that could survive only for finite times. Such a profound effect on dynamics by zero modes is shown with two-time correlators by Refs.~\cite{2017JSMTE..06.3105K,PhysRevX.7.041062}. Hence our results are an intuitive extension of this effect to the dynamics of information scrambling and OTOCs. In this regard, our results demonstrate that the scrambling could be slowed down in nonintegrable systems, introducing a two-step timescale to scrambling dynamics, with the name \emph{prescrambling}, analogizes with prethermalization as the name correctly implies. This encourages us to question how much OTOCs are really different than their simpler cousins, e.g. two-time correlators. An immediate observation shows us that Figs.~\ref{Fig5a} and \ref{Fig5b} of the near-integrability model behave considerably different: the former has a positive-valued plateau, paralleling with the behavior of two-time correlators, whereas the latter has a negative-valued plateau. To better understand such distinct behavior appearing in OTOCs and further elaborate on related questions, we introduce a method in the next section.

\subsection{Dynamical decomposition method} 

In this section, we develop a framework that can provide us more insight about detecting topological order in generic systems via OTOCs, as well as the saturation regime of OTOCs in general. Since we can already derive the OTOC saturation value analytically in the non-interacting regime (Sec.~\ref{ssec:exact}), we need a framework that works in nonintegrable models; a limit that is in general not analytically tractable. This framework is an application of dynamical decomposition to OTOC \cite{PhysRevLett.123.140602} and we aim to calculate $\bar{F}$ with a term that becomes the dominant contribution in $\bar{F}$ and a correction to it, as we move away from the non-interacting limit. Dynamical decomposition method is previously utilized to find a leading-order term in $\bar{F}$ (of arbitrary bulk spins) at zero-temperature to probe zero-temperature symmetry-breaking phase transitions \cite{PhysRevLett.123.140602}. Here we generalize the idea to infinite temperature and put forward a conjecture in analogy to the Eigenstate Thermalization Hypothesis (ETH), as explained in the following. Our motivation for putting forward this method is two-fold: (i) this approach provides us an approximated solution of the saturation regime for a generic system; (ii) it also offers us a common ground to compare the saturation regime of OTOCs with the saturation regime of two-time correlators to further understand if they differ in reflecting the dynamics of zero modes. We note why the point (ii) is interesting for our purposes: OTOCs at infinite-temperature are well-known probes of quantum chaos \cite{2013JHEP...04..022L,Maldacena2016,2017JHEP...10..138H,2018arXiv180200801X,doi:10.1002/andp.201600332,PhysRevA.99.052322}, whereas two-time correlators seem to be featureless to reflect such property of the system \cite{2017NJPh...19f3001B,doi:10.1002/andp.201600318}. Even though intuitively related, thermalization and scrambling seem to be different from each other, motivated by their different timescales, Refs.~\cite{PhysRevB.98.045102,Maldacena2016,2017NJPh...19f3001B}. Hence finding where OTOC points to additional information about the system, and where it can be reduced to two-point correlators, could prove useful to understand the relations between scrambling and thermalization. In the cases where such a reduction is possible, reminding of Wick’s theorem but for OTOCs, the hope is that one can use two-point correlators instead of OTOCs to determine the scrambling in an experimental setting, because implementing an OTOC protocol is unarguably harder than measuring a two-point correlation function \cite{PhysRevX.7.031011,articleRey,Landsman_2019,PhysRevA.94.040302,PhysRevA.99.052322,PhysRevLett.123.115701}. In the opposite situation  where OTOCs provide additional information, we could know how scrambling dynamics differ from thermalization, at least for the model under study.

By utilizing the energy eigenstates as a complete basis of the Hilbert space, OTOC at infinite-temperature can be written as
\begin{align}
F(t) =\frac{1}{M} \sum_{\alpha,\beta,\gamma,\delta} &W_{\alpha \beta} V_{\beta\gamma} W_{\gamma\delta} V_{\delta\alpha} e^{i (E_{\alpha}-E_{\beta} + E_{\gamma}-E_{\delta})t}
%\notag\\&\times \text{exp}\left[i (E_{\alpha}-E_{\beta} + E_{\gamma}-E_{\delta})t\right].
\label{OTOCdynamicsM}
\end{align}
where $W_{\alpha\beta}$ and $V_{\alpha\beta}$ are defined as $W_{\alpha\beta}=\langle \psi_{\alpha} |W|\psi_{\beta}\rangle$ and $V_{\alpha\beta}=\langle \psi_{\alpha} |V|\psi_{\beta}\rangle$ with $\Ket{\psi_{\alpha}}$ and $\Ket{\psi_{\beta}}$ being the energy eigenstates with associated energies $E_{\alpha}$, $\ldots$, $E_{\delta}$. To keep the notation simpler, we do not explicitly specify the degeneracies in Eq.~\eqref{OTOCdynamicsM}. 

In the long time limit ($t\rightarrow \infty$), only the static terms with $E_{\alpha}-E_{\beta} + E_{\gamma}-E_{\delta}=0$ contribute to the saturation value, while the rest of the terms dephase. Then the saturation value, and equivalently the long time-average $\bar{F}$, of OTOC \cite{PhysRevLett.123.140602} reads,
\begin{eqnarray}
\bar{F}&=&\frac{1}{M}\left(\sum_{\substack{E_{\alpha}=E_{\beta}, \\ E_{\gamma}=E_{\delta}}}+\sum_{\substack{E_{\alpha}=E_{\delta}, \\ E_{\beta}=E_{\gamma}}}-\sum_{\substack{E_{\alpha}=E_{\beta}= \\ E_{\gamma}=E_{\delta}}} + \sum_{\substack{E_{\alpha}\neq E_{\beta}\neq \\ E_{\gamma}\neq E_{\delta}}}\right) \notag \\
&\times & W_{\alpha \beta} V_{\beta\gamma} W_{\gamma\delta} V_{\delta\alpha}, \label{saturationEqM} 
\end{eqnarray}
where $\sum_{E_{\alpha}=E_{\beta},\hspace{1mm} E_{\gamma}=E_{\delta}}$ implies that we take the operator matrix elements that satisfy the corresponding energy condition $E_{\alpha}=E_{\beta}, \hspace{1mm} E_{\gamma}=E_{\delta}$. Since we look for a dominant contribution to Eq.~\eqref{saturationEqM} as the interaction strength increases, the most suitable dynamical decomposition is through a conjecture where $\bar{F}$ is dominated by the diagonal contribution. This corresponds to the contribution with the energy condition $E_{\alpha}=E_{\beta}=E_{\gamma}=E_{\delta}$ on the spectrum. A way to see why we expect our conjecture to hold is via remembering ETH. ETH, up to exceptions \cite{PhysRevLett.105.250401,PhysRevA.97.023603}, holds for nonintegrable systems whereas it fails for integrable systems \cite{2008Natur.452..854R}. One of the conditions of ETH is that the off-diagonal elements are suppressed compared to diagonal elements of the local observable written in the eigenbasis of the Hamiltonian. Therefore, based on the literature of ETH, we know that a local operator should dominantly populate its diagonal entries when the Hamiltonian is nonintegrable. In parallel with this argument, we numerically observe that our conjecture is indeed valid when an ansatz on the matrix elements of $W$ and $V$ is satisfied. This ansatz demands that the off-diagonal elements of the operators (in the eigenbasis) are suppressed with respect to the diagonal elements when the spectrum is explicitly degenerate; and can be formulated as $|W_{E_{\alpha}\neq E_{\beta}}|^2 \ll |W_{E_{\alpha}=E_{\beta}}|^2$ for both $W$ and $V$, as well as  $|V_{E_{\alpha}\neq E_{\beta}}|^2 \ll |W_{E_{\alpha}=E_{\beta}}|^2$ and vice versa. When the ansatz is satisfied, $\bar{F}$ simplifies to the diagonal contribution $F_{diag}$, 
\begin{eqnarray}
F_{diag}=\frac{1}{M}\sum_{\substack{E_{\alpha}=E_{\beta}= \\ E_{\gamma}=E_{\delta}}} W_{\alpha \beta} V_{\beta\gamma} W_{\gamma\delta} V_{\delta\alpha}.
\label{topologicalOTOC}
\end{eqnarray}
We note that the operator ansatz is the generalization of ETH's aforementioned criteria \cite{2008Natur.452..854R,2016AdPhy..65..239D,2016RPPh...79e6001G} to a degenerate spectrum. However, since we do not need to assume that the diagonal elements of the operator matrix are a smooth function of energy $W_{E_{\alpha}=E_{\beta}} = g(E_{\alpha})$, the other criteria of ETH \cite{2008Natur.452..854R} does not need to be followed, hence our conjecture does not require thermalization. This is reasonable, given that for a quantum system to thermalize strictly (in ETH sense) the saturation value should be predictable by the microcanonical ensemble in a narrow energy window on the spectrum \cite{2008Natur.452..854R}. There is not such a requirement for the saturation value of OTOCs. In conclusion, we can anticipate that our conjecture should be applicable for a wider range of systems e.g. including integrable but interacting systems.

If $W$ and $V$ are Majorana operators, i.e. $a_{2i-1}$, the only contribution to $F_{diag}$ comes from the degenerate energy levels which contain two eigenstates with opposite fermion parity. Since the two-fold degeneracy arises in the entire spectrum, a finite $F_{diag}$ is expected in the topologically non-trivial phase. However in the topologically trivial phase, although it could arise accidentally for some energy levels, two-fold degeneracy is generically not expected implying $\bar{F}_{diag} \sim 0$. Hence $\bar{F}_{diag}$ directly probes topological degeneracy in any system with $Z_2$ symmetry. Our conjecture can be rigorously proven for two-time correlation functions, where the off-diagonal contribution does not satisfy the corresponding energy condition $E_{\alpha}-E_{\beta}=0$ and thus, must vanish in long time. Hence, the saturation value for a two-time correlator,
\begin{eqnarray}
\bar{C} &=& \text{Tr}\left( W(t)W \right) = \frac{1}{M} \sum_{\substack{E_{\alpha}=E_{\beta}}} W_{\alpha \beta} V_{\beta\alpha}, \label{twoPoint}
\end{eqnarray}
already consists of only diagonal contribution with no need to introduce an operator ansatz, unlike OTOC. For OTOC, if the operator ansatz does not hold and hence the conjecture fails, other contributions to $\bar{F}$ might exist (Eq.~\eqref{saturationEqM}), which we call \emph{off-diagonal contribution}. Such cases, e.g. non-interacting model, clearly make the saturation regime of OTOC distinct than the saturation regime of two-time correlators, because
the off-diagonal contribution becomes comparable to the diagonal contribution, and even dominates $\bar{F}$. On the other hand when the conjecture holds, and hence off-diagonal contribution sums up to $\sim 0$, $F_{diag}$ becomes the approximated solution to $\bar{F}$; and since $F_{diag}$ (Eq.~\eqref{topologicalOTOC}) is related to $\bar{C}$ (Eq.~\eqref{twoPoint}), $\bar{F}$ might be predicted by $\bar{C}$. 

How $F_{diag}$ relates to $\bar{C}$ can be seen better in the non-interacting limit. At infinite temperature, $\bar{C}$ could be utilized to straightforwardly come up with an analytical expression for $F_{diag}$: We calculate matrix elements of the edge operator $W$,
\begin{eqnarray}
W_{\alpha \beta}\big \vert_{E_{\alpha}=E_{\beta}} &=&\Bra{\psi_{\alpha}}f(h)\gamma_1\left(\frac{\gamma_1+i\gamma_2}{\sqrt{2}}\right) \Ket{\psi_{\alpha}}  \notag\\
&=& \frac{2f(h)}{\sqrt{2}} = \sqrt{1-h^2}, \label{MajoranaFdiagEqn}
\end{eqnarray}
in the topologically non-trivial phase; $W_{\alpha \beta}\big \vert_{E_{\alpha}=E_{\beta}}=0$ otherwise. Here $f(h)$ is a smooth function of magnetic field $h$, that can be extracted numerically for finite size systems, whereas by using $\bar{C}$ \cite{PhysRevB.97.235134} we can determine an analytical expression $f(h)=\sqrt{2(1-h^2)}/2$ in the thermodynamic limit. Hence $F_{diag}=(1-h^2)^2$ can be written, while $\bar{C}=1-h^2$ \cite{PhysRevB.97.235134}. See Appendix C for details and the numerical demonstration of this relation.  

Now we calculate $\bar{F}_{diag}$ for three different scenarios: i) strongly interacting but integrable case (XXZ model), ii) nonintegrable models with different interaction strengths and iii) non-interacting limit; and numerically determine the bounds of our conjecture.

\subsubsection{Strongly interacting but integrable case}

We revisit the Fig.~\ref{Fig1} of the XXZ model in Sec.~\ref{sec2}. $F_{diag}$ is shown for an edge-spin $\sigma^z_1$ (obc) with red-squares; whereas the $F_{diag}$ of bulk-spins $\sigma^z_1$ (pbc) and $\sigma^z_7$ (obc) operators are with purple-dots and light-blue right-pointing triangles, respectively. We observe that the diagonal contribution could be used to approximate $\bar{F}$ at the edge in the Ising phases, confirming the conjecture. Even though this model has interactions between Majorana fermions Eq.~\eqref{XXZMajorana}, it is still an integrable system which might explain why $\bar{F}$ does not completely reduce to its diagonal contribution in the long-time limit. However, the qualitative behavior is the same. The diagonal (and hence topological) contribution in the XY-phase becomes zero which is consistent with a gapless phase. Hence the sole contribution in the XY-phase is the corrections, which shows a steady non-zero residue $\bar{F} \neq 0$. This residue seems to be a consequence of the rotational symmetry of the system, $[H,S_z]=0$ and could be expected to vanish away in the thermodynamic limit (Appendix F). Since the topological order is not visible to bulk degrees of freedom, we see $F_{diag} \sim 0$ for bulk operators.  

\subsubsection{From nonintegrable cases to non-interacting limit}

A generic Ising model could be introduced as,
\begin{eqnarray}
H &=& -J \sum_{j=1}^{N-1} \sigma^z_{j}\sigma^z_{j+1}-\Delta \sum_{j=1}^{N-2} \sigma^z_{j}\sigma^z_{j+2}+h\sum_{j=1}^N\sigma^x_j,\label{HNonIIsing} \\
&=& -iJ\sum_{j=1}^{N-1}  a_{2j}a_{2j+1}+\Delta \sum_{j=1}^{N-2}a_{2i}a_{2i+1}a_{2i+2}a_{2i+3}\notag \\
&-&ih\sum_{j=1}^N  a_{2j-1}a_{2j},
\label{HNonIIsingM}
\end{eqnarray}
where $\Delta$ is the next-nearest neighbor coupling between spins in Eq. \ref{HNonIIsing} and breaks the integrability of the model. The strength $\Delta$ introduces interactions between Majorana fermions in Eq. \ref{HNonIIsingM}. We focus on three different $\Delta$ values in our numerical analysis from weak to strong integrability-breaking terms (i) $\Delta/J=-0.1$, (ii) $\Delta/J=-0.5$ and (iii) $\Delta/J=-2$.
\begin{figure}
\centering
\subfloat[]{\label{Fig4a}\includegraphics[width=0.24\textwidth]{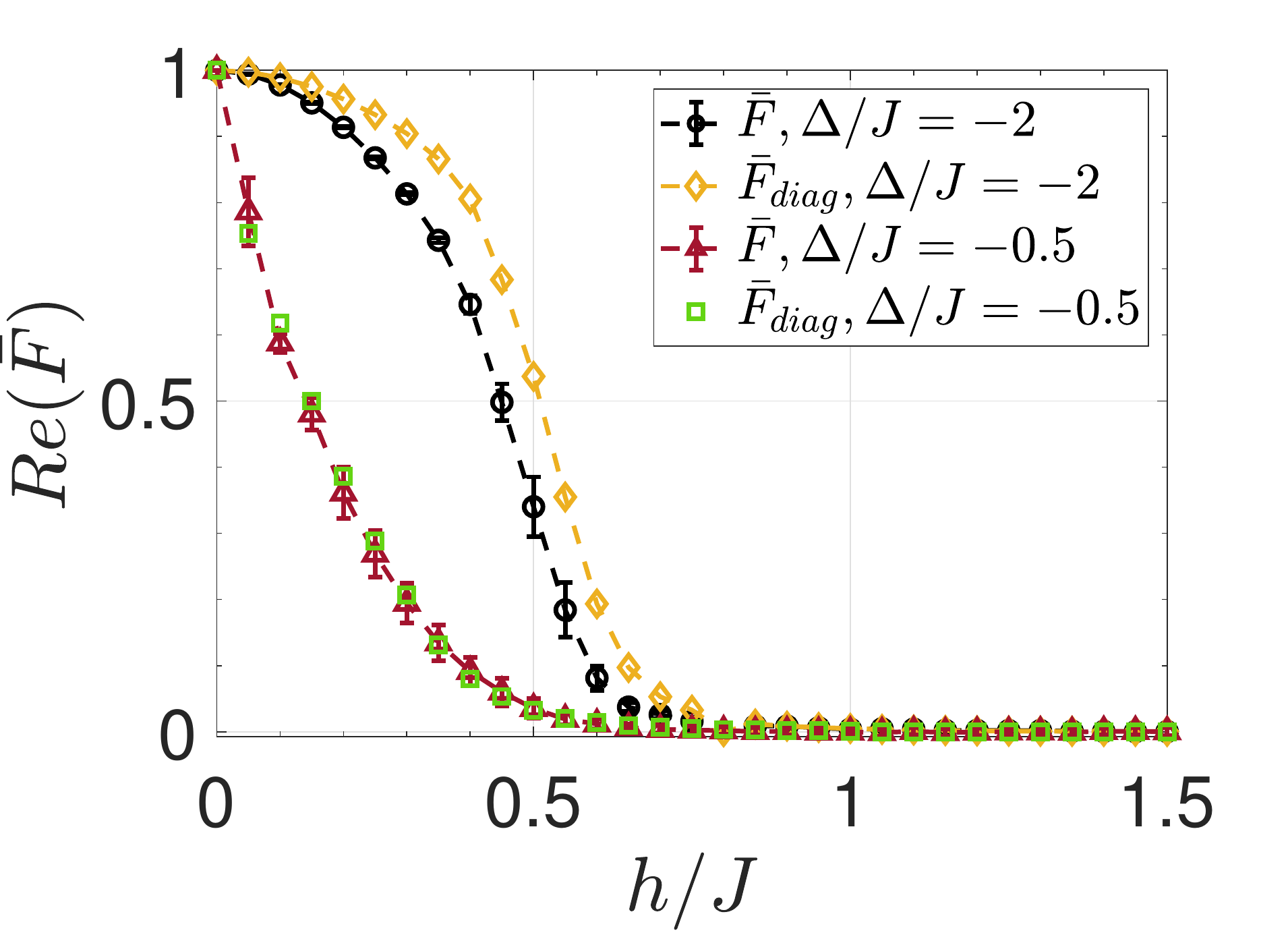}}\hfill 
\subfloat[]{\label{Fig4b}\includegraphics[width=0.24\textwidth]{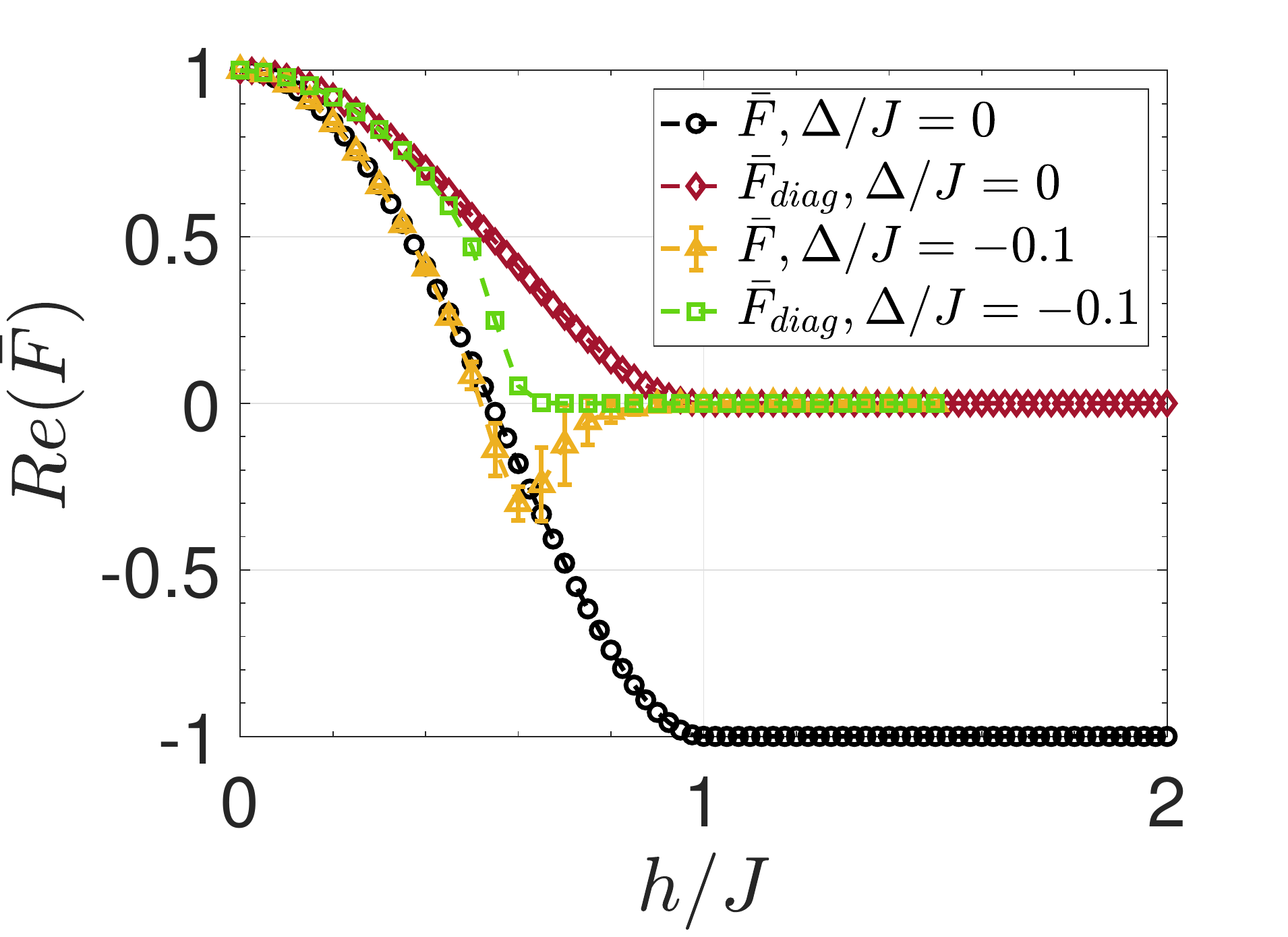}}\hfill 
\caption{Comparison of $\bar{F}$ and its diagonal contribution $\bar{F}_{diag}$ at different non-integrability breaking term strength $\Delta/J$. (a) For a time interval of $tJ=8\times 10^2$ and size $N=14$, $\bar{F}$ (red-triangles) and $\bar{F}_{diag}$ (green-squares) of $\Delta/J=-0.5$; and $\bar{F}$ (black-circles) and $\bar{F}_{diag}$ (yellow-diamonds) of $\Delta/J=-2$. Hence $\bar{F} \sim \bar{F}_{diag}$ holds for a generic nonintegrable system. (b) $\bar{F}$ (yellow-triangles) and $\bar{F}_{diag}$ (green-squares) of $\Delta/J=-0.1$ for a time interval of $tJ=2\times 10^3$ and size $N=14$; and $\bar{F}$ (black-circles) and $\bar{F}_{diag}$ (red-diamonds) of non-interacting fermion model for a size of $N=200$ at the infinite-time limit. At the vicinity of the non-interacting limit, off-diagonal contribution starts to be significant.}
\label{Fig6}
\end{figure}

As we increase the interaction strength, $\bar{F} \sim \bar{F}_{diag}$ holds as expected from the conjecture. Fig.~\ref{Fig4a} compares the dynamic phase diagrams of $\Delta/J=-0.5$ and $\Delta/J=-2$ where time of averaging is fixed to $tJ=800$ for a system size of $N=14$. On the other hand, at the vicinity of the non-interacting limit $\Delta/J=-0.1$, $\bar{F}$ differs from its diagonal contribution  $\bar{F}_{diag}$ considerably (yellow-triangles and green-circles Fig.~\ref{Fig4b}). Consistently, the operator ansatz in the non-interacting limit fails, leading to $\bar{F} \neq \bar{F}_{diag}$. Black-circles and red-diamonds in Fig.~\ref{Fig4b} show $\bar{F}$ and $\bar{F}_{diag}$ calculated at $N=200$ in the infinite-time limit, respectively. Note that the difference is the off-diagonal contribution, which increases towards the phase boundary $h/J \rightarrow 1$ and clearly is not bounded. The off-diagonal contribution is robust, i.e. it does not vanish at infinite-time in thermodynamic limit (Fig.~\ref{Fig4b}). The off-diagonal contribution also shows up in a generic model at near-integrability limit ($\Delta/J=-0.1$), seen in the observation that $\bar{F}$ diverges from $\bar{F}_{diag}$ (Sec.~\ref{ssec:preS} and App.~\ref{AppD}).

\subsubsection{Outlook}

In conclusion, deep in the interacting and/or nonintegrable limit, our conjecture holds and hence $\bar{F} \sim \bar{F}_{diag} \propto \bar{C}$. In near-integrability, OTOC starts to exhibit distinct behavior from two-time correlators and this becomes more apparent in the non-interacting model. We revisit Figs.~\ref{Fig5a} and \ref{Fig5b} where the former is a point deep in the non-trivial phase with $\bar{F}\sim \bar{F}_{diag}$ (Fig.~\ref{Fig4b}) and hence shows similar behavior to $\bar{C}$ with a positive-valued plateau. Whereas Fig.~\ref{Fig5b} demonstrating a closer point to $h_c$ gives $\bar{F}_{diag} \sim 0$, hence the OTOC time-average is mainly contributed by the off-diagonal contribution $|\bar{F}| \gg \bar{F}_{diag} \propto \bar{C}$, resulting in a negative-valued plateau.

\subsection{Effect of scrambling on dynamic phase diagrams}

\begin{figure}
\centering
\subfloat[]{\label{Fig6a}\includegraphics[width=0.24\textwidth]{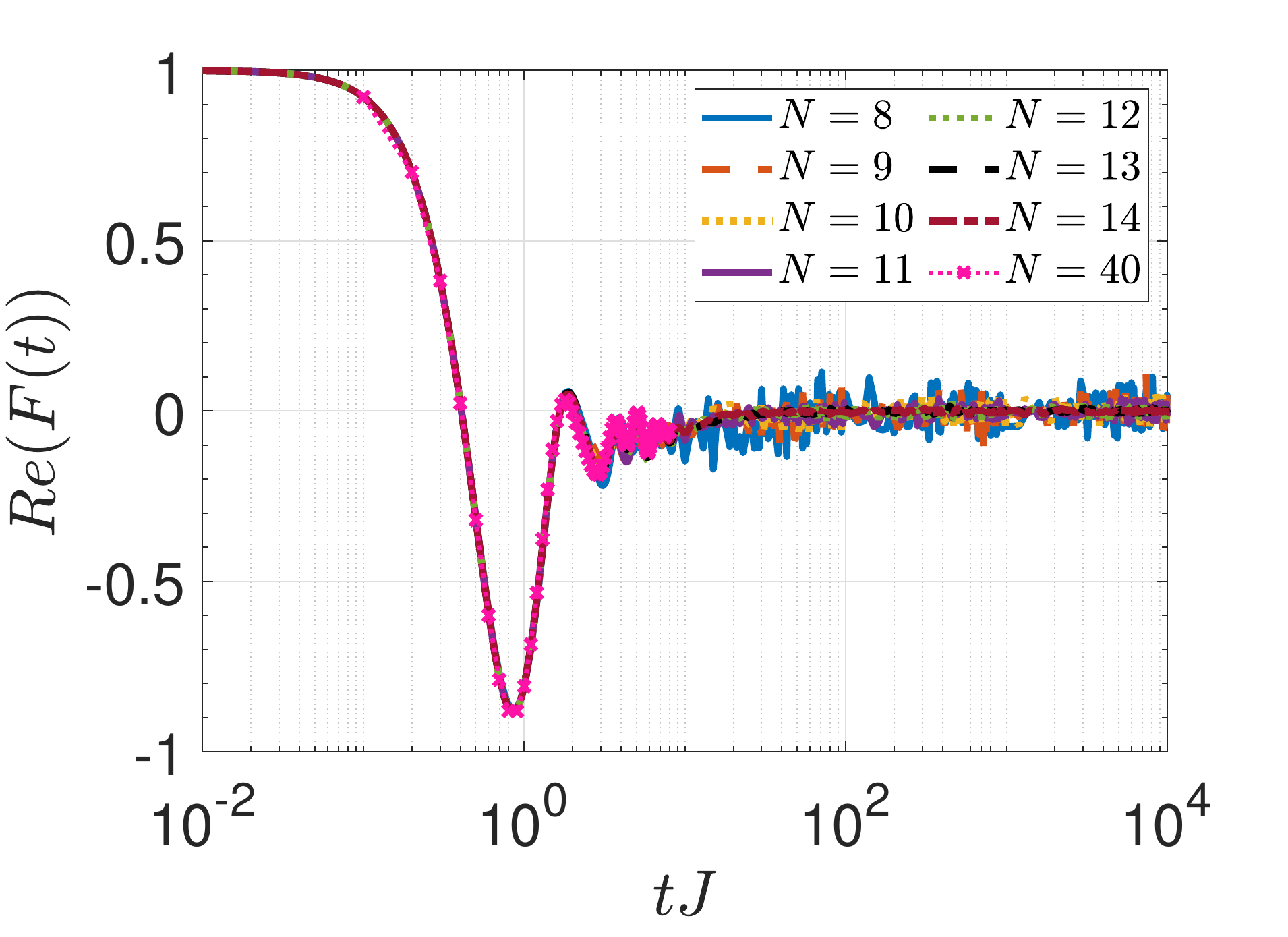}}\hfill
\subfloat[]{\label{Fig6b}\includegraphics[width=0.24\textwidth]{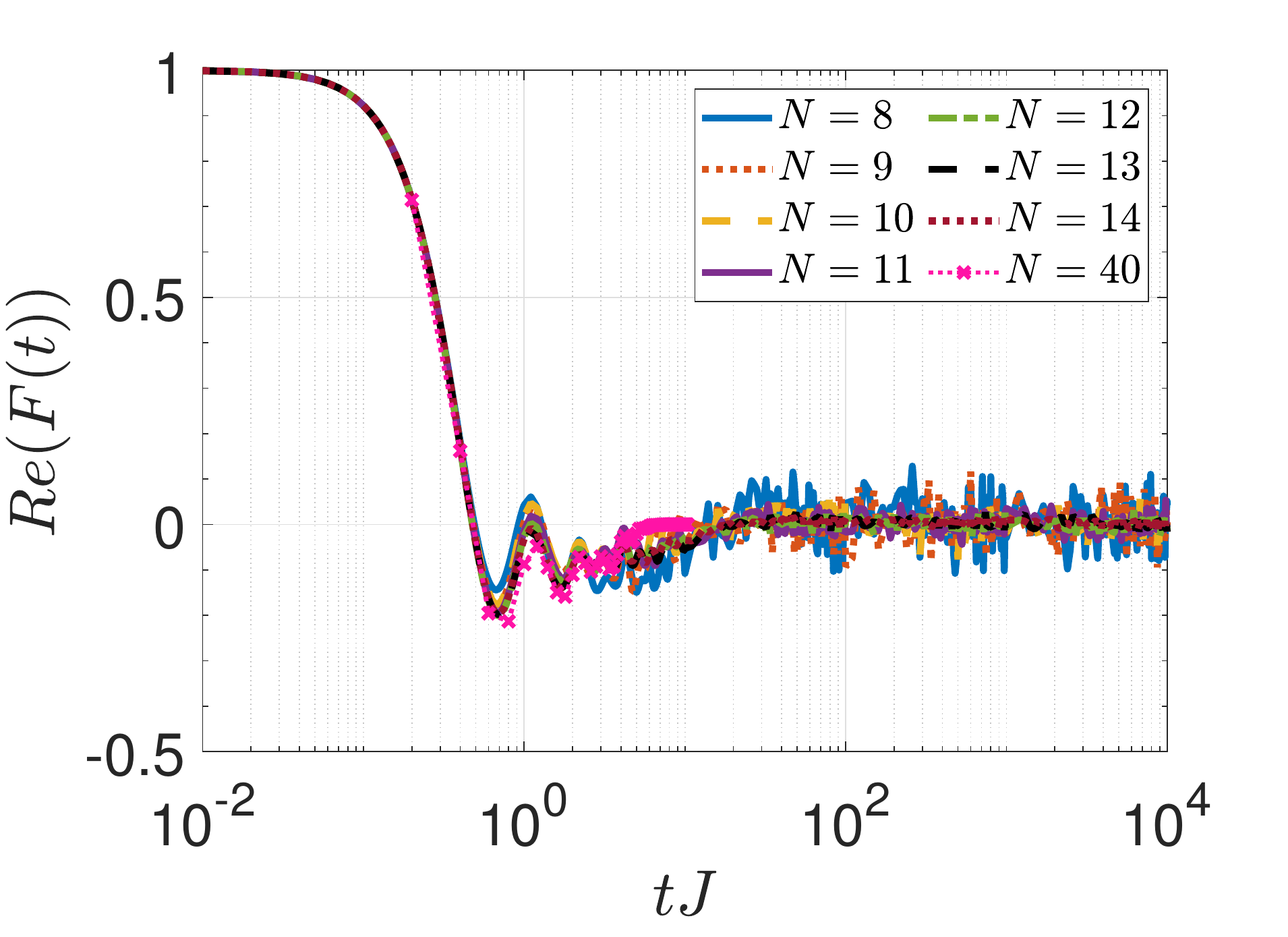}}\hfill
\caption{Coherence times of the edge spins based on OTOC of (a) $\Delta/J=-0.5$ and (b) $\Delta/J=-2$ closer to the critical point in their respective topological phases at $h/J=1$ for different system sizes. The size $N=40$ in both sub-figures is calculated via t-DMRG by averaging $10$ different random product states.}
\label{MFig6}
\end{figure}
The topological transition for $\Delta/J=-0.5$ and $\Delta/J=-2$ occurs at $h/J\sim 1.7$ and $h/J \sim 3.78$, respectively (Appendix E). On the other hand, Fig.~\ref{Fig4a} demonstrates the dynamic transition boundaries early on, $h_{dc}/J < 1$. Even though one might argue for finite-size effects, such a dramatic shift begs for additional reasons. The observation that prescrambling plateau has a finite lifetime in a nonintegrable model also suggests that the dynamic phase diagrams would significantly depend on the interval of the time-averaging (Appendix D for demonstration). Hence it is not clear even if a dynamical phase transition boundary could be well-defined. Given such technical problems, instead of finite-size scaling to mark a transition point, we aim to bound the dynamic phase boundaries in these models. Figs.~\ref{MFig6} demonstrate very limited prescrambling plateaus whose lifetimes are around $tJ\sim 20$ for $\Delta J=-0.5$ and $\Delta/J=-2$ at $h/J=1$. The curves of multiple system sizes collapse on each other in a computation performed with both ED (exact diagonalization) and DMRG. Hence we can state that the dynamic phase boundary over a relatively long period of time is bounded to $h_{dc}/J < 1$, indeed suggesting a significant shift from the zero-temperature phase boundaries.

\begin{figure}
\centering
\subfloat[]{\label{Fig7a}\includegraphics[width=0.24\textwidth]{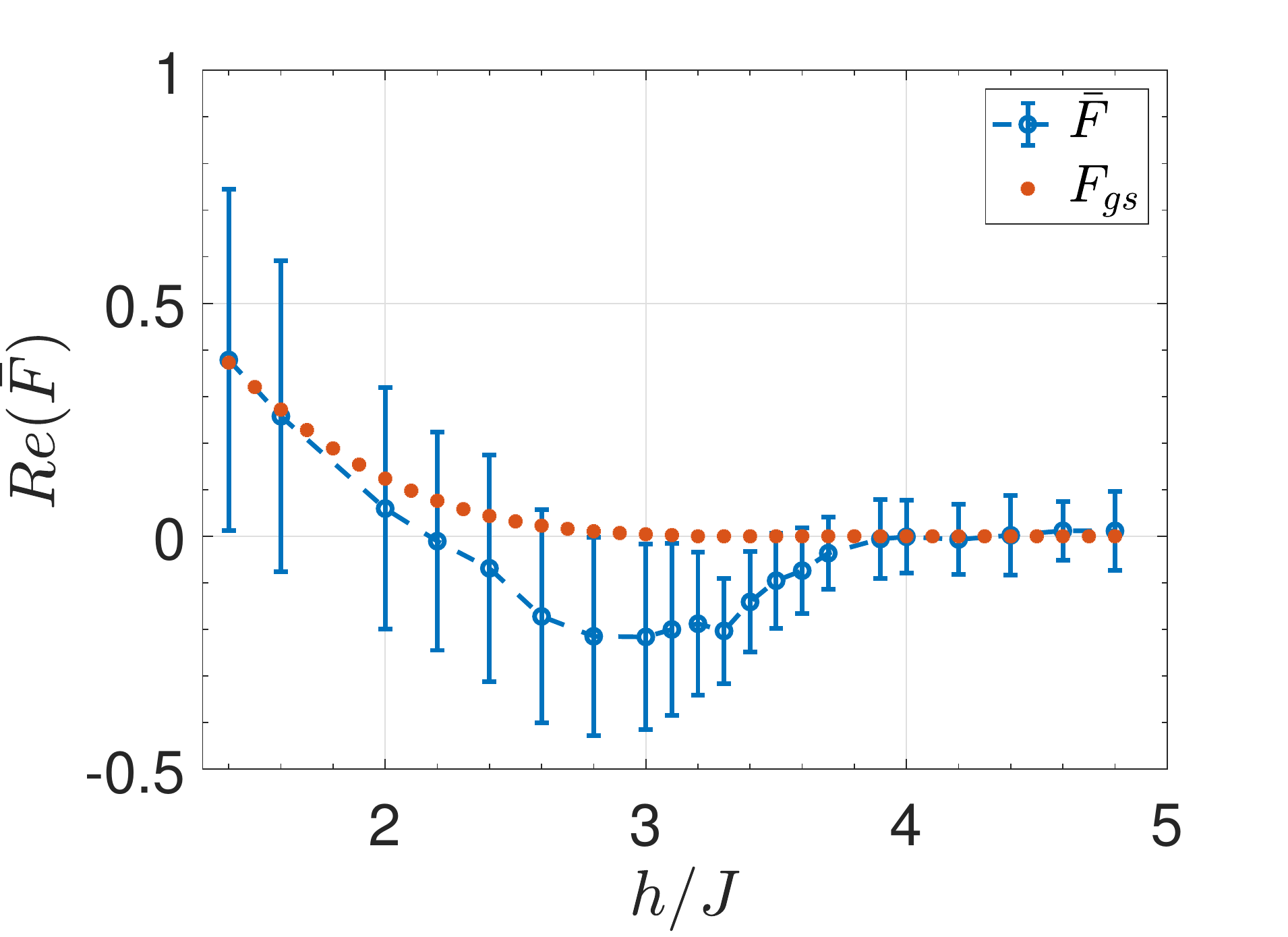}}\hfill 
\subfloat[]{\label{Fig7b}\includegraphics[width=0.24\textwidth]{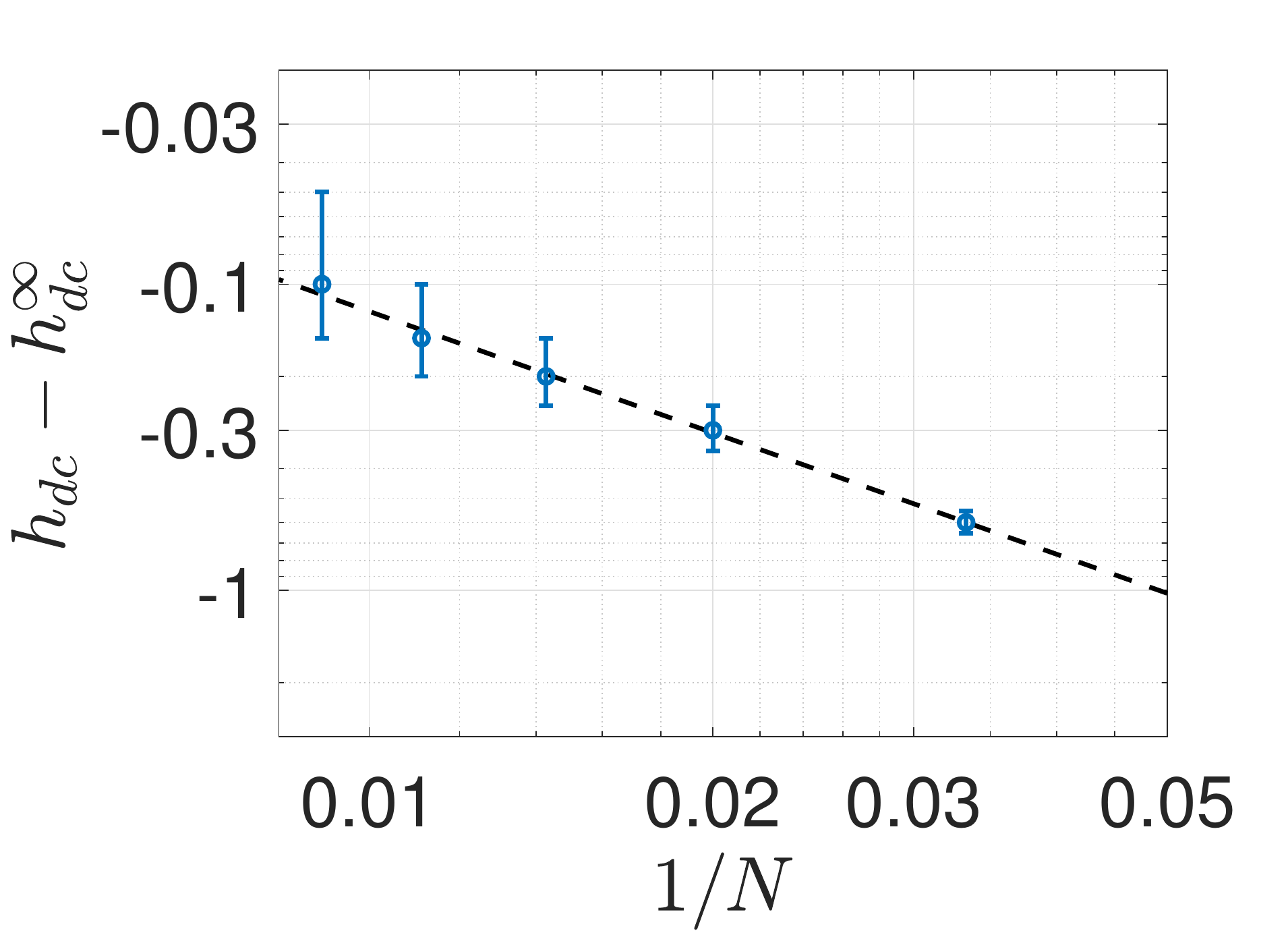}}\hfill 
\caption{(a) OTOC time-average of edge spin for the nonintegrable Ising model with $\Delta/J=-2$ at zero temperature and $N=30$ system size. Blue-circles and orange-diamonds show $\bar{F}$ real-time average over $tJ=N=30$ and the ground-state subspace contribution $F_{gs}$. (b) The system-size scaling of the critical point determined by $F_{gs}$ as $h_{dc}^{\infty} = 3.7\pm 0.05$. All computations in (a)-(b) are done either with t-DMRG or DMRG.}
\label{Fig7}
\end{figure}

Such phase boundary shifts, although more mild than demonstrated here, in dynamical phase diagrams with corresponding symmetry-breaking transitions and that are initiated with polarized states in near-integrable Ising chain have been recently discussed \cite{PhysRevLett.123.115701}. These shifts seem to be linked to exciting the system to higher energy levels when quenched from a polarized state. Hence we can anticipate that working at infinite-temperature possibly maximizes the amount of shift from the zero-temperature phase boundary. Therefore, we lower the temperature to zero and compute $\bar{F}$ and its diagonal contribution which is simply the ground state contribution $\bar{F}_{gs}$ in Fig.~\ref{Fig7a} at $N=30$ and over a time interval of $tJ=30$. The correspondence between $\bar{F}$ and $\bar{F}_{gs}$ motivates us to apply system-size scaling on $\bar{F}_{gs}$. Fig.~\ref{Fig7b} demonstrates this system-size scaling which determines the critical point as $h_{dc}^{\infty} = 3.7 \pm 0.05$. Therefore the dynamical phase boundary is very close to $h_{c}^{\infty} \sim 3.78(2)$ that is determined by two independent methods (Appendix E). Hence the dynamical phase diagram based on OTOC matches fairly well with the topological phase transition boundary in low temperature, suggesting that the shift observed in Fig.~\ref{Fig4a} is indeed an effect from the excited state spectrum. This is perhaps not too surprising, given the discussion on easy spin flips in Sec.~\ref{ssec:preS}. Since increasing the transverse field strength $h$ (linked to spin flip operator) enhances the effect of easy spin flips on the spectrum \cite{2017JSMTE..06.3105K}, the dynamical signature of the topological order is lost well before the field value reaches the critical transition boundary $h_c$. 

In conclusion, we demonstrate the effect of almost-strong zero modes on a dynamic phase diagram based on OTOC showing significant shift in the phase boundaries. Whether it is possible to find a functional dependence of the $h_{dc}$ on temperature is an interesting question that can be studied systematically in future studies. 

\section{Conclusions and Discussions}

We put forward a numerical observation on the XXZ model, where we showed the infinite-temperature OTOC, namely a correlator that probes the quantum chaos in interacting many-body systems, is also susceptible to ground-state phase transitions. The origin of this observation is demonstrated to be Majorana edge modes existing in the system with a systematic study of different models. This suggests the appearance of strong zero modes in the dynamics of information scrambling and OTOCs. We marked the topological phase transition in the non-interacting limit via $\bar{F}$. We further numerically studied the coherence times of the prescrambling plateaus in the nonintegrable models and demonstrated the effect of prescrambling in dynamic phase diagrams. We found that $\bar{F}$ continues to be an order parameter for the topologically non-trivial phase even in the nonintegrable limit where the dynamic phase boundary is significantly altered by the temperature. The dynamical decomposition of infinite-temperature OTOC into diagonal and off-diagonal contribution exhibits the differences and similarities between scrambling and thermalization dynamics affected by (almost-)strong zero modes.

The observations on finite topological order detected via OTOC point to edge spins that remain local for long times in generic systems. Hence the scrambling of the edge spins with the rest of the system is negligible when the $Z_2$ topological order exists. Therefore, we demonstrate how topologically-protected degrees of freedom fight against being scrambled, either completely preventing (integrable systems) or restricting (generic systems) the operator spreading and thus exhibiting a clear interplay between the topological order and scrambling. Nonintegrable systems at infinite temperature are almost always expected to scramble down to zero where the decay rate depends on the symmetries existing in the Hamiltonian. However, we see that this is not always the case and the scrambling can be severely hindered by the topological protection of information. Motivated by these observations, we introduced a two-step scrambling process with the new timescale being prescrambling time $\tau_{presc}$ and the associated process, \emph{topologically induced prescrambling}. Our conclusions in principle can be generalized to higher dimensions for topological states with similar fraction excitations and topological degeneracy \cite{PhysRevX.7.041062}, although the numerical verification is yet to be found.

In principle, this probe allows experimental detection of topological states without a need to cool down the system to ultra-low temperatures whether it is the OTOCs, Eq.~\eqref{OTOCEq} or two-time correlators Eq.~\eqref{twoPoint}, when the control parameter is sufficiently away from the zero-temperature phase boundary. In particular, the infinite-temperature OTOCs are experimentally more appealing than zero-temperature OTOCs \cite{2018arXiv181111191S}, since it can be challenging to prepare a ground state as the initial state in certain experimental platforms.

Although surprising, the interplay between information scrambling and topological order is an intuitive observation. Beside the notion of strong-zero modes affecting the thermalization dynamics \cite{PhysRevX.7.041062}, the entanglement entropy of a ground state has a universal topological contribution in topologically non-trivial phases \cite{PhysRevLett.96.110404,PhysRevLett.96.110405,2012NatPh...8..902J}. Moreover, the connection between OTOCs and the entanglement entropy of the time-evolved states has been introduced too \cite{FAN2017707,PhysRevX.8.021013}. Hence here we make another connection that relates a dynamical quantity to a static property of the Hamiltonian.

\section{Acknowledgements}

C.B.D. thanks Frank Pollmann for helpful suggestions, Chaitanya Murthy for intriguing and fruitful discussions, and P. Myles Eugenio for helpful discussions on the manuscript and work. This work was supported by National Science Foundation under Grant EFRI-1741618.

\appendix

\section{Methods Explained \label{AppA}}

To determine the degeneracy in the spectrum, we need to characterize the uncertainty in energy, $\Delta E$. This means that we define an energy window around each energy level with $\Delta E$ as $[E_m-\Delta E,E_m+\Delta E]$ where we assume that the states remain in this window are degenerate with the state whose associated energy is $E_m$. This process defines an energy resolution and in a way coarse-grains the energy spectrum. 

As discussed in Ref. \cite{PhysRevLett.123.140602}, the energy resolution is related to the interval of the time-evolution. Longer time-evolution translates to finer energy resolution, resolving the smallest energy differences in the spectrum, $\mathcal{T} \Delta E \sim 1$, where $\mathcal{T}$ is the total time of the evolution. Hence anytime we simulate a system with a finite time interval, we define an energy resolution as $\Delta E = \frac{\pi}{4\mathcal{T}}$. In return, the parameter $\Delta E$ determines the degenerate subspaces in the spectrum and hence helps us to determine the diagonal contribution $\bar{F}_{diag}$ in OTOC time-average. Note that this reverse relation between the time interval and energy resolution also implies that any degeneracy lifting will be eventually captured by a long-time evolution. 

We call an equation derived by the dynamical decomposition as a framework equation. If the operator in the eigenbasis $W_{\alpha\beta}$ can be calculated analytically for an integrable system, that would present us the analytical expression of its OTOC saturation value. However, one can numerically derive the matrix elements $W_{\alpha\beta}$ too and use them in the framework of dynamical decomposition. Any brute force calculation of the OTOC saturation value requires an estimation on the time-dependent part in the dynamical Eq. \ref{OTOCdynamicsM}, e.g. which energy pairs are equal to each other. The energy resolution $\Delta E$ is used here to define a threshold so that we could exert the degenerate subspaces on the OTOC calculation. Crudely speaking, this threshold determines whether the saturation value is contributed by the found energy set $\left\lbrace E_{\alpha},E_{\beta}, E_{\gamma},E_{\delta} \right\rbrace$. In the end, the numerical incorporation of a finite energy resolution into our framework equation that analytically determines the saturation value, also provides us the time-average of OTOC over any time interval up to dramatic transient features \cite{PhysRevLett.123.140602}. Hence we equivalently call $\bar{F}$ both for long-time saturation value and the time-average of OTOC.

When we numerically calculate the OTOC saturation value, we do the summations in Eq. \ref{saturationEqM}. This introduces an approximation to the final OTOC saturation value in our numerical result. We set a threshold where any term greater than the threshold is found and summed over. We determine our threshold based on the dimension of the Hilbert space, $\sim 1/M^2$, where $M$ is the dimension of the Hilbert space. This generally bounds the error on the order of $\sim 10^{-2}$ (we remind the reader that $|F| \leq 1$). 
%This approximation is the reason why the line with red-squares in Fig. \ref{Fig4a} slightly diverges from the line with blue-circles.

We utilize ITensor platform in C$++$ environment and MPS (matrix product states) for DMRG computations \cite{ITensor}. To prepare infinite temperature states in MPS format, we average over random product states. We restrict the bond numbers to $m \lesssim 100$. Since the bond numbers increase rapidly as the system evolves in time, this results less accuracy for the later times. Therefore, we restrict our time-evolution with MPS at infinite-temperature to $tJ \lesssim 10$. The t-DMRG of OTOC in low temperatures or zero temperature present modest bond numbers, hence we are able to simulate OTOC at zero temperature for longer times.

\section{Derivation of Fermionic OTOC \label{AppB}}

In order to (both analytically and numerically) solve Kitaev chain, we double the Hilbert space of single-particles and generate the BdG Hamiltonian. This Hamiltonian gives us a symmetric spectrum around energy $E=0$ where there are two states at $E=0$ when the chain is open due to the localized Majorana fermions at two ends. Therefore, if we derive an equation for OTOC in terms of single-particle states, via summing over only $E=0$ states (Majorana zero modes) due to Eq.~\eqref{OTOCEq3M}, we can calculate the OTOC in the infinite-time limit. 

We work with the fermion operator in doubled space, that is, in addition to $d_i=c_i$ we also have $d_{i+N}=c_i^{\dag}$, hence $d_i$ has a dimension of $2N$ where $N$ is the dimension of the free fermionic system without pairing terms. Note that in addition to the familiar anti-commutation relation $\left\lbrace d_i, d_j^{\dag} \right\rbrace = \delta_{ij}$, we have $\left\lbrace d_i, d_{j+N} \right\rbrace = \delta_{ij}$ and $\left\lbrace d_i^{\dag}, d_{j+N}^{\dag} \right\rbrace = \delta_{ij}$. Hence, a Majorana operator can be defined as $a_{2i-1} = c_i+c_i^{\dag} = \left(d_i+d_i^{\dag}+d_{i+N}+d_{i+N}^{\dag}\right)/2$. With this algebra in mind, we can derive
\begin{eqnarray}
F_{2i-1,2i-1}(t)&=&\frac{1}{2^{N}} \text{Tr}\left( a_{2i-1}(t) a_{2i-1} a_{2i-1}(t) a_{2i-1} \right).
\end{eqnarray}
After the substitution of $d_i$ operators,
\begin{widetext}
\begin{eqnarray}
F_{2i-1,2i-1}(t)&=& \frac{1}{2^{2N}} \frac{1}{2^2} \text{Tr}\bigg ( d_i(t)a_{2i-1} d_i(t) a_{2i-1} +  d_i^{\dag}(t)a_{2i-1} d_i^{\dag}(t) a_{2i-1} + d_{i+N}(t)a_{2i-1} d_{i+N}(t) a_{2i-1} \notag \\
&+&  d_{i+N}^{\dag}(t)a_{2i-1} d_{i+N}^{\dag}(t) a_{2i-1} +2\left( d_i(t)a_{2i-1} d_i^{\dag}(t) a_{2i-1} + d_i(t)a_{2i-1} d_{i+N}(t) a_{2i-1} + d_i(t)a_{2i-1} d_{i+N}^{\dag}(t) a_{2i-1}\right)  \notag
\end{eqnarray}

\begin{eqnarray}
&+& 2\left( d_i^{\dag}(t)a_{2i-1} d_{i+N}(t) a_{2i-1} + d_i^{\dag}(t)a_{2i-1} d_{i+N}^{\dag}(t) a_{2i-1} + d_{i+N}^{\dag}(t)a_{2i-1} d_{i+N}(t) a_{2i-1}\right) \bigg).\label{OTOCEq1}
\end{eqnarray}
\end{widetext}

Since the dimension of the Hilbert space is $2^{2N}$, the following identities hold:  
\begin{eqnarray}
\text{Tr}\left( d_i d_i^{\dag}+  d_i^{\dag} d_i \right) &=& 2^{2N} \rightarrow \text{Tr}\left( d_i d_i^{\dag}\right) = 2^{2N-1}. \notag \\
\text{Tr}\left( d_{i+N} d_{i+N}^{\dag} \right) &=& \text{Tr}\left( d_{i} d_{i+N} \right) \\
&=& \text{Tr}\left( d^{\dag}_{i} d^{\dag}_{i+N} \right) = 2^{2N-1}. \notag \\
\text{Tr}\left( d_i d_i^{\dag} \left(d_i^{\dag} d_i + d_i d_i^{\dag}\right) \right) &=& 2^{2N-1} \rightarrow \text{Tr}\left( d_i d_i^{\dag} d_i d_i^{\dag}\right) = 2^{2N-1}. \notag \\
\text{Tr}\left( d_{i+N} d_{i+N}^{\dag} d_{i+N} d_{i+N}^{\dag}\right) &=& \text{Tr}\left( d_{i} d_{i+N} d_{i} d_{i+N}\right) \notag \\
&=& \text{Tr}\left( d_i^{\dag} d_{i+N}^{\dag} d_{i}^{\dag} d_{i+N}^{\dag}\right) = 2^{2N-1}.\notag
\end{eqnarray}
Eq.~\eqref{OTOCEq1} takes a form of
\begin{eqnarray}
& & F_{2i-1,2i-1}(t) = \\
& &\frac{1}{2^{2N}} \frac{1}{2^2} \sum_{k,l}^{2N} \bigg [ (G_{ik}(t)G_{il}(t) + G_{i+N,k}(t)G_{i+N,l}(t) \notag \\
&+& 2 G_{ik}(t)G_{i+N,l}(t))  \text{Tr}( d_k a_{2i-1} d_l a_{2i-1})  + \text{h.c.}  \bigg] \notag
\end{eqnarray}

\begin{eqnarray}
&+& \frac{2}{2^{2N}} \frac{1}{2^2} \sum_{k,l}^{2N} \bigg [ ( G_{ik}(t)G^*_{il}(t) + G_{ik}(t)G^*_{i+N,l}(t) ) \notag \\
&\times &\text{Tr}( d_k a_{2i-1} d^{\dag}_l a_{2i-1}) \notag \\
&+& \left(G^*_{ik}(t)G_{i+N,l}(t)+G^*_{i+N,k}(t)G_{i+N,l}(t) \right) \notag \\
&\times &\text{Tr}( d^{\dag}_k a_{2i-1} d_l a_{2i-1}) \bigg ], \notag 
\end{eqnarray}
in terms of the matrix elements of the single-particle propagators $G(t)=\exp\left( -iH_{\text{BdG}}t \right)$. 

The term $\text{Tr}( d_k a_{2i-1} d_l a_{2i-1})$ is non-zero only when $k=l=i$ or $k=l=i+N$ where in both cases $\text{Tr}( d_k a_{2i-1} d_l a_{2i-1}) = 2^{2N}$. The term $\text{Tr}( d_k a_{2i-1} d^{\dag}_l a_{2i-1})$, on the other hand, vanishes for $k=l=i$ and $k=l=i+N$, however survives for $k=l\neq i$ and $k=l\neq i+N$. In this case, $\text{Tr}( d_k a_{2i-1} d^{\dag}_l a_{2i-1}) = -2^{2N}$. Note that none of these terms survives if $k=i, l=i+N$ and vice versa. Therefore we end up with

\begin{widetext}
\begin{eqnarray}
F_{2i-1,2i-1}(t)&=& \frac{1}{2^2} \left[ \left(G_{ii}(t)\right)^2  +2 \left(G_{i,i+N}(t)\right)^2 + \left(G_{i+N,i+N}(t)\right)^2 + 2\left(G_{ii}(t)G_{i+N,i}(t) + G_{i,i+N}(t)G_{i+N,i+N}(t)\right) + \text{c.c}\right]\notag \\
&-& \frac{1}{2}\sum_{k\neq i, k\neq i+N}^{2N} \left(|G_{ik}(t)|^2+ |G_{i+N,k}(t)|^2 + G_{ik}(t)G^*_{i+N,k}(t) + G^*_{ik}(t)G_{i+N,k}(t) \right).
\end{eqnarray}
\end{widetext}

The unitarity condition reads $\sum_{k}^{2N} |G_{ik}|^2=1$, then
\begin{eqnarray}
\sum_{k\neq i, k\neq i+N}^{2N} |G_{ik}(t)|^2 &=& 1- |G_{ii}(t)|^2 - |G_{i,i+N}(t)|^2. 
\end{eqnarray}
Furthermore, we utilize the relation $\sum_{k=1}^{2N} G_{ik}G^*_{i+N,k}=0$ which leads to
\begin{eqnarray}
\sum_{k\neq i, k\neq i+N}^{2N} G_{ik}(t)G^*_{i+N,k}(t) &=& \\
- G_{ii}(t)G^*_{i+N,i}(t) &-& G_{i,i+N}(t)G^*_{i+N,i+N}(t).\notag
\end{eqnarray}
When these relations are utilized, one can write the final result as
\begin{eqnarray}
F_{2i-1,2i-1}(t)&=& \left(\text{Re}\left(G_{ii}(t)\right) + \text{Re}\left(G_{i,i+N}(t)\right)\right)^2  \\
&+& \left(\text{Re}\left(G_{i,i+N}(t)\right) + \text{Re}\left(G_{i+N,i+N}(t)\right)\right)^2-1, \notag
\end{eqnarray}
for OTOC for a Majorana fermion of type $a_{2i-1}$.  Given $G_{ij}(t)=\sum_{\alpha} \exp\left( -iE_{\alpha}t \right)\braket{\psi_{\alpha,j} | \psi_{\alpha,i}}$ where $\psi_{\alpha,i}$ means the $i^{\text{th}}$ element of the eigenstate $\alpha$ of $H_{\text{BdG}}$, this result should eventually lead to the result stated in the main text,
\begin{eqnarray}
&F&_{2i-1,2i-1}(t)= \label{OTOCEq3}\\
&& \left[\sum_{\alpha=1}^{2N} \left(|\psi_{i \alpha}|^2+ \psi_{i \alpha} \psi_{i+N, \alpha}^* \right) \cos \left(\epsilon_{\alpha}t\right)\right]^2 \notag \\
&+& \left[\sum_{\alpha=1}^{2N} \left(|\psi_{i+N, \alpha}|^2+ \psi_{i+N, \alpha} \psi_{i, \alpha}^* \right) \cos \left(\epsilon_{\alpha}t\right)\right]^2 -1. \notag
\end{eqnarray}

\section{The relation between OTOCs and two-time correlators \label{AppC}}

\renewcommand{\thefigure}{C\arabic{figure}}
\setcounter{figure}{0}  %  this will re-count eq from 1

Eq.~\eqref{twoPoint} shows that the saturation value of a two-time correlator will always be governed by the diagonal elements in the operator $W$. Then $W_{\alpha\beta}=\Bra{\psi_{\alpha}}W\Ket{\psi_{\beta}}$ can be straightforwardly calculated in the non-interacting limit. Here, $\Ket{\psi_{\beta}}$ and $\Ket{\psi_{\alpha}}$ are even and odd parity states in a doubly-degenerate subspace that is dictated by the Majorana zero modes. We note that $\Ket{\psi_{\gamma}}=d\Ket{\psi_{\alpha}}=f(h)\left(\frac{\gamma_1+i\gamma_2}{\sqrt{2}}\right)\Ket{\psi_{\alpha}}$, where $f(h)$ is a function of magnetic field $h$ and $f(h=0)=1/\sqrt{2}$, however decreases as $h\rightarrow 1$. The quantity that we need to calculate becomes $\Bra{\psi_{\alpha}}Wf(h)\left(\gamma_1+i\gamma_2\right) \Ket{\psi_{\alpha}}/ \sqrt{2}$. The effect appears when we use edge spins, hence
\begin{eqnarray}
W &=& \sigma^z_1=\left(c_1+c_1^{\dag}\right)=\gamma_1 \label{W1}\\
W &=& \sigma^z_N=\prod_{j<N}\left(1-2c_j^{\dag}c_j\right)\left(c_N+c_N^{\dag}\right)\notag \\
&=& \mathbb{P}\left(c_N-c_N^{\dag}\right) = i\mathbb{P} \gamma_2,\label{W2}
\end{eqnarray}
where $\mathbb{P}=\prod_{j}^N\left(1-2c_j^{\dag}c_j\right)$ is the parity operator. Eqs. \ref{W1}-\ref{W2} show the operator $W$ in Ising, Dirac and Majorana bases, respectively. If we work with the operator Eq. \ref{W1},
\begin{eqnarray}
\Bra{\psi_{\alpha}}f(h)\gamma_1\left(\frac{\gamma_1+i\gamma_2}{\sqrt{2}}\right) \Ket{\psi_{\alpha}} &=& \frac{2f(h)}{\sqrt{2}},
\end{eqnarray}
where we utilized $\left(\gamma_i\right)^2=\mathbb{I}$ and $-i\gamma_1\gamma_2\Ket{\psi_{\alpha}}=-\Ket{\psi_{\alpha}}$ since $\Ket{\psi_{\alpha}}$ is an odd-parity state. Similarly for Eq. \ref{W2},
\begin{eqnarray}
if(h)\Bra{\psi_{\alpha}}\mathbb{P} \gamma_2\left(\frac{\gamma_1+i\gamma_2}{\sqrt{2}}\right) \Ket{\psi_{\alpha}} &=& \frac{2f(h)}{\sqrt{2}},
\end{eqnarray}
where we additionally use $\mathbb{P}\Ket{\psi_{\alpha}}=-\Ket{\psi_{\alpha}}$. Given each degenerate subspace contributes equally, we write $\bar{C}=2f(h)^2$. A simple functional form of Eq. \ref{twoPoint} is calculated as $\bar{C} =1-h^2$ for $h<J$ and $\bar{C}=0$ for $h>J$ in Ref.~\cite{PhysRevB.97.235134}. We substitute this analytical result into Eq. \eqref{twoPoint} and obtain $W_{\alpha\beta}=\sqrt{1-h^2}$ for $h>J$ in the topologically non-trivial phase. Hence we observe that the diagonal contribution of OTOC is a direct dynamical probe of topological order, giving a non-zero $F_{\text{ex}}^{mj}=\left(1-h^2\right)^2$ in the non-trivial phase.

\begin{figure}
\centerline{\includegraphics[width=0.35\textwidth]{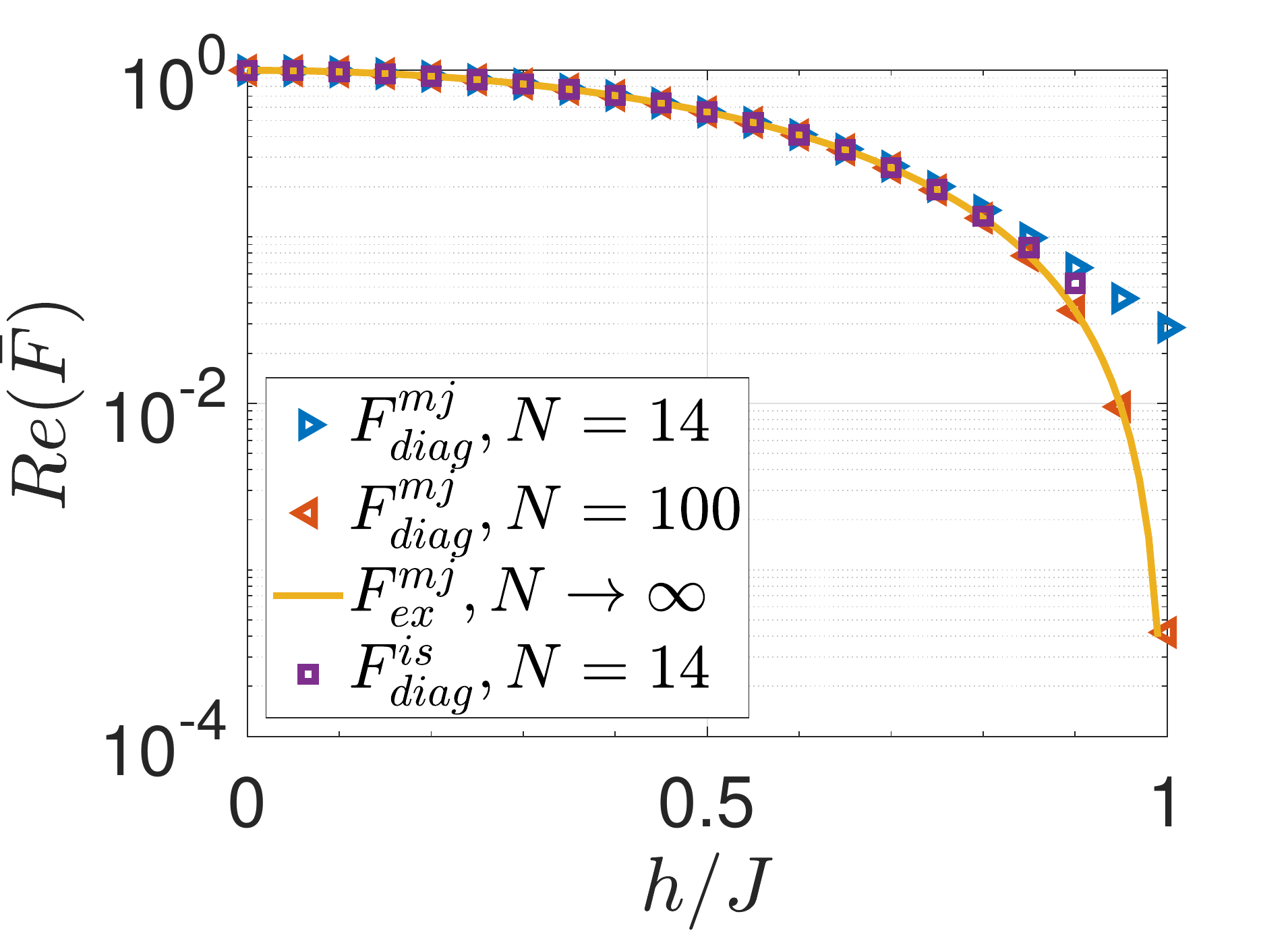}}
\caption{Diagonal contribution in the Ising model and non-interacting fermionic model after dynamical decomposition is applied. Purple-circles show the diagonal contribution Eq.~\eqref{topologicalOTOC} at $N=14$ in the Ising model (for a time interval $tJ=\frac{\pi}{4}10\sim 7.85$), while the blue right-pointing triangles ($N=14$) and red left-pointing triangles ($N=100$) show Eq.~\eqref{topologicalOTOC} for $H_{\text{BdG}}$ in non-interacting fermion system at infinite-time limit. The exact form is derived from the two-time correlators of Majorana fermions (solid-orange).} \label{Fig4S}
\end{figure}

To demonstrate how $\bar{F}_{diag}$ of Ising model can match with Eq.~\eqref{MajoranaFdiagEqn} of non-interacting fermionic system whose calculation is purely based on Majorana zero modes, we plot Fig.~\ref{Fig4S}. Blue right-pointing triangles and orange left-pointing triangles show $\bar{F}_{diag}^{mj}$ numerically computed via Majorana zero modes from BdG Hamiltonian for system sizes of $N=14$ and $N=100$, respectively. Note that $\bar{F}_{diag}^{is}$ of the Ising model (purple-squares) computed at $N=14$ for a time interval of $tJ \sim 8$ matches well with $F_{diag}^{mj}$ at the same size, implying that $F_{diag}^{is}$ could be used to detect the presence/absence of Majorana zero modes. The difference between $N=14$ and $N=100$ sizes of $\bar{F}_{diag}^{mj}$ shows how finite size effects show up near the transition point due to the divergent length scale associated with the quantum critical point. Additionally we compare $\bar{F}_{diag}^{mj}$ at $N=100$ with the analytically derived result $\bar{F}_{\text{ex}}^{mj}$ that is denoted by solid-orange line in Fig.~\ref{Fig4S} and observe that they match perfectly.

\section{Further results on the Ising Model \label{AppD}}

\renewcommand{\thefigure}{D\arabic{figure}}
\setcounter{figure}{0}  %  this will re-count eq from 1

\begin{figure}
\centerline{\includegraphics[width=0.35\textwidth]{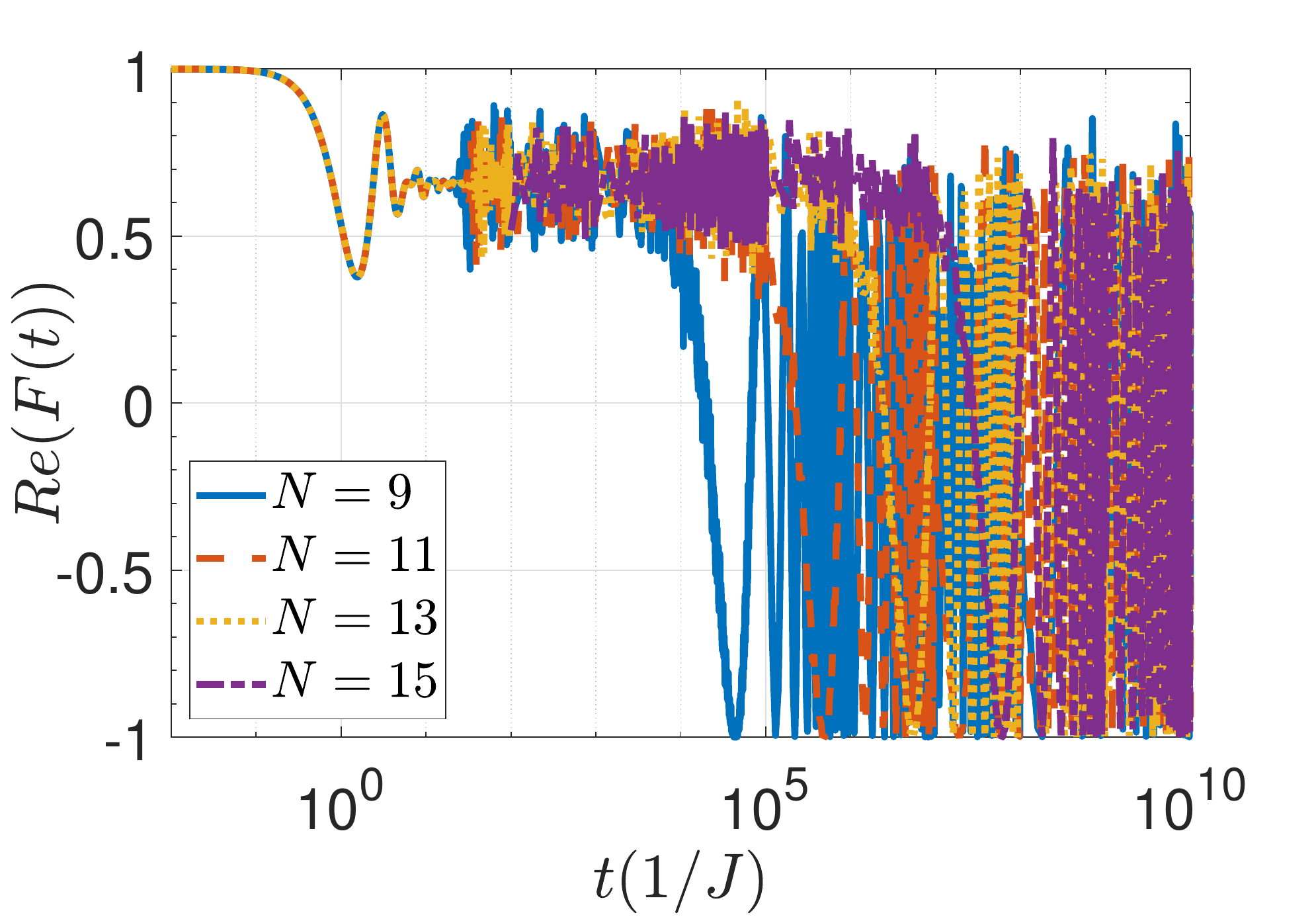}}
\caption{Coherence time computation of the integrable Ising model deep in the non-trivial phase $h/J=0.3$. The coherence times exhibit exponential increase with the system size which implies that prescrambling lasts indefinitely.}
\label{AFig1}
\end{figure}

Fig.~\ref{AFig1} shows that the prescrambling time-scale scales with the system size in the Ising model. Hence, in the thermodynamic limit, prescrambling continues to survive, giving a finite OTOC saturation (time-average) $\bar{F} \neq 0$ at the infinite-time limit. 

\begin{figure}
\centerline{\includegraphics[width=0.35\textwidth]{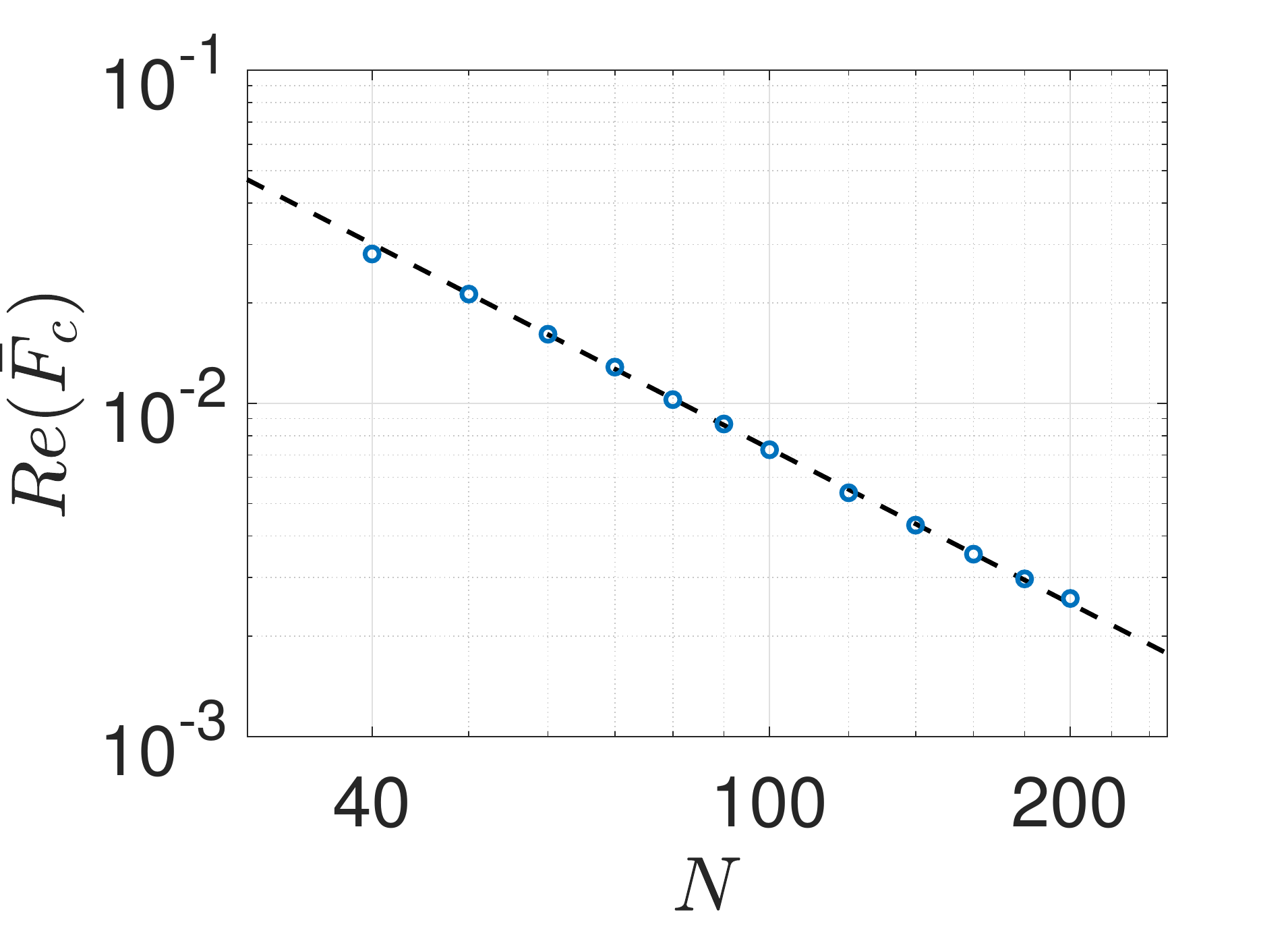}}
\caption{The scaling of OTOC, $F$ with the system size $N$ at the transition point determined by the second derivative of the OTOC (see main text). The scaling parameters are: $F^{\infty}\sim N^{-1.5452} - 1$ with $R^2=0.9994$.}
\label{AFig2}
\end{figure}

Fig.~\ref{AFig2} shows the system-size scaling of fermionic OTOC time-average at the phase transition point that is also determined by OTOC itself. The scaling parameters of the phase transition point was already given in the main text. Here we provide the scaling parameters of the OTOC amplitude with respect to system size: $F^{\infty}\sim N^{-1.5452} - 1$, meaning the OTOC in thermodynamic limit should saturate at $F^{\infty} = -1$ in the transition point.

\begin{figure}
\centering
\subfloat[]{\label{AFig6a}\includegraphics[width=0.24\textwidth]{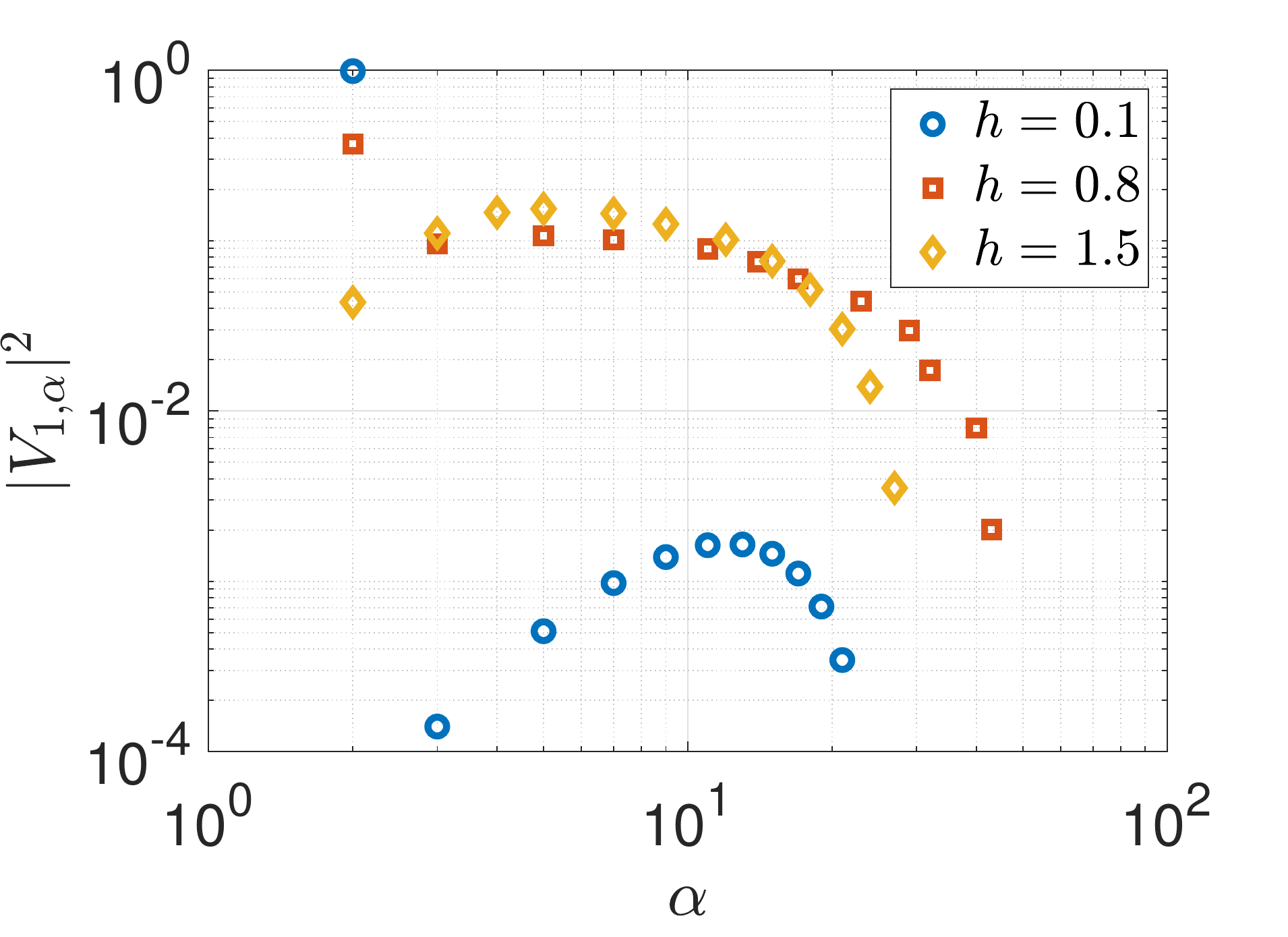}}\hfill 
\subfloat[]{\label{AFig6d}\includegraphics[width=0.24\textwidth]{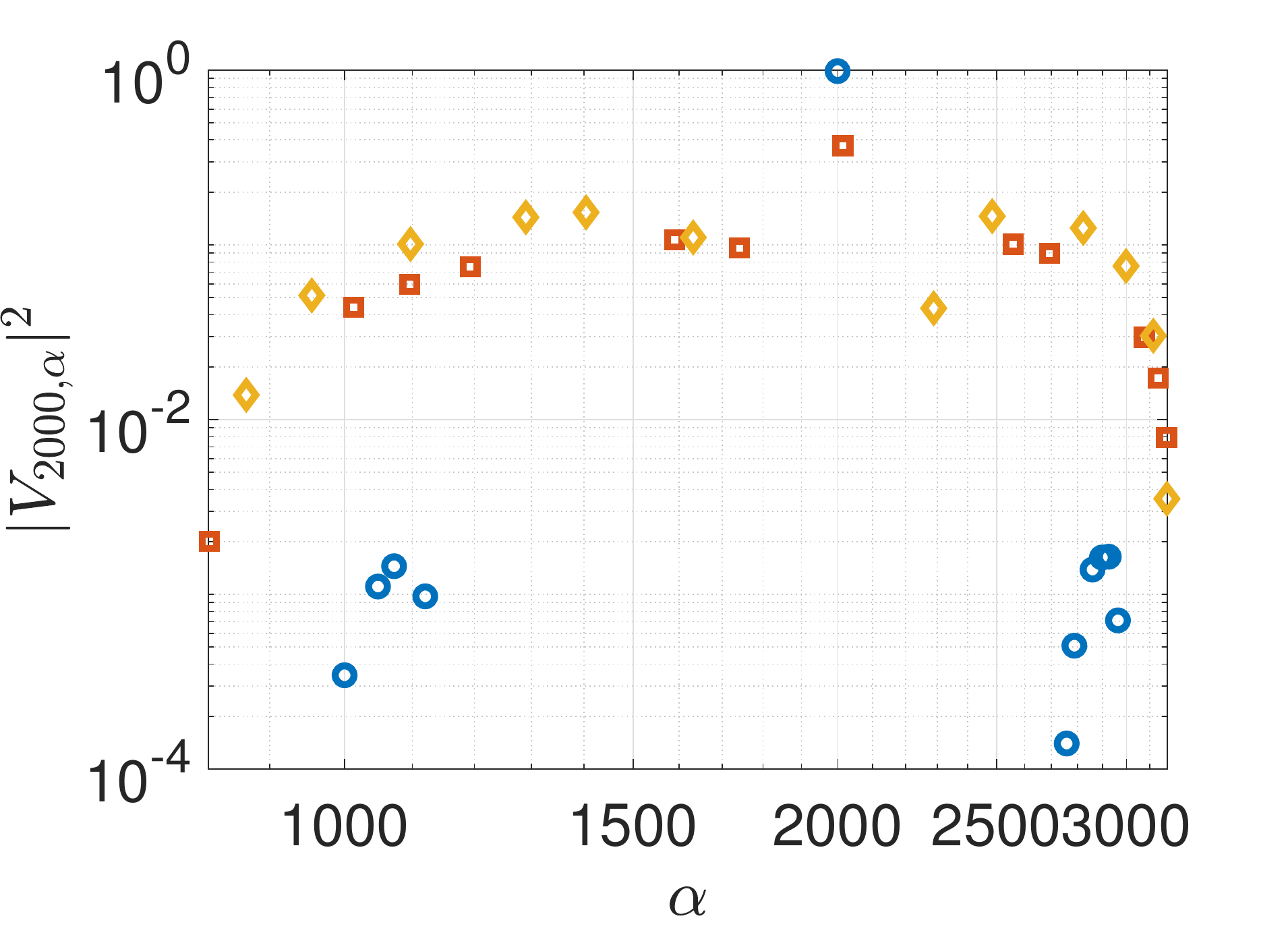}}\hfill
\subfloat[]{\label{AFig6e}\includegraphics[width=0.24\textwidth]{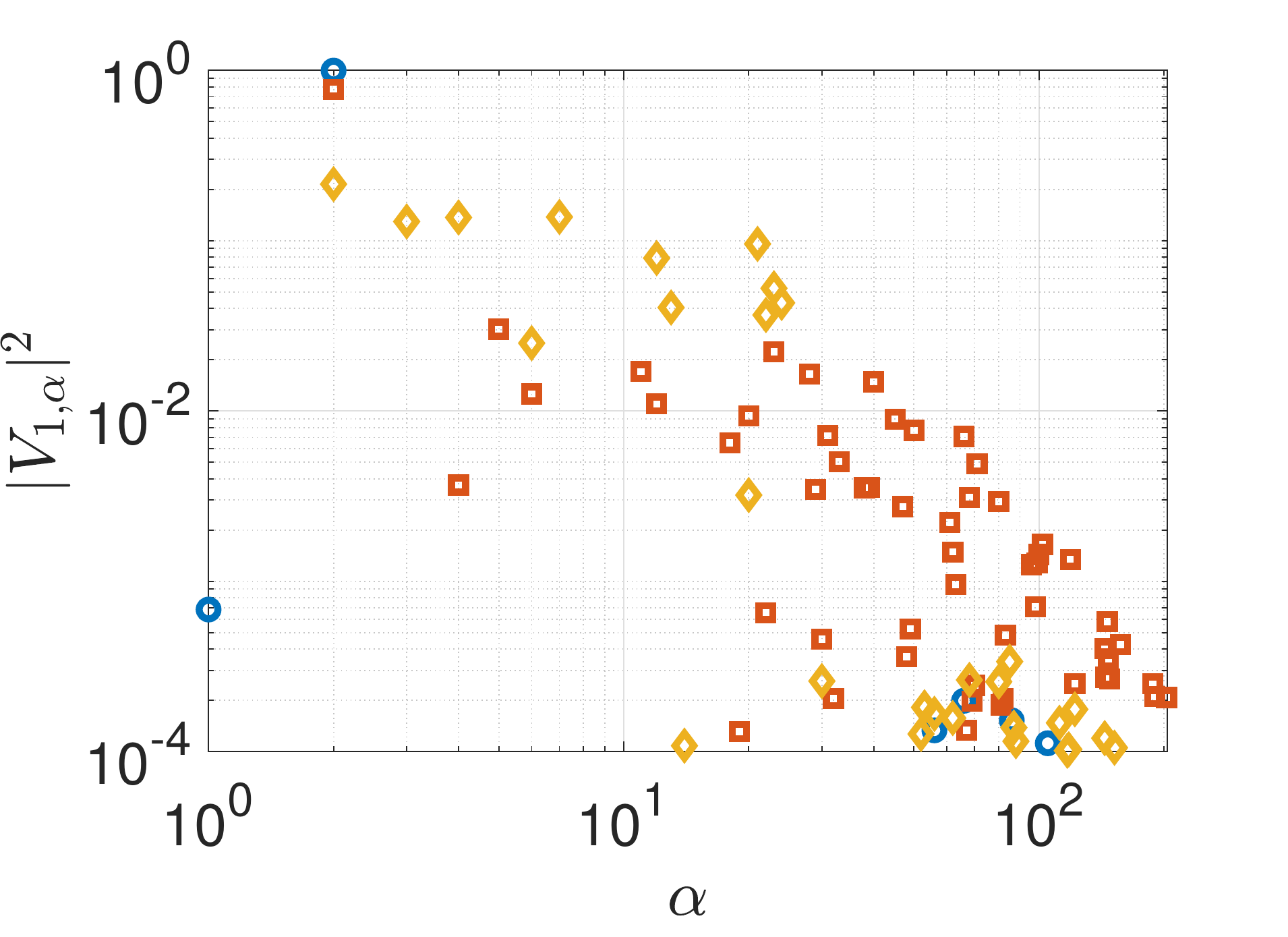}}\hfill 
\subfloat[]{\label{AFig6h}\includegraphics[width=0.24\textwidth]{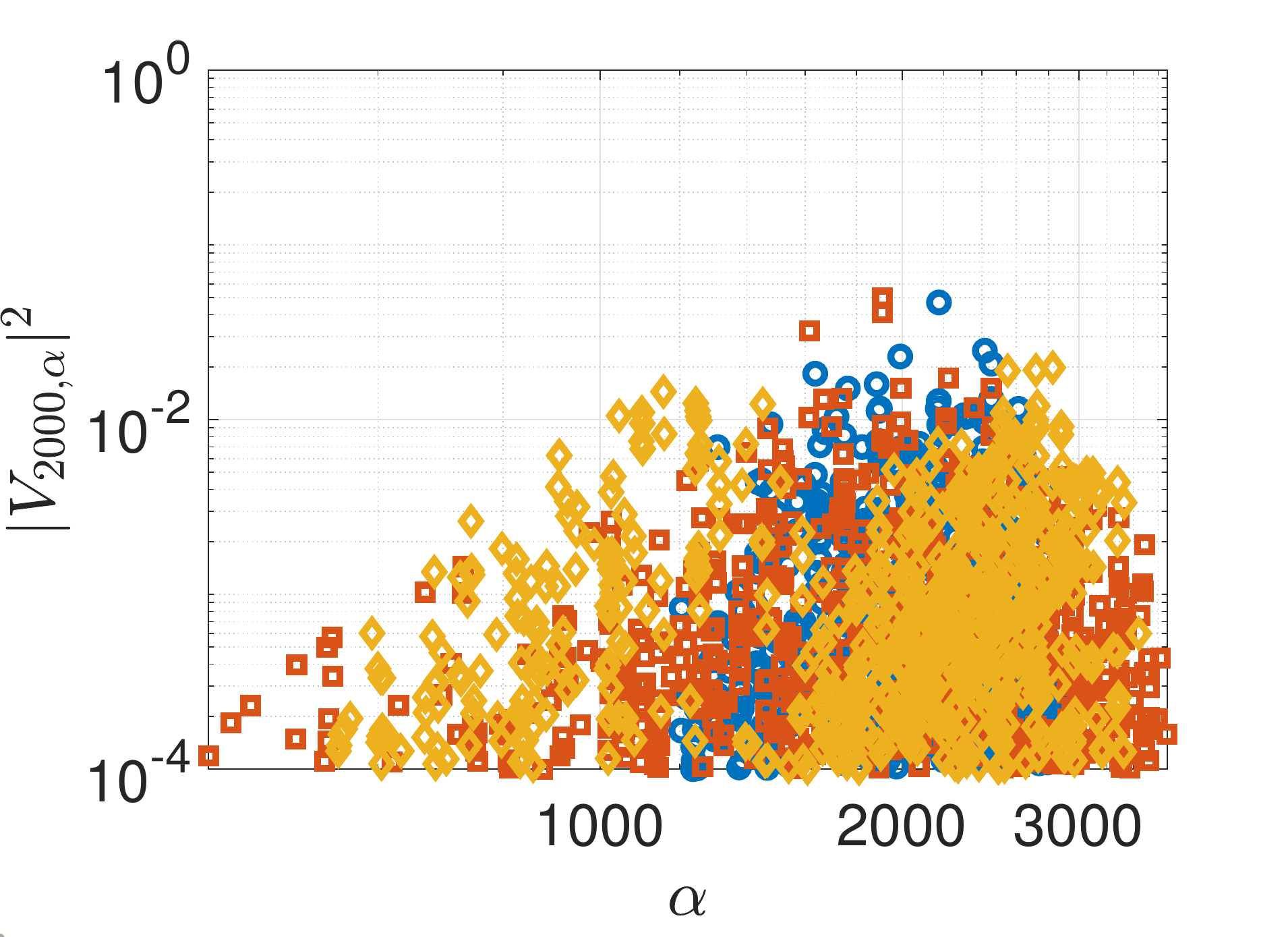}}\hfill
\caption{The operator ansatz tested on the Ising model. Matrix elements $|V_{\beta \alpha}|^2$ are plotted for (a) $\beta=1$ (b) $\beta=2000$ with respect to $\alpha$ for an edge operator $\sigma^z_1$ (open boundary); same $\beta$ (c)-(d) for a bulk operator (periodic boundary) at a size $N=12$. Blue-circles, red-squares and orange-diamonds stand for field strength $h/J=0.1$, $h/J=0.8$ and $h/J=1.5$, respectively for all subfigures.}
\label{AFig6}
\end{figure}

Now we explicitly demonstrate how operator ansatz is satisfied or violated in the integrable Ising model. For this, we plot the matrix elements $|V_{\beta \alpha}|^2$ for various $\beta$ in the spectrum at different $h$ values in Fig.~\ref{AFig6}. Note that $\Ket{\psi_{\beta}}$ and $\Ket{\psi_{\alpha}}$ in $|V_{\beta \alpha}|^2$ denote states sorted according to their energies.

The first two subfigures (a)-(b) are for an edge spin operator $\sigma^z_1$, whereas the rest (c)-(d) are for a bulk spin operator. We sample the ground state (a)-(c) and a state in the middle of the spectrum (b)-(d) in these subfigures. Deep in the topologically non-trivial phase, $h/J=0.1$, we see that the operator ansatz is satisfied $|V_{E_{\alpha}=E_{\beta}}|^2 \gg |V_{E_{\alpha}\neq E_{\beta}}|^2$ for an edge spin (blue-circles). For a bulk spin, the operator ansatz is valid only in the ground state subspace with $E_{\alpha}=E_{gs}$, the condition put forward by Ref.~\cite{PhysRevLett.123.140602} for the dynamical detection of symmetry-breaking phase transitions via OTOCs. This is how the edge spins preserve the topological order in the OTOC throughout the spectrum, while the bulk spins can preserve only the symmetry-breaking order. Closer to the transition point, e.g. $h/J=0.8$,  the order $|V_{E_{\alpha}=E_{\beta}}|^2$, expectantly, decreases while the off-diagonal elements $|V_{E_{\alpha}\neq E_{\beta}}|^2$ grow, which is a signature of integrability at this special non-interacting limit. Hence the operator ansatz, still in the topologically non-trivial phase, breaks down explaining how the OTOC saturation starts to be dominated by off-diagonal contribution (Fig.~\ref{Fig4b} where $\bar{F} \neq \bar{F}_{diag}$ in the non-trivial phase). Note that this breakdown of the operator ansatz in the ordered phase does not happen for the bulk spin that is in its ground state, Fig.~\ref{AFig6e}. The operator ansatz in the topologically trivial phase, e.g. $h/J=1.5$, continues to fail (compare orange-diamonds with blue-circles in Figs.~\ref{AFig6a}-\ref{AFig6d}). Eventually this causes a non-vanishing OTOC time-average $\bar{F} \neq 0$ in the trivial phase, even though this time average value has nothing to do with topological order (Sec.~\ref{ssec:exact}).

\section{Further results on the nonintegrable Ising models \label{AppE}}

\renewcommand{\thefigure}{E\arabic{figure}}
\setcounter{figure}{0}  %  this will re-count eq from 1

\begin{figure}
\centering
\subfloat[]{\label{AFig4a}\includegraphics[width=0.24\textwidth]{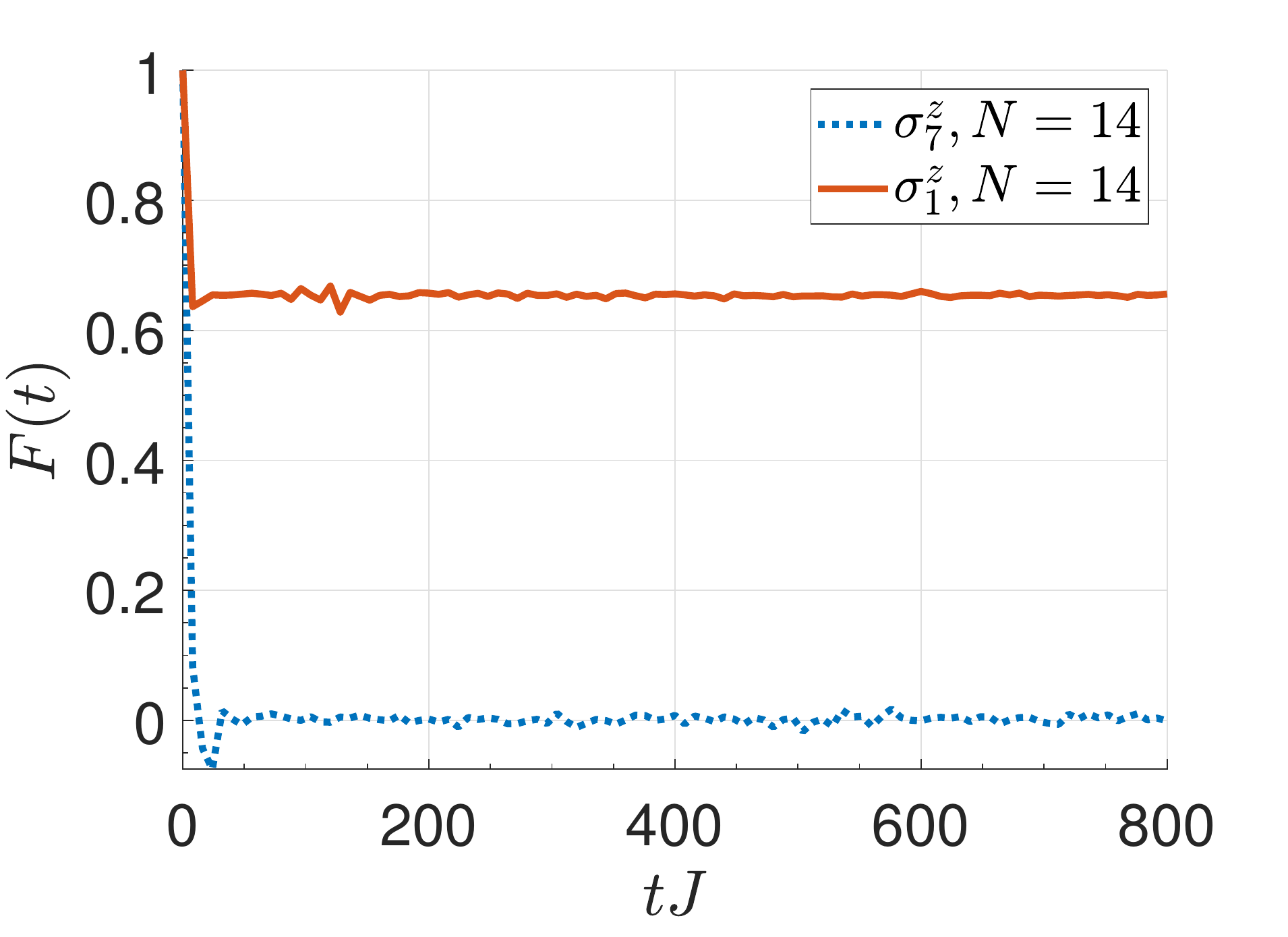}}\hfill 
\subfloat[]{\label{AFig4b}\includegraphics[width=0.24\textwidth]{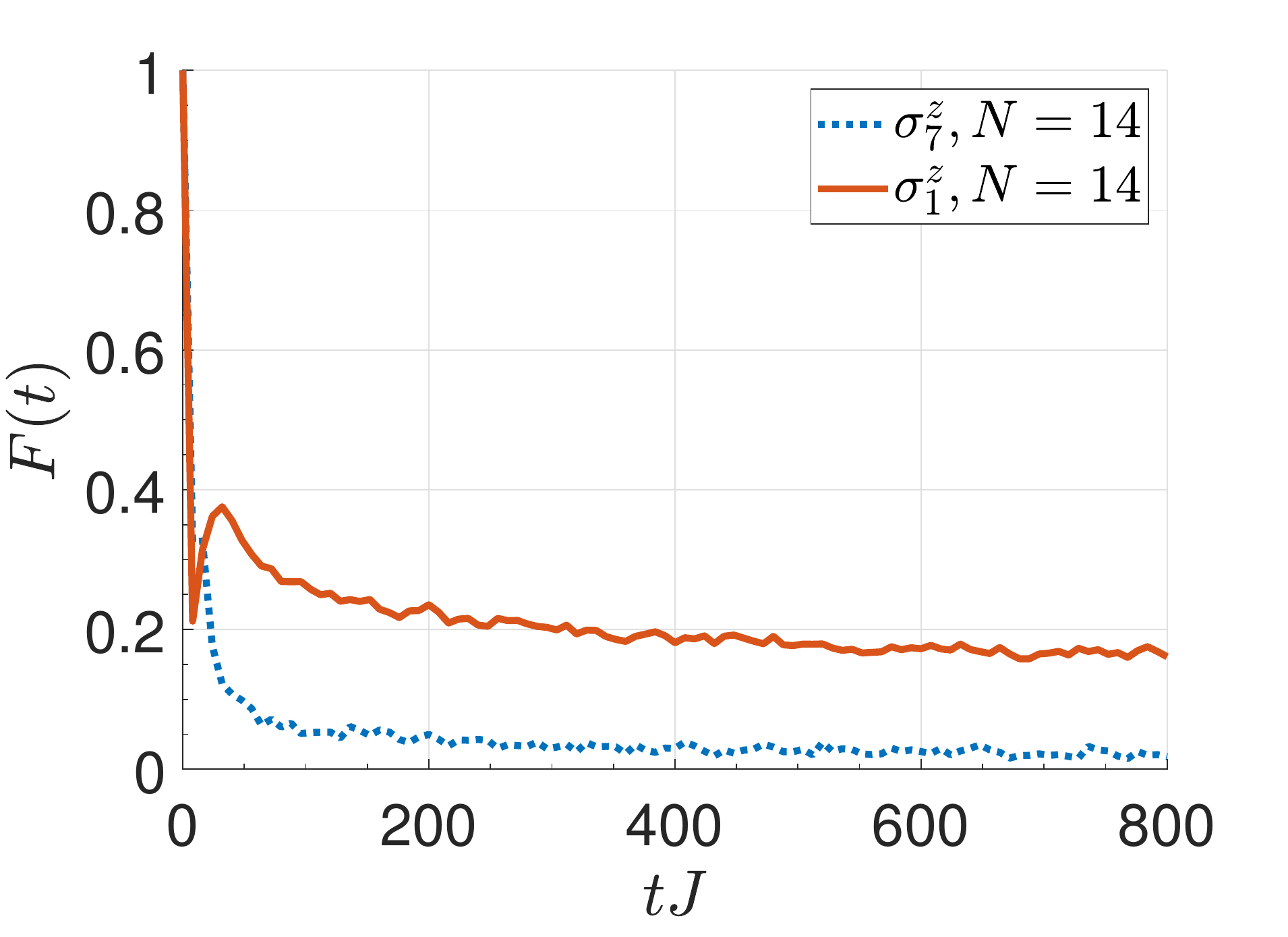}}\hfill 
\caption{Real time dynamics of OTOC with both edge (red-solid) and bulk (blue-dotted) spins in nonintegrable transverse-field Ising model at $h/J=0.3$ for (a) $\Delta/J=-0.1$ and (b) $\Delta/J=-0.5$ with size $N=14$.}
\label{AFig4}
\end{figure}

We first compare the scrambling dynamics of edge (red-solid) and bulk (blue-dotted) spins in real time, Fig.~\ref{AFig4} in the regimes of near-integrability $\Delta/J=-0.1$ and far from integrability $\Delta/J=-0.5$. The edge and bulk spins behave drastically different for significantly long times, even though the size is considerably small, $N=14$. Hence, we can still observe the effect of zero modes in nonintegrable models, however as discussed in the main text, in a weaker form than in integrable models.

\begin{figure}
\centering
\subfloat[]{\label{AFig3a}\includegraphics[width=0.24\textwidth]{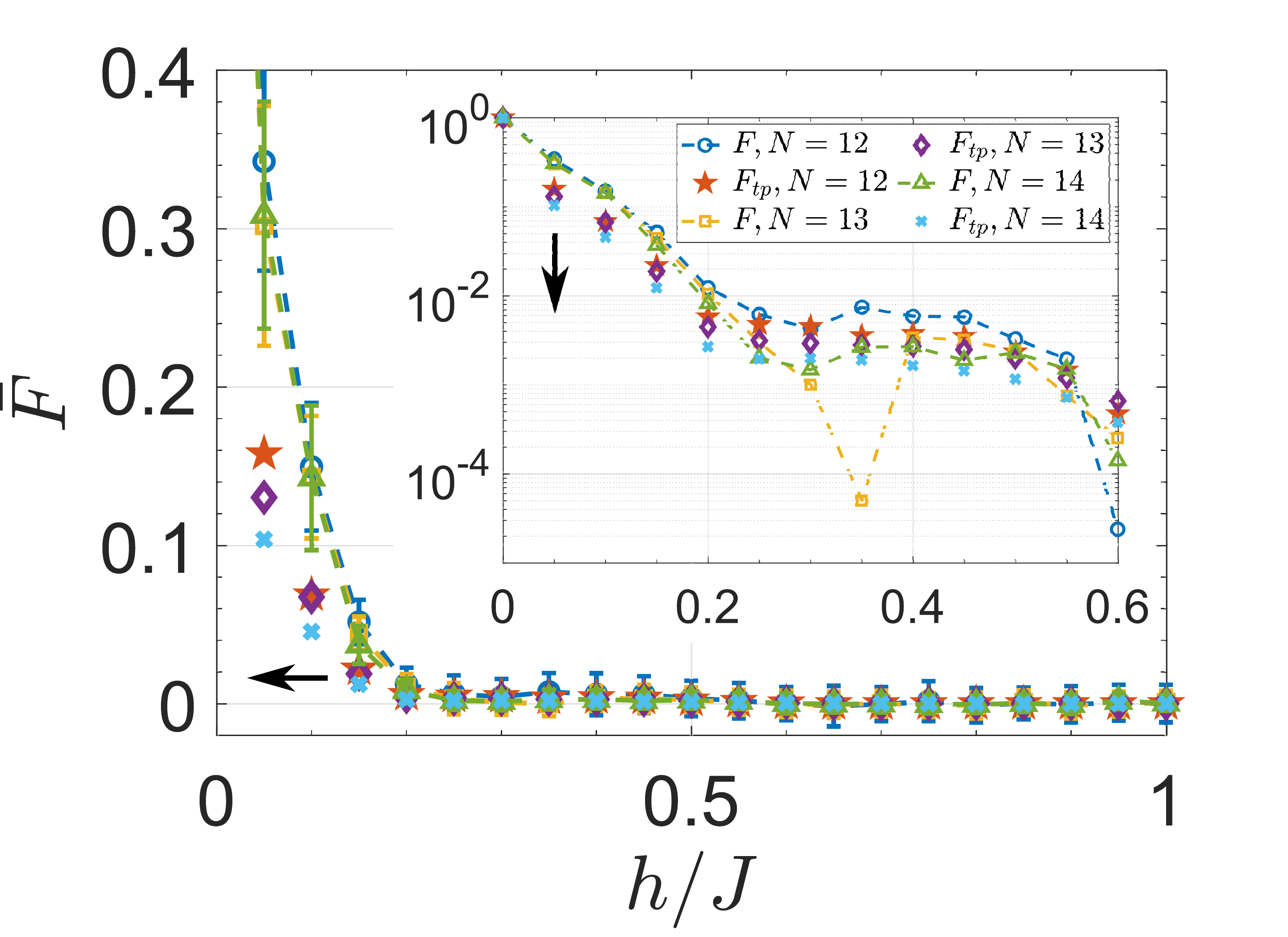}}\hfill 
\subfloat[]{\label{AFig3b}\includegraphics[width=0.24\textwidth]{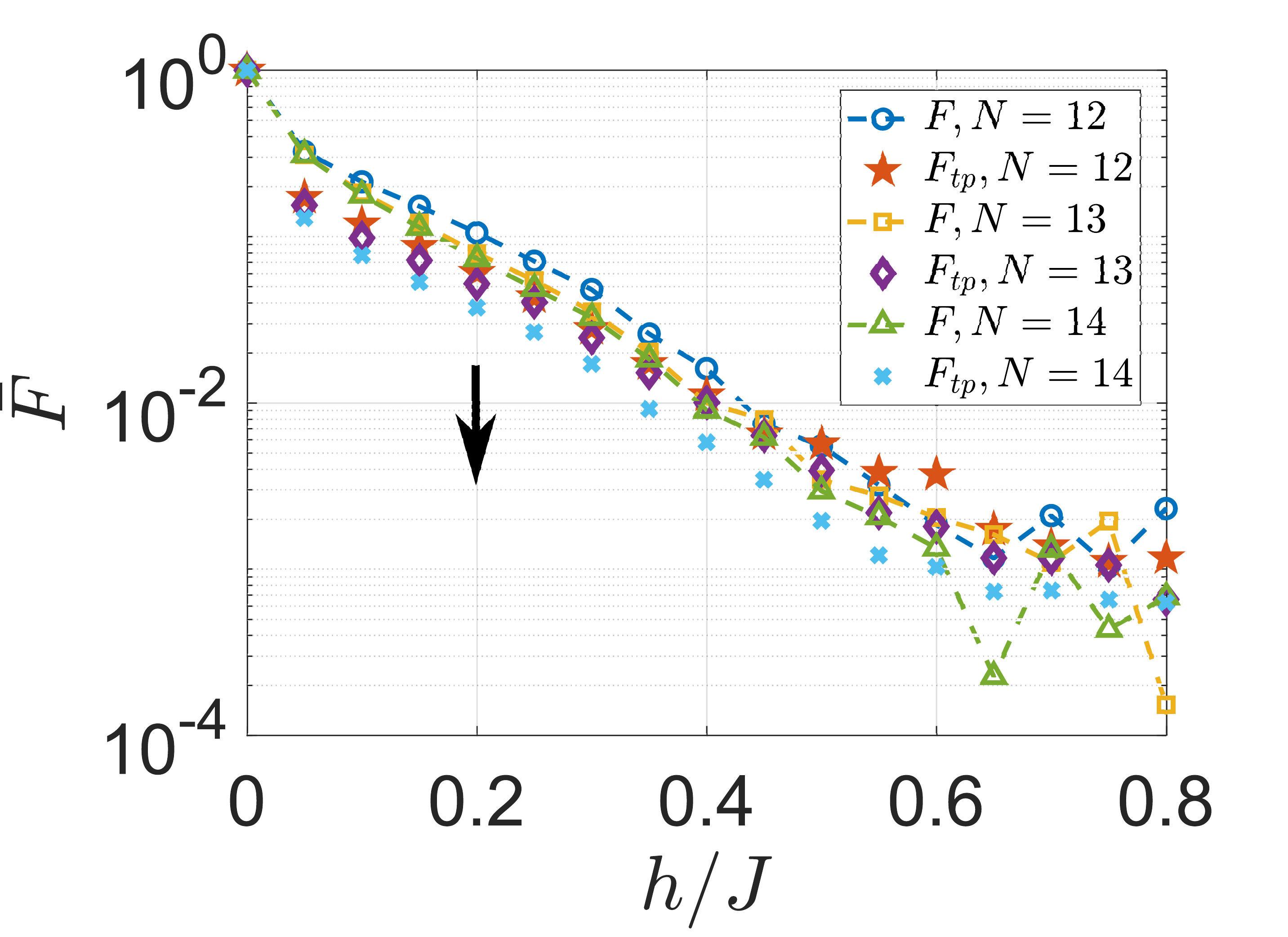}}\hfill 
\caption{Nonintegrable transverse-field Ising model. OTOC time-average of bulk spins in (a) small integrability breaking term $\Delta/J=-0.1$ in linear and logarithmic (inset) scales. Red pentagrams, purple diamonds and light-blue crosses show $\bar{F}_{diag}$ whereas the blue circles, yellow squares and green triangles show $\bar{F}$ for $N=12$, $N=13$ and $N=14$, respectively. (b) The case of $\Delta/J=-0.5$ integrability breaking term. $\bar{F}$ and $\bar{F}_{diag}$ for $N=12$ (blue-circles and red-pentagrams), $N=13$ (yellow-squares and purple-diamonds) and $N=14$ (green-triangles and light-blue crosses). All curves have open boundary conditions and a time interval of $tJ\sim 800$.}
\label{AFig3}
\end{figure}

Now we plot a dynamic phase diagram for a bulk spin in Figs.~\ref{AFig3} and observe it is drastically different than of an edge spin: as we increase the system size, both $\bar{F}$ and $\bar{F}_{diag}$ approach to zero for all $h$, and hence gets even farther away from the transition point. Figs.~\ref{AFig3a} and \ref{AFig3b} show the OTOC of bulk spins in the models with $\Delta/J=-0.1$ and $\Delta/J=-0.5$, respectively. 

\begin{figure}
\centering
\subfloat[]{\label{AFig5e}\includegraphics[width=0.24\textwidth]{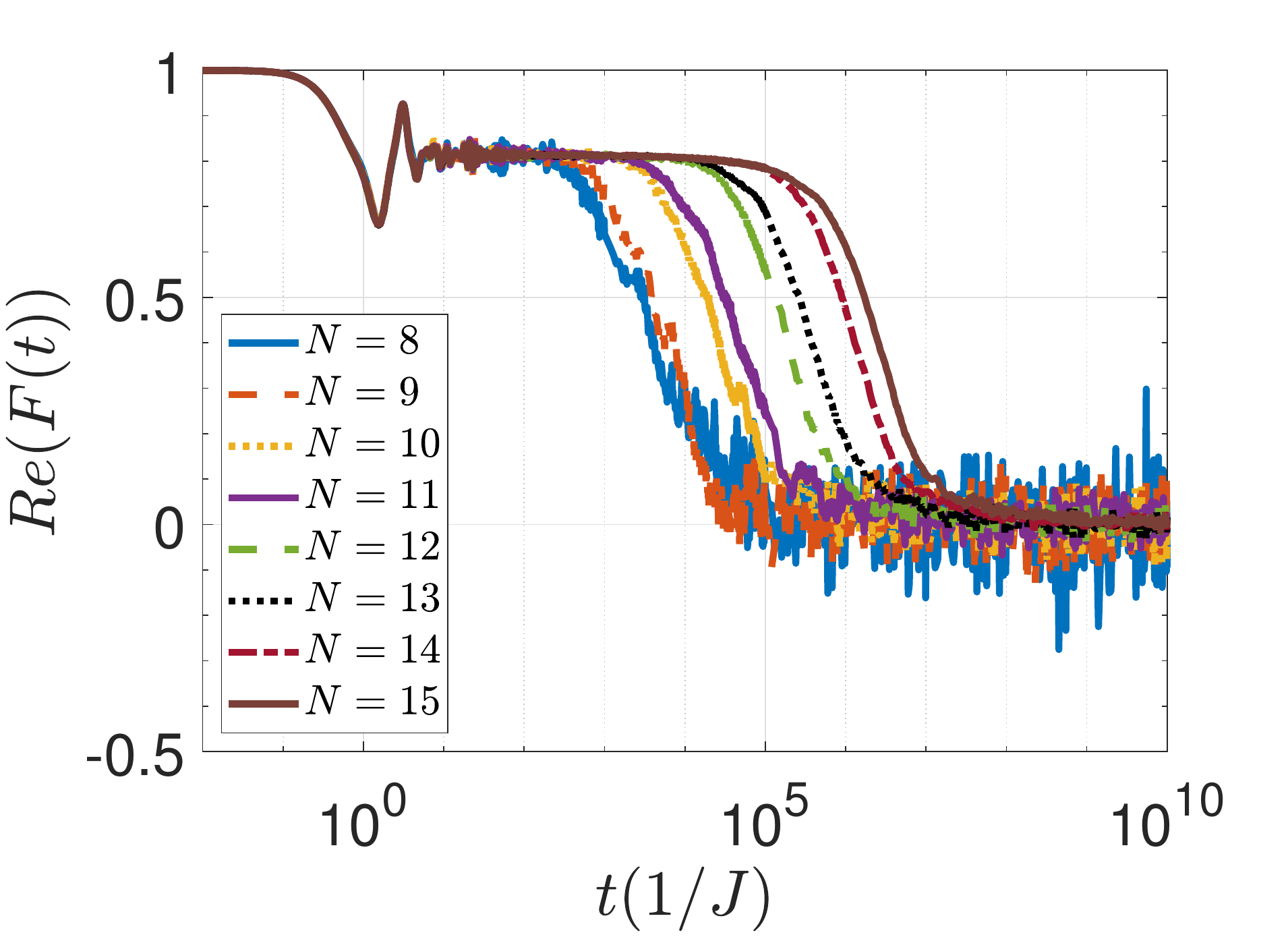}}\hfill  
\subfloat[]{\label{AFig5f}\includegraphics[width=0.24\textwidth]{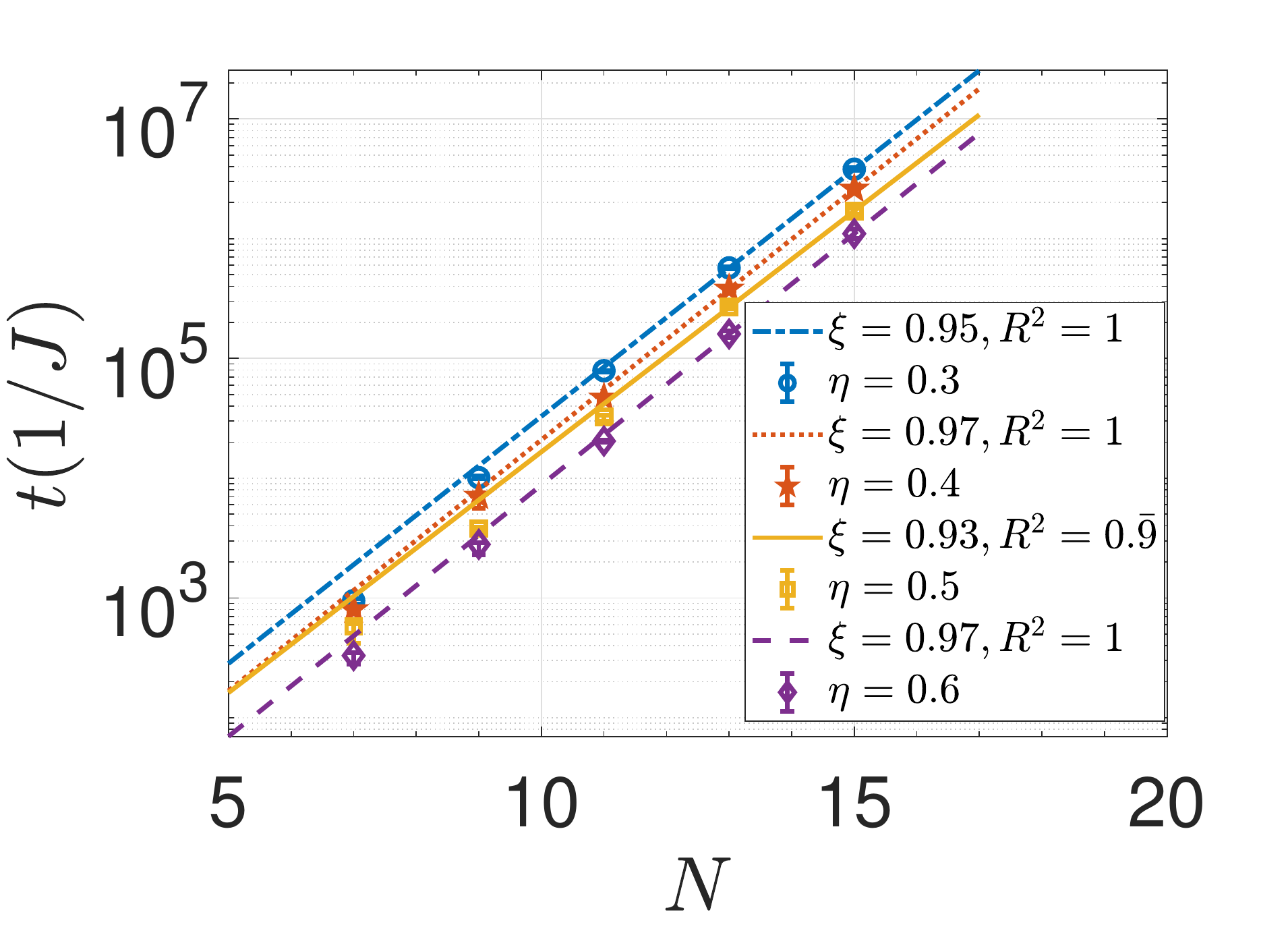}}\hfill
\caption{(a) Coherence times of the edge spins based on OTOC at $\Delta/J=-2$, deep in the topologically non-trivial phase $h/J=0.3$ and (b) the system-size scaling of the coherence times in (a). Note that different curves correspond to different threshold values $\eta$ where we look for the times that provide $F(t)=\eta$. $\xi$ is the exponent in the exponential scaling and all of them are around $\xi \sim 1$.}
\label{FigA5}
\end{figure}
The coherence times of the edge spins at $\Delta/J=-2$ deep in the non-trivial phase (Fig.~\ref{AFig5e}) exhibit exponential increase with the system size in Fig.~\ref{AFig5f} up to an apparent odd-even effect. All different scaling samples collapse at around $\xi\sim1$ for the exponent of the exponential scaling. While it is highly expected that this increase should slow down with bigger system sizes, based on our available data we cannot state that this behaviour is an example of prescrambling, instead it looks like a finite-size effect up until $N=15$ system size. Hence it is not always easy to extract a curve collapse to demonstrate prescrambling in systems with small sizes.

\begin{figure}
\centerline{\includegraphics[width=0.35\textwidth]{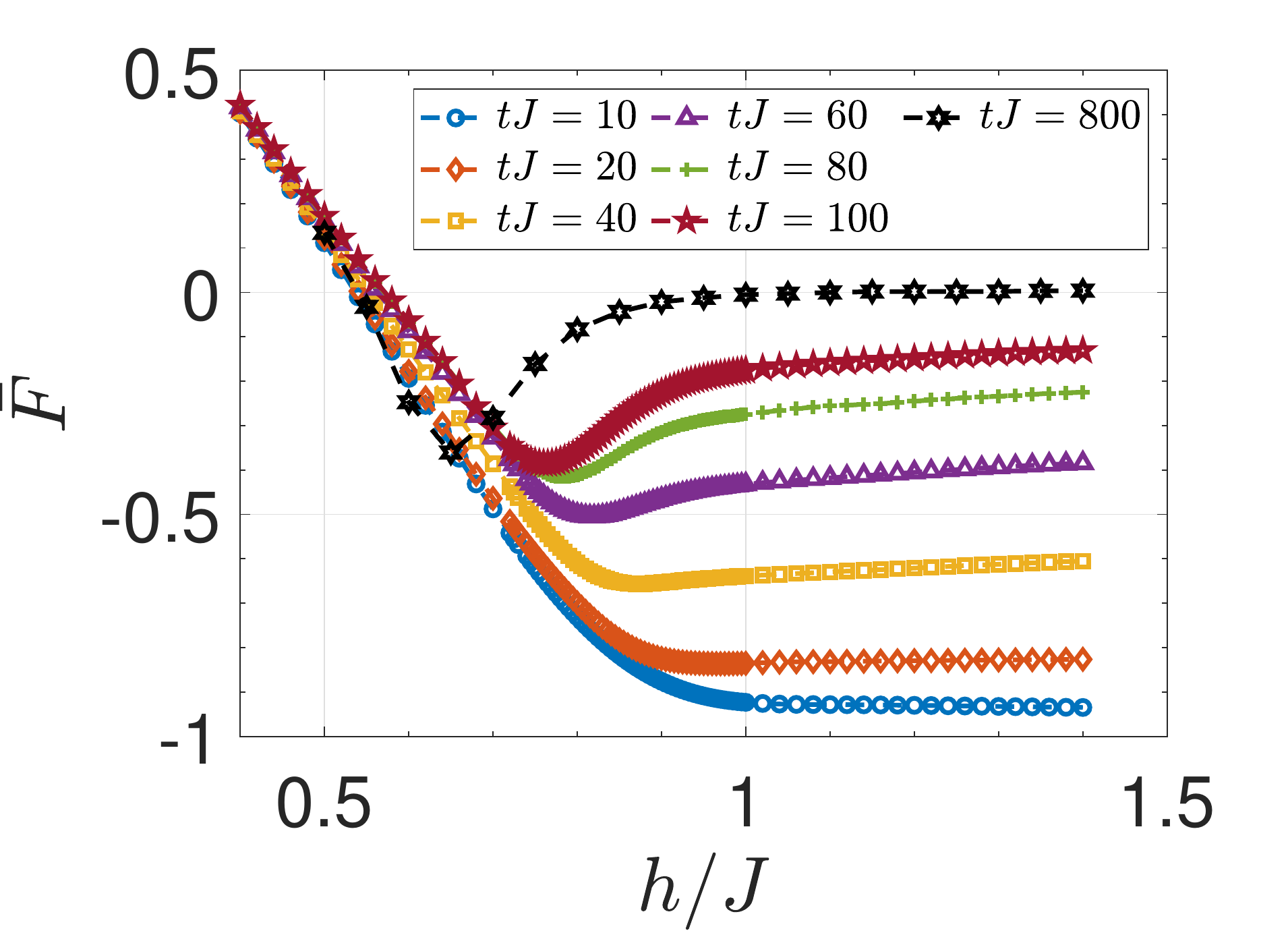}}
\caption{Demonstration of the time-dependence of the phase diagram for the model with $\Delta/J=-0.1$ at $N=14$ system size. Blue circles, orange diamonds, yellow squares, purple triangles, green pluses, red pentagrams and black hexagrams stand for $tJ=10, 20,40,60,80,100,800$, respectively.}
\label{AFig12}
\end{figure}
Fig.~\ref{AFig12} demonstrates the dependence of a dynamic phase diagram on the interval of time averaging. The data is for the model at near-integrability. The result with blue-circles that is computed in a short time interval of $tJ=10$ converges to the OTOC of non-interacting limit, while increasing the averaging time from $tJ=10$ to later times causes the phase diagram to change significantly. Hence in the short-time limit, the coherence times of the prescrambling plateau are significantly contributed not only by the diagonal contribution, but also the off-diagonal contribution. This additional contribution, that is specific to OTOC, in fact survives until very long times, e.g. $t \gtrsim 2\times 10^3$ (Fig.~\ref{Fig5b} in main text). However, farther away from the non-interacting limit the off-diagonal contribution vanishes faster, whereas the diagonal contribution remains for longer times.

\begin{figure}
\centering
\subfloat[]{\label{AFig10a}\includegraphics[width=0.24\textwidth]{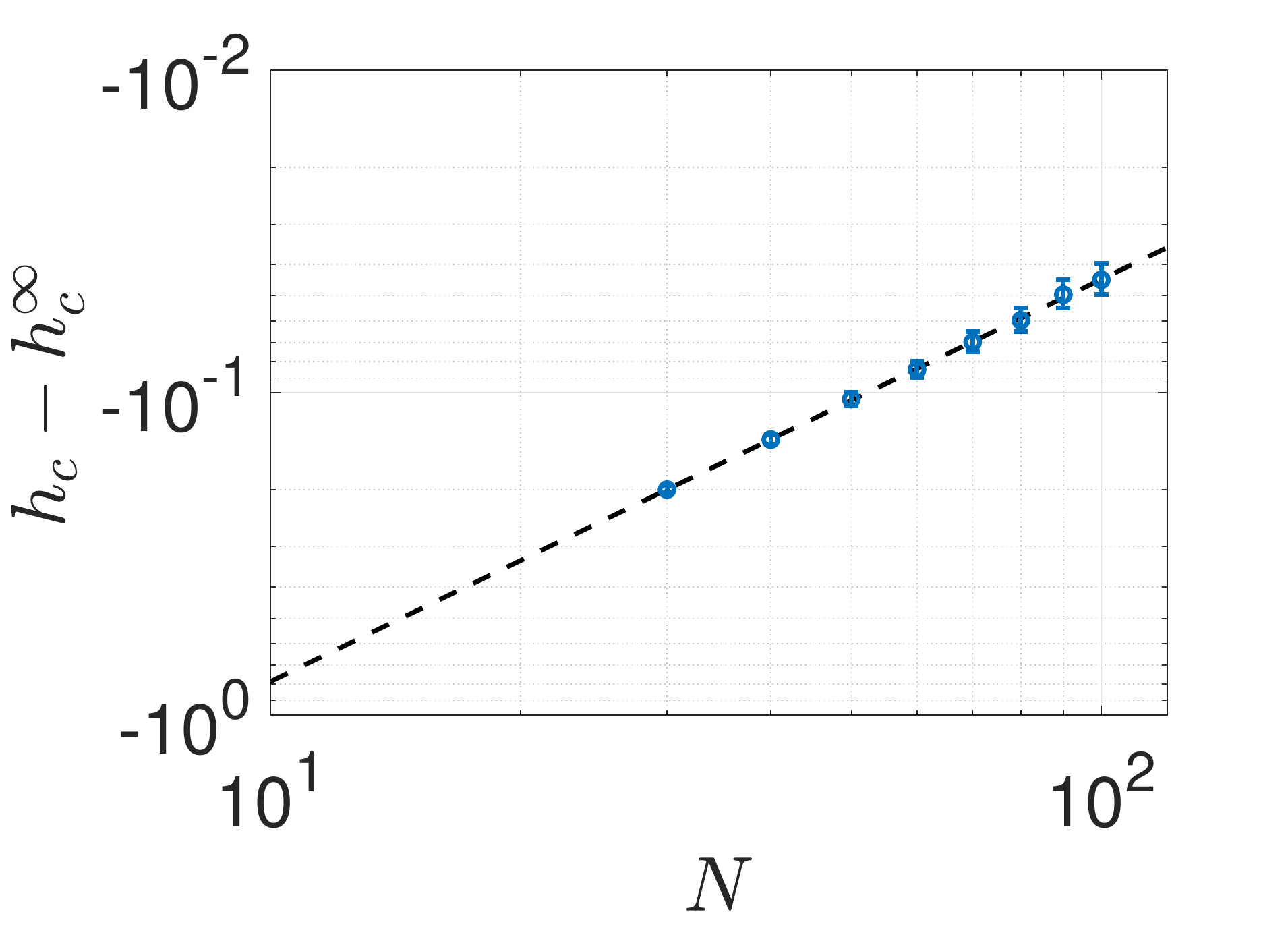}}\hfill 
\subfloat[]{\label{AFig10b}\includegraphics[width=0.24\textwidth]{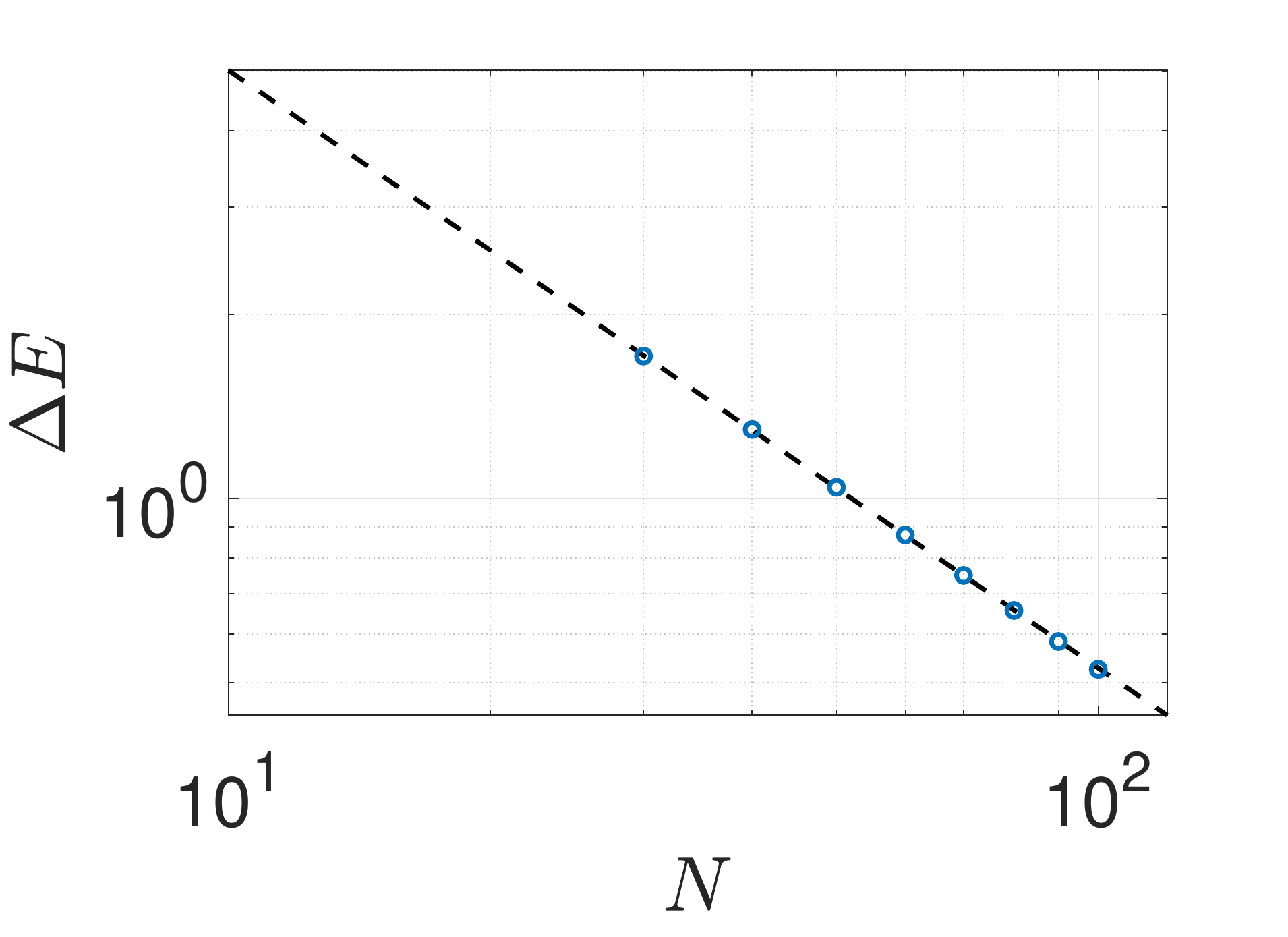}}\hfill 
\caption{The scaling parameters for the ground state phase transition of the model with $\Delta/J=-2$, calculated via DMRG. (a) The system-size scaling of the critical point, giving $h_c^{\infty}=3.7746$ in the thermodynamic limit. (b) The system-size scaling of the energy gap, giving an exponent of $\sim -1$ and showing that the gap closes in the thermodynamic limit.}
\label{AFig10}
\end{figure}
\begin{figure}
\centerline{\includegraphics[width=0.35\textwidth]{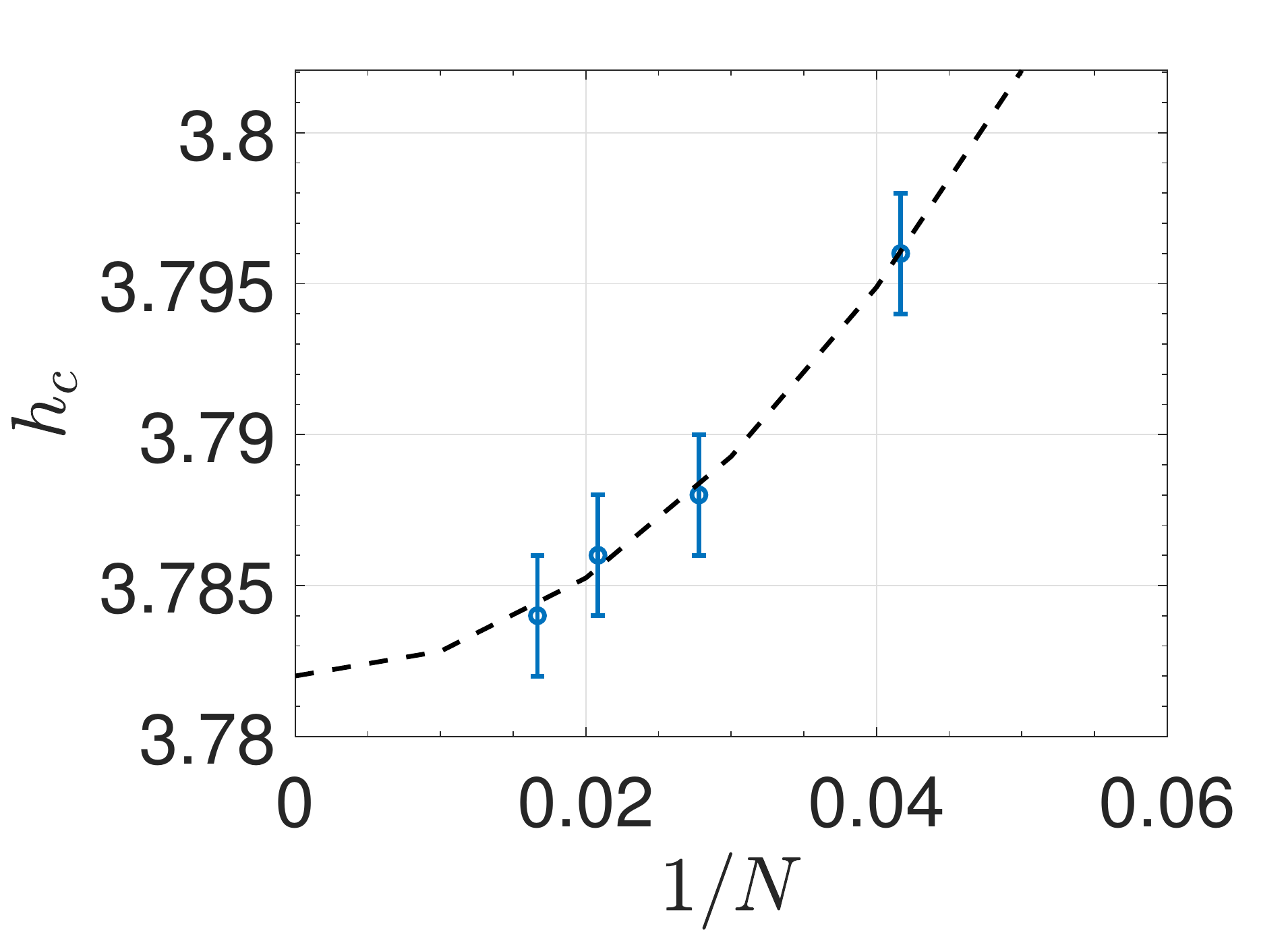}}
\caption{The Binder cumulant calculated for the Ising model with $\Delta/J=-2$. The system size scaling gives $h_c^{\infty}=3.782$.}
\label{AFig9}
\end{figure}
We mark the ground state phase transition point in the model with $\Delta/J=-2$ via (i) minimizing the energy gap at the transition point; and (ii) Binder cumulant. We first present (i): The scaling parameters for the transition point read $h_c \sim N^{-1.2467} + 3.7746$ where the transition point in the thermodynamic limit is found $h_c^{\infty}=3.7746$ with $R^2=0.9997$. The scaling parameters for the energy gap read $\Delta E \sim N^{-0.9775}$ with $R^2=0.9999$. So the system-size scaling exponent for the energy gap is close to $-1$. See Figs.~\ref{AFig10} for the scaling figures. (ii) Fig.~\ref{AFig9} shows the system size scaling of Binder cumulant, 
\begin{eqnarray}
U=\frac{3}{2}\left(1-\frac{1}{3}\frac{\Braket{S_z^4}}{\Braket{S_z^2}^2}  \right),
\end{eqnarray}
where $S_z=\sum_i^N \sigma^z_i$, the total magnetization operator. This method marks the phase boundary as $h_c^{\infty}=3.782$.

\section{Further results on the XXZ model \label{AppF}}

\renewcommand{\thefigure}{F\arabic{figure}}
\setcounter{figure}{0}  %  this will re-count eq from 1

\begin{figure}
\centering
\subfloat[]{\label{AFig8a}\includegraphics[width=0.24\textwidth]{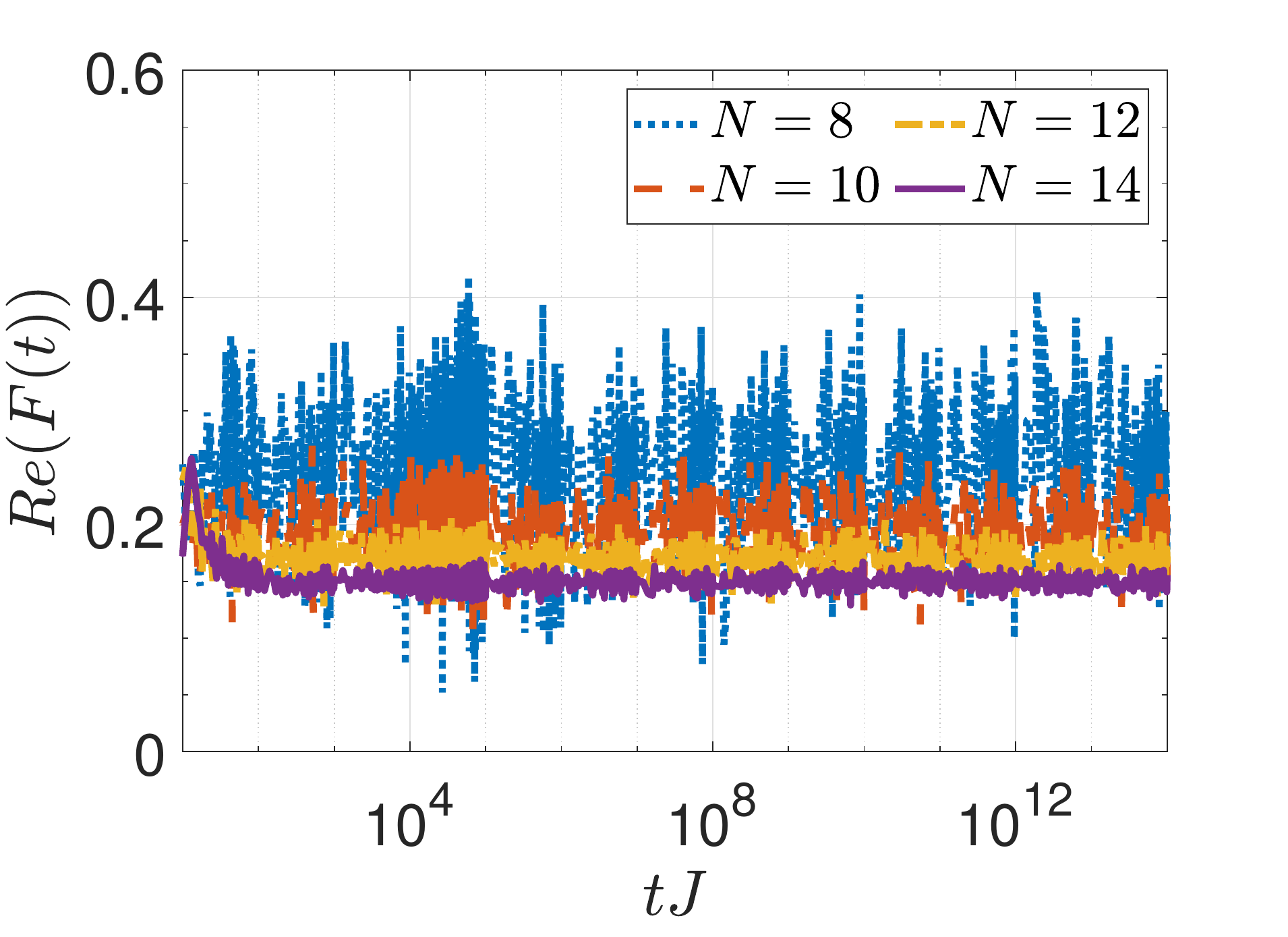}}\hfill 
\subfloat[]{\label{AFig8b}\includegraphics[width=0.24\textwidth]{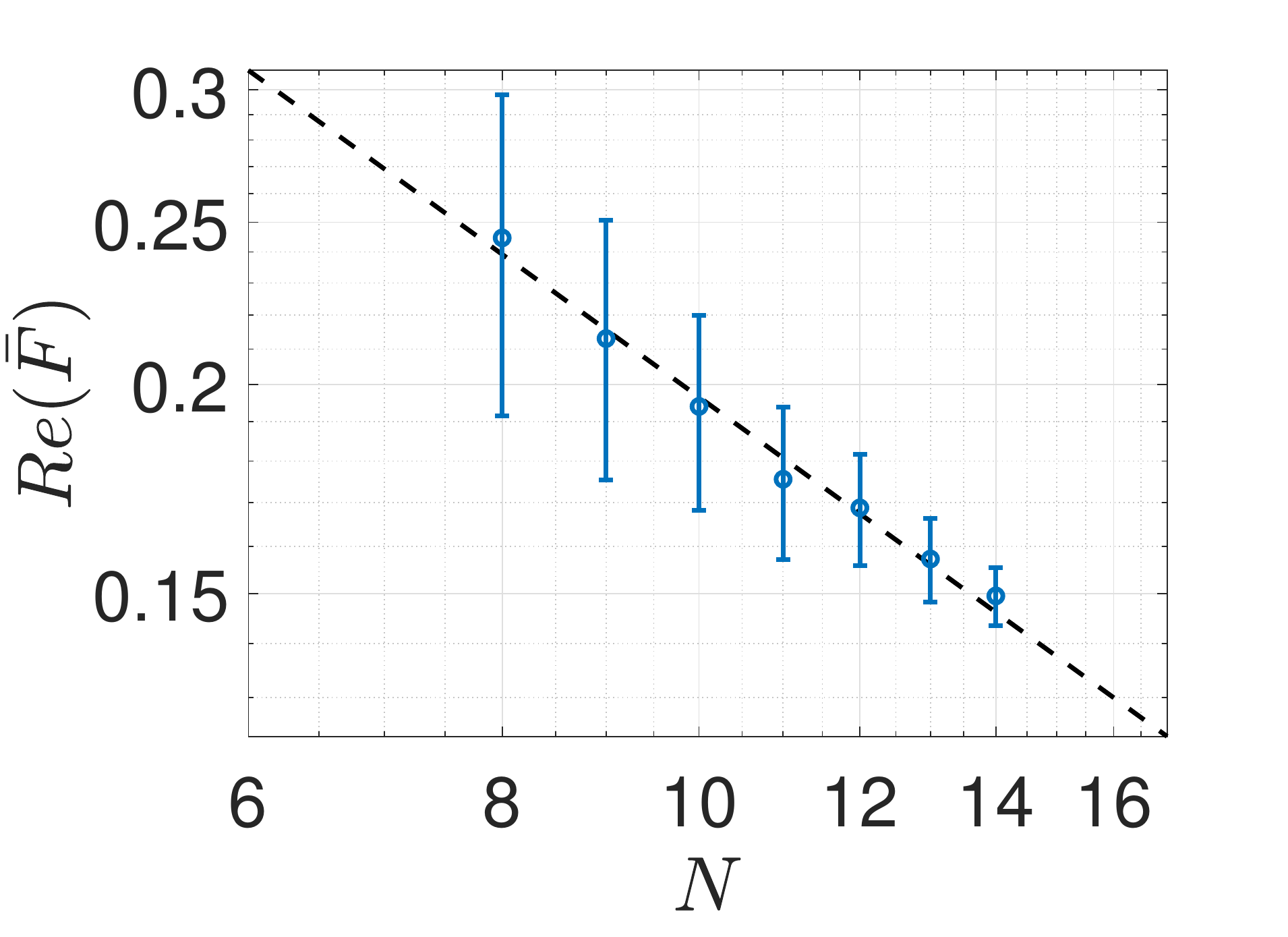}}\hfill 
\caption{(a) The saturation value for long times and different system sizes ($N=8$ to $N=14$) are plotted for the gapless phase of the XXZ model. (b) The system size scaling of the saturation value where the error bars show the extend of the oscillations around the average of the signals in (a). The scaling has a form of $Re(\bar{F}) \propto N^{-\xi}$ where $\xi \sim 0.9$.}
\label{AFig8}
\end{figure}

Fig.~\ref{AFig8} shows long-time dynamics of OTOC in the gapless phase of the XXZ model and how the time-average of this signal scales with the system size. We see the scaling has a form of $Re(\bar{F}) \propto N^{-\xi}$ where $\xi \sim 0.9$. Hence in the thermodynamic limit we expect $\bar{F} \rightarrow 0$ in the gapless phase. 

\begin{figure}
\centerline{\includegraphics[width=0.35\textwidth]{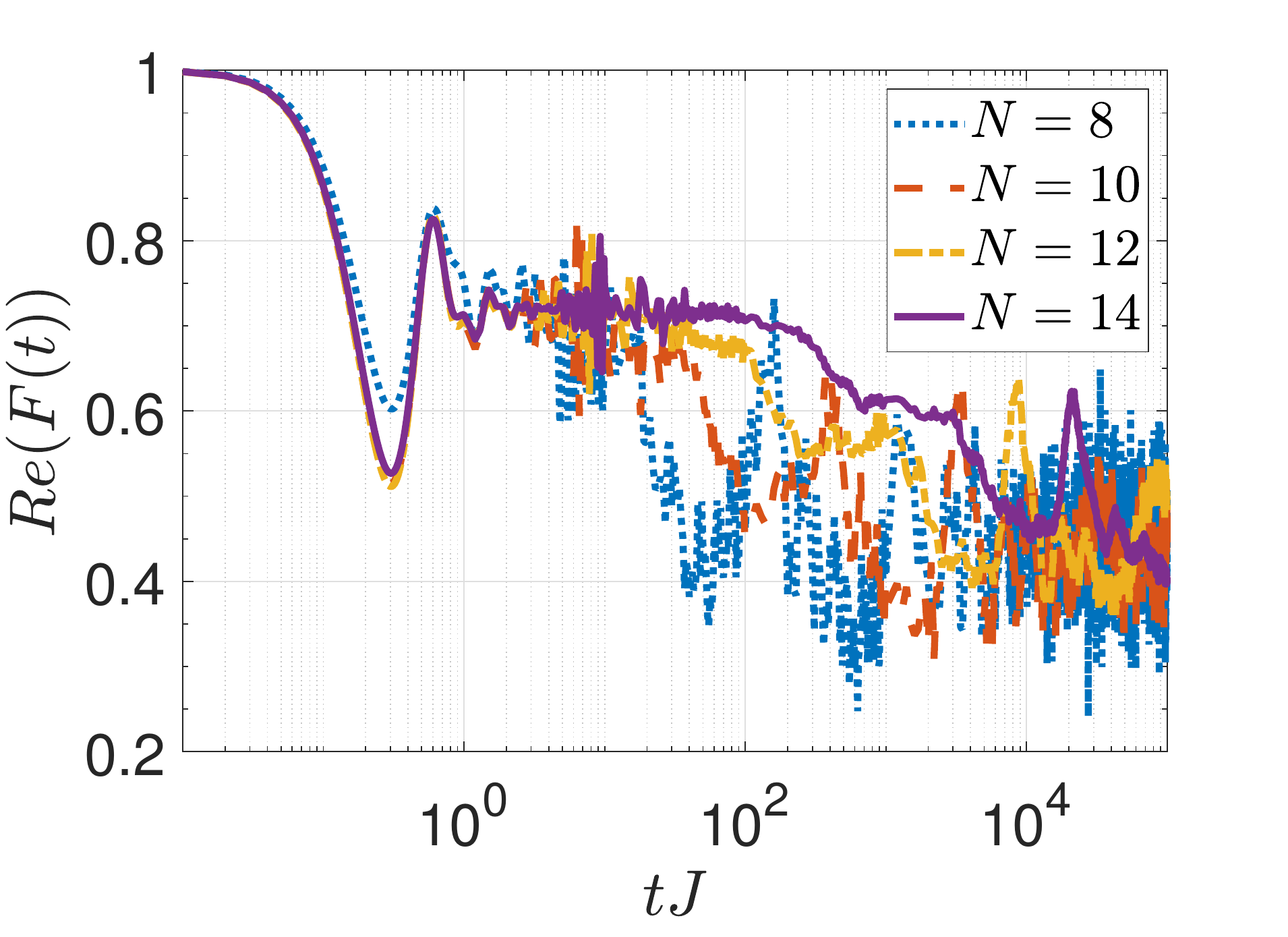}}
\caption{The coherence times of prescrambling in the gapped phase of the XXZ model, $J_z/J=5$ for different system sizes. The exponential increase in the prescrambling time intervals with the system size suggests that the scrambling seen is a finite-size effect.}
\label{AFig11}
\end{figure}

Fig.~\ref{AFig11} shows prescrambling time scales exponentially increase with the system size, a similar figure to Fig.~\ref{Fig5d} in the main text, however much closer to the transition boundary. The exponential increase in system size implies that the scrambling is a finite-size effect, hence in thermodynamic limit, prescrambling plateau should persist, giving $\bar{F} \neq 0$ in the topologically non-trivial gapped phase. 

\bibliographystyle{apsrev4-1}
%\bibliography{Bibliography} % The references (bibliography) information are stored in the file named "Bibliography.bib"


\end{document}